\documentclass[aps,prd,twocolumn,superscriptaddress,preprintnumbers,nofootinbib]{revtex4-2}
\usepackage{amsmath}
\usepackage{amssymb}
\usepackage{natbib}
\usepackage{graphicx}
\usepackage{epsf}
\usepackage{subfigure}
\usepackage{color}
\usepackage{threeparttable}
\usepackage{dcolumn}
\usepackage{comment}
\usepackage{epsfig}
\usepackage{xspace}
\usepackage{multirow}
\usepackage{makecell}
\usepackage{mathtools}
\usepackage{appendix}
\usepackage{booktabs}
\usepackage[
colorlinks=True,
allcolors=blue,
linktocpage=true,
]{hyperref}
\usepackage{latexsym}
\usepackage{array}
\usepackage{orcidlink}
\date{\today}

\DeclareGraphicsExtensions{.jpg,.pdf,.png,.eps,.ps}

\newcommand*{\TT}{\ensuremath{TT}}
\newcommand*{\EE}{\ensuremath{E\!E}}
\newcommand*{\TE}{\ensuremath{T\!E}}
\newcommand*{\EETE}{\ensuremath{T\!E/E\!E}}
\newcommand*{\TTTEEE}{\ensuremath{TT/T\!E/E\!E}}
\newcommand*{\lcdm}{$\Lambda$CDM}
\newcommand*{\planck}{\textit{Planck}}
\newcommand*{\WMAP}{\textit{WMAP}}
\newcommand*{\polarbear}{\textsc{polarbear}}
\newcommand*{\cosmopower}{\textit{CosmoPower}}

\newcommand*{\Neff}{\ensuremath{N_{\mathrm{eff}}}}
\newcommand*{\Yp}{\ensuremath{Y_{\mathrm{P}}}}

\newcommand*{\Hubble}{\ensuremath{H_0}}
\newcommand*{\kmsmpc}{\ensuremath{\mathrm{km\,s^{-1}\,Mpc^{-1}}}}

\newcommand{\RNum}[1]{\uppercase\expandafter{\romannumeral #1\relax}}

\newcommand*{\sqdeg}{\ensuremath{{\rm deg}^2}}

\defcitealias{dutcher21}{D21}
\defcitealias{riess20}{R20}

\definecolor{amber}{rgb}{1.0, 0.49, 0.0}

\newcommand{\skipt}[1]{}

\newcolumntype{C}[1]{>{\centering\let\newline\\\arraybackslash\hspace{0pt}}m{#1}}

\begin{document}

\title{A Measurement of the CMB Temperature Power Spectrum and Constraints on Cosmology from the SPT-3G 2018 \TTTEEE{} Data Set}

\affiliation{School of Physics, University of Melbourne, Parkville, VIC 3010, Australia}
\affiliation{Joseph Henry Laboratories of Physics, Jadwin Hall, Princeton University, Princeton, NJ 08544, USA}
\affiliation{Mullard Space Science Laboratory, University College London, Holmbury St. Mary, Dorking, Surrey, RH5 6NT, UK}
\affiliation{Sorbonne Universit\'{e}, CNRS, UMR 7095, Institut d'Astrophysique de Paris, 98 bis bd Arago, 75014 Paris, France}
\affiliation{School of Physics and Astronomy, Cardiff University, Cardiff CF24 3YB, United Kingdom}
\affiliation{Fermi National Accelerator Laboratory, MS209, P.O. Box 500, Batavia, IL, 60510, USA}
\affiliation{Kavli Institute for Cosmological Physics, University of Chicago, 5640 South Ellis Avenue, Chicago, IL, 60637, USA}
\affiliation{Department of Astronomy and Astrophysics, University of Chicago, 5640 South Ellis Avenue, Chicago, IL, 60637, USA}
\affiliation{Department of Astronomy, University of Illinois Urbana-Champaign, 1002 West Green Street, Urbana, IL, 61801, USA}
\affiliation{Center for AstroPhysical Surveys, National Center for Supercomputing Applications, Urbana, IL, 61801, USA}
\affiliation{High-Energy Physics Division, Argonne National Laboratory, 9700 South Cass Avenue., Lemont, IL, 60439, USA}
\affiliation{Kavli Institute for Particle Astrophysics and Cosmology, Stanford University, 452 Lomita Mall, Stanford, CA, 94305, USA}
\affiliation{Department of Physics, Stanford University, 382 Via Pueblo Mall, Stanford, CA, 94305, USA}
\affiliation{SLAC National Accelerator Laboratory, 2575 Sand Hill Road, Menlo Park, CA, 94025, USA}
\affiliation{Enrico Fermi Institute, University of Chicago, 5640 South Ellis Avenue, Chicago, IL, 60637, USA}
\affiliation{Department of Physics, University of Chicago, 5640 South Ellis Avenue, Chicago, IL, 60637, USA}
\affiliation{Department of Physics, University of California, Berkeley, CA, 94720, USA}
\affiliation{California Institute of Technology, 1200 East California Boulevard., Pasadena, CA, 91125, USA}
\affiliation{High Energy Accelerator Research Organization (KEK), Tsukuba, Ibaraki 305-0801, Japan}
\affiliation{Department of Physics and McGill Space Institute, McGill University, 3600 Rue University, Montreal, Quebec H3A 2T8, Canada}
\affiliation{Canadian Institute for Advanced Research, CIFAR Program in Gravity and the Extreme Universe, Toronto, ON, M5G 1Z8, Canada}
\affiliation{Department of Astrophysical and Planetary Sciences, University of Colorado, Boulder, CO, 80309, USA}
\affiliation{Department of Physics, University of Illinois Urbana-Champaign, 1110 West Green Street, Urbana, IL, 61801, USA}
\affiliation{Department of Physics and Astronomy, University of California, Los Angeles, CA, 90095, USA}
\affiliation{Department of Physics, Case Western Reserve University, Cleveland, OH, 44106, USA}
\affiliation{CASA, Department of Astrophysical and Planetary Sciences, University of Colorado, Boulder, CO, 80309, USA }
\affiliation{Department of Physics, University of Colorado, Boulder, CO, 80309, USA}
\affiliation{Department of Physics \& Astronomy, University of California, One Shields Avenue, Davis, CA 95616, USA}
\affiliation{Physics Division, Lawrence Berkeley National Laboratory, Berkeley, CA, 94720, USA}
\affiliation{Steward Observatory and Department of Astronomy, University of Arizona, 933 N. Cherry Ave., Tucson, AZ 85721, USA}
\affiliation{Dunlap Institute for Astronomy \& Astrophysics, University of Toronto, 50 St. George Street, Toronto, ON, M5S 3H4, Canada}
\affiliation{David A. Dunlap Department of Astronomy \& Astrophysics, University of Toronto, 50 St. George Street, Toronto, ON, M5S 3H4, Canada}
\affiliation{Materials Sciences Division, Argonne National Laboratory, 9700 South Cass Avenue, Lemont, IL, 60439, USA}
\affiliation{Three-Speed Logic, Inc., Victoria, B.C., V8S 3Z5, Canada}
\affiliation{Harvard-Smithsonian Center for Astrophysics, 60 Garden Street, Cambridge, MA, 02138, USA}
\affiliation{Department of Physics and Astronomy, Michigan State University, East Lansing, MI 48824, USA}
\author{L.~Balkenhol\,\orcidlink{0000-0001-6899-1873}}
\email[Corresponding author: ]{lbalkenhol@student.unimelb.edu.au}
\affiliation{School of Physics, University of Melbourne, Parkville, VIC 3010, Australia}
\author{D.~Dutcher\,\orcidlink{0000-0002-9962-2058}}
\affiliation{Joseph Henry Laboratories of Physics, Jadwin Hall, Princeton University, Princeton, NJ 08544, USA}
\author{A.~Spurio Mancini\,\orcidlink{0000-0001-5698-0990}}
\affiliation{Mullard Space Science Laboratory, University College London, Holmbury St. Mary, Dorking, Surrey, RH5 6NT, UK}
\author{A.~Doussot}
\affiliation{Sorbonne Universit\'{e}, CNRS, UMR 7095, Institut d'Astrophysique de Paris, 98 bis bd Arago, 75014 Paris, France}
\author{K.~Benabed}
\affiliation{Sorbonne Universit\'{e}, CNRS, UMR 7095, Institut d'Astrophysique de Paris, 98 bis bd Arago, 75014 Paris, France}
\author{S.~Galli}
\affiliation{Sorbonne Universit\'{e}, CNRS, UMR 7095, Institut d'Astrophysique de Paris, 98 bis bd Arago, 75014 Paris, France}
\author{P.~A.~R.~Ade}
\affiliation{School of Physics and Astronomy, Cardiff University, Cardiff CF24 3YB, United Kingdom}
\author{A.~J.~Anderson\,\orcidlink{0000-0002-4435-4623}}
\affiliation{Fermi National Accelerator Laboratory, MS209, P.O. Box 500, Batavia, IL, 60510, USA}
\affiliation{Kavli Institute for Cosmological Physics, University of Chicago, 5640 South Ellis Avenue, Chicago, IL, 60637, USA}
\affiliation{Department of Astronomy and Astrophysics, University of Chicago, 5640 South Ellis Avenue, Chicago, IL, 60637, USA}
\author{B.~Ansarinejad}
\affiliation{School of Physics, University of Melbourne, Parkville, VIC 3010, Australia}
\author{M.~Archipley\,\orcidlink{0000-0002-0517-9842}}
\affiliation{Department of Astronomy, University of Illinois Urbana-Champaign, 1002 West Green Street, Urbana, IL, 61801, USA}
\affiliation{Center for AstroPhysical Surveys, National Center for Supercomputing Applications, Urbana, IL, 61801, USA}
\author{A.~N.~Bender\,\orcidlink{0000-0001-5868-0748}}
\affiliation{High-Energy Physics Division, Argonne National Laboratory, 9700 South Cass Avenue., Lemont, IL, 60439, USA}
\affiliation{Kavli Institute for Cosmological Physics, University of Chicago, 5640 South Ellis Avenue, Chicago, IL, 60637, USA}
\affiliation{Department of Astronomy and Astrophysics, University of Chicago, 5640 South Ellis Avenue, Chicago, IL, 60637, USA}
\author{B.~A.~Benson\,\orcidlink{0000-0002-5108-6823}}
\affiliation{Fermi National Accelerator Laboratory, MS209, P.O. Box 500, Batavia, IL, 60510, USA}
\affiliation{Kavli Institute for Cosmological Physics, University of Chicago, 5640 South Ellis Avenue, Chicago, IL, 60637, USA}
\affiliation{Department of Astronomy and Astrophysics, University of Chicago, 5640 South Ellis Avenue, Chicago, IL, 60637, USA}
\author{F.~Bianchini\,\orcidlink{0000-0003-4847-3483}}
\affiliation{Kavli Institute for Particle Astrophysics and Cosmology, Stanford University, 452 Lomita Mall, Stanford, CA, 94305, USA}
\affiliation{Department of Physics, Stanford University, 382 Via Pueblo Mall, Stanford, CA, 94305, USA}
\affiliation{SLAC National Accelerator Laboratory, 2575 Sand Hill Road, Menlo Park, CA, 94025, USA}
\author{L.~E.~Bleem\,\orcidlink{0000-0001-7665-5079}}
\affiliation{High-Energy Physics Division, Argonne National Laboratory, 9700 South Cass Avenue., Lemont, IL, 60439, USA}
\affiliation{Kavli Institute for Cosmological Physics, University of Chicago, 5640 South Ellis Avenue, Chicago, IL, 60637, USA}
\author{F.~R.~Bouchet\,\orcidlink{0000-0002-8051-2924}}
\affiliation{Sorbonne Universit\'{e}, CNRS, UMR 7095, Institut d'Astrophysique de Paris, 98 bis bd Arago, 75014 Paris, France}
\author{L.~Bryant}
\affiliation{Enrico Fermi Institute, University of Chicago, 5640 South Ellis Avenue, Chicago, IL, 60637, USA}
\author{E.~Camphuis\,\orcidlink{0000-0003-3483-8461}}
\affiliation{Sorbonne Universit\'{e}, CNRS, UMR 7095, Institut d'Astrophysique de Paris, 98 bis bd Arago, 75014 Paris, France}
\author{J.~E.~Carlstrom}
\affiliation{Kavli Institute for Cosmological Physics, University of Chicago, 5640 South Ellis Avenue, Chicago, IL, 60637, USA}
\affiliation{Enrico Fermi Institute, University of Chicago, 5640 South Ellis Avenue, Chicago, IL, 60637, USA}
\affiliation{Department of Physics, University of Chicago, 5640 South Ellis Avenue, Chicago, IL, 60637, USA}
\affiliation{High-Energy Physics Division, Argonne National Laboratory, 9700 South Cass Avenue., Lemont, IL, 60439, USA}
\affiliation{Department of Astronomy and Astrophysics, University of Chicago, 5640 South Ellis Avenue, Chicago, IL, 60637, USA}
\author{T.~W.~Cecil\,\orcidlink{0000-0002-7019-5056}}
\affiliation{High-Energy Physics Division, Argonne National Laboratory, 9700 South Cass Avenue., Lemont, IL, 60439, USA}
\author{C.~L.~Chang}
\affiliation{High-Energy Physics Division, Argonne National Laboratory, 9700 South Cass Avenue., Lemont, IL, 60439, USA}
\affiliation{Kavli Institute for Cosmological Physics, University of Chicago, 5640 South Ellis Avenue, Chicago, IL, 60637, USA}
\affiliation{Department of Astronomy and Astrophysics, University of Chicago, 5640 South Ellis Avenue, Chicago, IL, 60637, USA}
\author{P.~Chaubal}
\affiliation{School of Physics, University of Melbourne, Parkville, VIC 3010, Australia}
\author{P.~M.~Chichura\,\orcidlink{0000-0002-5397-9035}}
\affiliation{Department of Physics, University of Chicago, 5640 South Ellis Avenue, Chicago, IL, 60637, USA}
\affiliation{Kavli Institute for Cosmological Physics, University of Chicago, 5640 South Ellis Avenue, Chicago, IL, 60637, USA}
\author{T.-L.~Chou}
\affiliation{Department of Physics, University of Chicago, 5640 South Ellis Avenue, Chicago, IL, 60637, USA}
\affiliation{Kavli Institute for Cosmological Physics, University of Chicago, 5640 South Ellis Avenue, Chicago, IL, 60637, USA}
\author{A.~Coerver}
\affiliation{Department of Physics, University of California, Berkeley, CA, 94720, USA}
\author{T.~M.~Crawford\,\orcidlink{0000-0001-9000-5013}}
\affiliation{Kavli Institute for Cosmological Physics, University of Chicago, 5640 South Ellis Avenue, Chicago, IL, 60637, USA}
\affiliation{Department of Astronomy and Astrophysics, University of Chicago, 5640 South Ellis Avenue, Chicago, IL, 60637, USA}
\author{A.~Cukierman\,\orcidlink{0000-0002-7471-719X}}
\affiliation{Kavli Institute for Particle Astrophysics and Cosmology, Stanford University, 452 Lomita Mall, Stanford, CA, 94305, USA}
\affiliation{SLAC National Accelerator Laboratory, 2575 Sand Hill Road, Menlo Park, CA, 94025, USA}
\affiliation{Department of Physics, Stanford University, 382 Via Pueblo Mall, Stanford, CA, 94305, USA}
\affiliation{California Institute of Technology, 1200 East California Boulevard., Pasadena, CA, 91125, USA}
\author{C.~Daley\,\orcidlink{0000-0002-3760-2086}}
\affiliation{Department of Astronomy, University of Illinois Urbana-Champaign, 1002 West Green Street, Urbana, IL, 61801, USA}
\author{T.~de~Haan}
\affiliation{High Energy Accelerator Research Organization (KEK), Tsukuba, Ibaraki 305-0801, Japan}
\author{K.~R.~Dibert}
\affiliation{Department of Astronomy and Astrophysics, University of Chicago, 5640 South Ellis Avenue, Chicago, IL, 60637, USA}
\affiliation{Kavli Institute for Cosmological Physics, University of Chicago, 5640 South Ellis Avenue, Chicago, IL, 60637, USA}
\author{M.~A.~Dobbs}
\affiliation{Department of Physics and McGill Space Institute, McGill University, 3600 Rue University, Montreal, Quebec H3A 2T8, Canada}
\affiliation{Canadian Institute for Advanced Research, CIFAR Program in Gravity and the Extreme Universe, Toronto, ON, M5G 1Z8, Canada}
\author{W.~Everett}
\affiliation{Department of Astrophysical and Planetary Sciences, University of Colorado, Boulder, CO, 80309, USA}
\author{C.~Feng}
\affiliation{Department of Physics, University of Illinois Urbana-Champaign, 1110 West Green Street, Urbana, IL, 61801, USA}
\author{K.~R.~Ferguson\,\orcidlink{0000-0002-4928-8813}}
\affiliation{Department of Physics and Astronomy, University of California, Los Angeles, CA, 90095, USA}
\author{A.~Foster\,\orcidlink{0000-0002-7145-1824}}
\affiliation{Department of Physics, Case Western Reserve University, Cleveland, OH, 44106, USA}
\author{A.~E.~Gambrel}
\affiliation{Kavli Institute for Cosmological Physics, University of Chicago, 5640 South Ellis Avenue, Chicago, IL, 60637, USA}
\author{R.~W.~Gardner}
\affiliation{Enrico Fermi Institute, University of Chicago, 5640 South Ellis Avenue, Chicago, IL, 60637, USA}
\author{N.~Goeckner-Wald}
\affiliation{Department of Physics, Stanford University, 382 Via Pueblo Mall, Stanford, CA, 94305, USA}
\affiliation{Kavli Institute for Particle Astrophysics and Cosmology, Stanford University, 452 Lomita Mall, Stanford, CA, 94305, USA}
\author{R.~Gualtieri\,\orcidlink{0000-0003-4245-2315}}
\affiliation{High-Energy Physics Division, Argonne National Laboratory, 9700 South Cass Avenue., Lemont, IL, 60439, USA}
\author{F.~Guidi\,\orcidlink{0000-0001-7593-3962}}
\affiliation{Sorbonne Universit\'{e}, CNRS, UMR 7095, Institut d'Astrophysique de Paris, 98 bis bd Arago, 75014 Paris, France}
\author{S.~Guns}
\affiliation{Department of Physics, University of California, Berkeley, CA, 94720, USA}
\author{N.~W.~Halverson}
\affiliation{CASA, Department of Astrophysical and Planetary Sciences, University of Colorado, Boulder, CO, 80309, USA }
\affiliation{Department of Physics, University of Colorado, Boulder, CO, 80309, USA}
\author{E.~Hivon\,\orcidlink{0000-0003-1880-2733}}
\affiliation{Sorbonne Universit\'{e}, CNRS, UMR 7095, Institut d'Astrophysique de Paris, 98 bis bd Arago, 75014 Paris, France}
\author{G.~P.~Holder\,\orcidlink{0000-0002-0463-6394}}
\affiliation{Department of Physics, University of Illinois Urbana-Champaign, 1110 West Green Street, Urbana, IL, 61801, USA}
\author{W.~L.~Holzapfel}
\affiliation{Department of Physics, University of California, Berkeley, CA, 94720, USA}
\author{J.~C.~Hood}
\affiliation{Kavli Institute for Cosmological Physics, University of Chicago, 5640 South Ellis Avenue, Chicago, IL, 60637, USA}
\author{N.~Huang}
\affiliation{Department of Physics, University of California, Berkeley, CA, 94720, USA}
\author{L.~Knox}
\affiliation{Department of Physics \& Astronomy, University of California, One Shields Avenue, Davis, CA 95616, USA}
\author{M.~Korman}
\affiliation{Department of Physics, Case Western Reserve University, Cleveland, OH, 44106, USA}
\author{C.-L.~Kuo}
\affiliation{Kavli Institute for Particle Astrophysics and Cosmology, Stanford University, 452 Lomita Mall, Stanford, CA, 94305, USA}
\affiliation{Department of Physics, Stanford University, 382 Via Pueblo Mall, Stanford, CA, 94305, USA}
\affiliation{SLAC National Accelerator Laboratory, 2575 Sand Hill Road, Menlo Park, CA, 94025, USA}
\author{A.~T.~Lee}
\affiliation{Department of Physics, University of California, Berkeley, CA, 94720, USA}
\affiliation{Physics Division, Lawrence Berkeley National Laboratory, Berkeley, CA, 94720, USA}
\author{A.~E.~Lowitz}
\affiliation{Kavli Institute for Cosmological Physics, University of Chicago, 5640 South Ellis Avenue, Chicago, IL, 60637, USA}
\affiliation{Steward Observatory and Department of Astronomy, University of Arizona, 933 N. Cherry Ave., Tucson, AZ 85721, USA}
\author{C.~Lu}
\affiliation{Department of Physics, University of Illinois Urbana-Champaign, 1110 West Green Street, Urbana, IL, 61801, USA}
\author{M.~Millea\,\orcidlink{0000-0001-7317-0551}}
\affiliation{Department of Physics, University of California, Berkeley, CA, 94720, USA}
\author{J.~Montgomery}
\affiliation{Department of Physics and McGill Space Institute, McGill University, 3600 Rue University, Montreal, Quebec H3A 2T8, Canada}
\author{Y.~Nakato}
\affiliation{Department of Physics, Stanford University, 382 Via Pueblo Mall, Stanford, CA, 94305, USA}
\author{T.~Natoli}
\affiliation{Kavli Institute for Cosmological Physics, University of Chicago, 5640 South Ellis Avenue, Chicago, IL, 60637, USA}
\affiliation{Department of Astronomy and Astrophysics, University of Chicago, 5640 South Ellis Avenue, Chicago, IL, 60637, USA}
\author{G.~I.~Noble\,\orcidlink{0000-0002-5254-243X}}
\affiliation{Dunlap Institute for Astronomy \& Astrophysics, University of Toronto, 50 St. George Street, Toronto, ON, M5S 3H4, Canada}
\affiliation{David A. Dunlap Department of Astronomy \& Astrophysics, University of Toronto, 50 St. George Street, Toronto, ON, M5S 3H4, Canada}
\author{V.~Novosad}
\affiliation{Materials Sciences Division, Argonne National Laboratory, 9700 South Cass Avenue, Lemont, IL, 60439, USA}
\author{Y.~Omori}
\affiliation{Department of Astronomy and Astrophysics, University of Chicago, 5640 South Ellis Avenue, Chicago, IL, 60637, USA}
\affiliation{Kavli Institute for Cosmological Physics, University of Chicago, 5640 South Ellis Avenue, Chicago, IL, 60637, USA}
\author{S.~Padin}
\affiliation{Kavli Institute for Cosmological Physics, University of Chicago, 5640 South Ellis Avenue, Chicago, IL, 60637, USA}
\affiliation{California Institute of Technology, 1200 East California Boulevard., Pasadena, CA, 91125, USA}
\author{Z.~Pan}
\affiliation{High-Energy Physics Division, Argonne National Laboratory, 9700 South Cass Avenue., Lemont, IL, 60439, USA}
\affiliation{Kavli Institute for Cosmological Physics, University of Chicago, 5640 South Ellis Avenue, Chicago, IL, 60637, USA}
\affiliation{Department of Physics, University of Chicago, 5640 South Ellis Avenue, Chicago, IL, 60637, USA}
\author{P.~Paschos}
\affiliation{Enrico Fermi Institute, University of Chicago, 5640 South Ellis Avenue, Chicago, IL, 60637, USA}
\author{K.~Prabhu}
\affiliation{Department of Physics \& Astronomy, University of California, One Shields Avenue, Davis, CA 95616, USA}
\author{W.~Quan}
\affiliation{Department of Physics, University of Chicago, 5640 South Ellis Avenue, Chicago, IL, 60637, USA}
\affiliation{Kavli Institute for Cosmological Physics, University of Chicago, 5640 South Ellis Avenue, Chicago, IL, 60637, USA}
\author{M.~Rahimi}
\affiliation{School of Physics, University of Melbourne, Parkville, VIC 3010, Australia}
\author{A.~Rahlin\,\orcidlink{0000-0003-3953-1776}}
\affiliation{Fermi National Accelerator Laboratory, MS209, P.O. Box 500, Batavia, IL, 60510, USA}
\affiliation{Kavli Institute for Cosmological Physics, University of Chicago, 5640 South Ellis Avenue, Chicago, IL, 60637, USA}
\author{C.~L.~Reichardt\,\orcidlink{0000-0003-2226-9169}}
\affiliation{School of Physics, University of Melbourne, Parkville, VIC 3010, Australia}
\author{M.~Rouble}
\affiliation{Department of Physics and McGill Space Institute, McGill University, 3600 Rue University, Montreal, Quebec H3A 2T8, Canada}
\author{J.~E.~Ruhl}
\affiliation{Department of Physics, Case Western Reserve University, Cleveland, OH, 44106, USA}
\author{E.~Schiappucci}
\affiliation{School of Physics, University of Melbourne, Parkville, VIC 3010, Australia}
\author{G.~Smecher}
\affiliation{Three-Speed Logic, Inc., Victoria, B.C., V8S 3Z5, Canada}
\author{J.~A.~Sobrin\,\orcidlink{0000-0001-6155-5315}}
\affiliation{Fermi National Accelerator Laboratory, MS209, P.O. Box 500, Batavia, IL, 60510, USA}
\affiliation{Kavli Institute for Cosmological Physics, University of Chicago, 5640 South Ellis Avenue, Chicago, IL, 60637, USA}
\author{A.~A.~Stark}
\affiliation{Harvard-Smithsonian Center for Astrophysics, 60 Garden Street, Cambridge, MA, 02138, USA}
\author{J.~Stephen}
\affiliation{Enrico Fermi Institute, University of Chicago, 5640 South Ellis Avenue, Chicago, IL, 60637, USA}
\author{A.~Suzuki}
\affiliation{Physics Division, Lawrence Berkeley National Laboratory, Berkeley, CA, 94720, USA}
\author{C.~Tandoi}
\affiliation{Department of Astronomy, University of Illinois Urbana-Champaign, 1002 West Green Street, Urbana, IL, 61801, USA}
\author{K.~L.~Thompson}
\affiliation{Kavli Institute for Particle Astrophysics and Cosmology, Stanford University, 452 Lomita Mall, Stanford, CA, 94305, USA}
\affiliation{Department of Physics, Stanford University, 382 Via Pueblo Mall, Stanford, CA, 94305, USA}
\affiliation{SLAC National Accelerator Laboratory, 2575 Sand Hill Road, Menlo Park, CA, 94025, USA}
\author{B.~Thorne}
\affiliation{Department of Physics \& Astronomy, University of California, One Shields Avenue, Davis, CA 95616, USA}
\author{C.~Tucker}
\affiliation{School of Physics and Astronomy, Cardiff University, Cardiff CF24 3YB, United Kingdom}
\author{C.~Umilta\,\orcidlink{0000-0002-6805-6188}}
\affiliation{Department of Physics, University of Illinois Urbana-Champaign, 1110 West Green Street, Urbana, IL, 61801, USA}
\author{J.~D.~Vieira}
\affiliation{Department of Astronomy, University of Illinois Urbana-Champaign, 1002 West Green Street, Urbana, IL, 61801, USA}
\affiliation{Department of Physics, University of Illinois Urbana-Champaign, 1110 West Green Street, Urbana, IL, 61801, USA}
\affiliation{Center for AstroPhysical Surveys, National Center for Supercomputing Applications, Urbana, IL, 61801, USA}
\author{G.~Wang}
\affiliation{High-Energy Physics Division, Argonne National Laboratory, 9700 South Cass Avenue., Lemont, IL, 60439, USA}
\author{N.~Whitehorn\,\orcidlink{0000-0002-3157-0407}}
\affiliation{Department of Physics and Astronomy, Michigan State University, East Lansing, MI 48824, USA}
\author{W.~L.~K.~Wu\,\orcidlink{0000-0001-5411-6920}}
\affiliation{Kavli Institute for Particle Astrophysics and Cosmology, Stanford University, 452 Lomita Mall, Stanford, CA, 94305, USA}
\affiliation{SLAC National Accelerator Laboratory, 2575 Sand Hill Road, Menlo Park, CA, 94025, USA}
\author{V.~Yefremenko}
\affiliation{High-Energy Physics Division, Argonne National Laboratory, 9700 South Cass Avenue., Lemont, IL, 60439, USA}
\author{M.~R.~Young}
\affiliation{Fermi National Accelerator Laboratory, MS209, P.O. Box 500, Batavia, IL, 60510, USA}
\affiliation{Kavli Institute for Cosmological Physics, University of Chicago, 5640 South Ellis Avenue, Chicago, IL, 60637, USA}
\author{J.~A.~Zebrowski}
\affiliation{Kavli Institute for Cosmological Physics, University of Chicago, 5640 South Ellis Avenue, Chicago, IL, 60637, USA}
\affiliation{Department of Astronomy and Astrophysics, University of Chicago, 5640 South Ellis Avenue, Chicago, IL, 60637, USA}
\affiliation{Fermi National Accelerator Laboratory, MS209, P.O. Box 500, Batavia, IL, 60510, USA}
\affiliation{Department of Physics, University of California, Berkeley, CA, 94720, USA}
\collaboration{SPT-3G Collaboration}
\noaffiliation

\begin{abstract}
We present a sample-variance-limited measurement of the temperature power spectrum (\TT{}) of the cosmic microwave background (CMB) using observations of a $\sim\! 1500\,\sqdeg$ field made by SPT-3G in 2018.
We report multifrequency power spectrum measurements at $95$, $150$, and $220\,\mathrm{GHz}$ covering the angular multipole range $750 \leq \ell < 3000$.
We combine this \TT{} measurement with the published polarization power spectrum measurements from the 2018 observing season and update their associated covariance matrix to complete the SPT-3G 2018 \TTTEEE{} data set.
This is the first analysis to present cosmological constraints from SPT \TT{}, \TE{}, and \EE{} power spectrum measurements jointly.
We blind the cosmological results and subject the data set to a series of consistency tests at the power spectrum and parameter level.
We find excellent agreement between frequencies and spectrum types and our results are robust to the modeling of astrophysical foregrounds.
We report results for \lcdm{} and a series of extensions, drawing on the following parameters: the amplitude of the gravitational lensing effect on primary power spectra $A_\mathrm{L}$, the effective number of neutrino species $\Neff{}$, the primordial helium abundance $\Yp{}$, and the baryon clumping factor due to primordial magnetic fields $b$.
We find that the SPT-3G 2018 \TTTEEE{} data are well fit by \lcdm{} with a probability-to-exceed of $15\%$.
For \lcdm{}, we constrain the expansion rate today to $H_0 = 68.3 \pm 1.5\,\kmsmpc{}$ and the combined structure growth parameter to $S_8 = 0.797 \pm 0.042$.
The SPT-based results are effectively independent of \planck{}, and the cosmological parameter constraints from either data set are within $<1\,\sigma$ of each other.
The addition of temperature data to the SPT-3G \EETE{} power spectra improves constraints by $8-27\%$ for each of the \lcdm{} cosmological parameters.
When additionally fitting $A_\mathrm{L}$, $\Neff{}$, or $\Neff{}+\Yp{}$, the posteriors of these parameters tighten by $5-24\%$.
In the case of primordial magnetic fields, complete \TTTEEE{} power spectrum measurements are necessary to break the degeneracy between $b$ and $n_s$, the spectral index of primordial density perturbations.
We report a 95\% confidence upper limit from SPT-3G data of $b < 1.0$.
The cosmological constraints in this work are the tightest from SPT primary power spectrum measurements to-date and the analysis forms a new framework for future SPT analyses.
\end{abstract}


\maketitle

\section{Introduction}
\label{sec:intro}

The temperature and polarization anisotropies imprinted in the cosmic microwave background (CMB) during recombination encode information on the contents and dynamics of the early universe.
High-precision measurements of the CMB power spectra by satellites and ground-based telescopes enable us to determine the six free parameters of the standard \lcdm{} model with exceptional precision and place tight limits on possible model extensions \citep{planck18-6, dutcher21, balkenhol21, choi20, aiola20}.
Improving measurements of the CMB anisotropies is a key science goal of ground-based CMB experiments such as the South Pole Telescope (SPT hereafter) \citep{carlstrom11}, the Atacama Cosmology Telescope (ACT hereafter) \citep{kosowsky03}, \polarbear{} \citep{kermish12}, and BICEP/\textit{Keck} \citep{keating03b, staniszewski12}.

The \planck{} satellite has mapped the CMB temperature anisotropies down to scales of approximately seven arcminutes to the cosmic-variance limit \citep{planck18-5} and contemporary interest is shifting to polarization data; precision measurements of small angular scale modes of the \TE{} and \EE{} spectra have significant cosmological constraining power \citep{galli14}.
Nevertheless, the \TT{} power spectrum is two orders of magnitude larger than the polarization spectra and temperature data dominate the constraining power of seminal CMB data sets \citep{reichardt09a, das11b, bennett13, story13, planck18-5}.
Complete \TTTEEE{} data sets have significantly more constraining power in \lcdm{} compared to \EETE{} data alone, based simply on a mode-counting argument.
Moreover, certain extensions to the standard model, e.g. primordial magnetic fields, can only be effectively constrained by full \TTTEEE{} data \citep{galli22} due to parameter degeneracies.

In this work, we present cosmological constraints from \TTTEEE{} power spectrum measurements obtained from observations of an approximately $1500\,\sqdeg$ region in the southern sky made by SPT-3G \citep{benson14}, the latest receiver installed on the SPT, in 2018.
The complete SPT-3G 2018 \TTTEEE{} data set comprises previously unpublished \TT{} data, which we present here, and the polarization power spectra presented by \citet[hereafter \citetalias{dutcher21}]{dutcher21} with an updated covariance matrix.
We present cosmological constraints on \lcdm{} and a series of extensions, drawing on the following parameters: the amplitude of the gravitational lensing effect on primary power spectra $A_\mathrm{L}$, the effective number of neutrino species $\Neff{}$, the primordial helium abundance $\Yp{}$, and the baryon clumping factor due to primordial magnetic fields $b$.
We describe our blinding procedure and present an in-depth assessment of the consistency between frequencies and spectrum types.

This paper is structured as follows.
In \S\ref{sec:recap} we summarize important aspects of the data and analysis pipeline of \citetalias{dutcher21} and highlight key changes we make.
In \S\ref{sec:like} we present the updated likelihood code including the foreground model used for temperature data, and details of the parameter fitting procedure.
We demonstrate the consistency of the SPT-3G 2018 data in \S\ref{sec:consistency} and show the \TTTEEE{} power spectra in \S\ref{sec:ps}.
We report cosmological constraints in \S\ref{sec:cosmo} and summarize our findings in \S\ref{sec:conclusion}.

\section{Data and Analysis}
\label{sec:recap}

\citet{sobrin22} present the SPT-3G instrument and \citetalias{dutcher21} detail the 2018 observations and describe the associated data processing pipeline.
These aspects of the analysis have not changed.
We briefly summarize key aspects here and refer the reader to \citetalias{dutcher21} and \citet{sobrin22} for complete discussions.

The data presented here were collected by SPT-3G during an observation period of four months in 2018.
The main SPT-3G survey field covers an area of $\sim\!1500\,\sqdeg$ in the southern sky divided into four subfields.
We calibrate the time-ordered data (TOD) using a series of calibration observations of galactic HII regions.
Sources brighter than $50\,\mathrm{mJy}$ at $150\,\mathrm{GHz}$ are masked and we filter the TOD using low- and high-pass filters, as well as a common-mode filter.
The filtered TOD are processed into maps with $2'$ square pixels using the Lambert azimuthal equal-area projection.
We form a set of $N=30$ temperature and polarization maps with approximately uniform noise properties, so-called ``bundles''.
We calculate cross-spectra between these bundles and bin them into ``band powers''.
We debias the band powers following the \textsc{MASTER} framework \citep{hivon02} using a suite of simulations, thereby accounting for the effects of the survey mask, the TOD filtering, as well as the instrument beam and the pixel window function.
Lastly, we derive absolute per-subfield and full-field calibrations through comparison with \planck{} data \citep{planck18-5}.

The analysis in \citetalias{dutcher21} is designed to maximize sensitivity to the polarization spectra on intermediate and small angular scales.
The common-mode filter applied to the TOD heavily suppresses temperature anisotropies on scales larger than a quarter of a degree.
We therefore set a minimum angular multipole for \TT{} spectra of $\ell^{TT}_{\rm min}=750$.

We make two updates to the calculation of the band power covariance matrix.
First, we account for correlated noise between frequencies in intensity.
For $\ell < 1000$, the atmospheric noise in the $150$ and $220\,\mathrm{GHz}$ data are highly correlated.
Because the noise in the $220\,\mathrm{GHz}$ data is an order of magnitude larger compared to the $150\,\mathrm{GHz}$ data, the former data require precision modeling of the noise correlation.
For this reason, we exclude the $150\times 220\,\mathrm{GHz}$ and $220\times 220\,\mathrm{GHz}$ spectra at $\ell < 1000$.
Second, we improve the treatment of bin-to-bin correlations induced by the flat-sky projection step.
We detail changes to the covariance matrix and their impact on the results reported in \citetalias{dutcher21} in Appendix \ref{app:changes}.

\subsection{Blinding}
\label{sec:blinding}

In a key change from \citetalias{dutcher21} and past SPT \TT{}, \TE{}, and \EE{} analysis, we blind parameter constraints until a series of consistency tests are passed, which we detail in \S\ref{sec:consistency}.
Our blinding procedure entails offsetting cosmological results by random vectors prior to plotting parameter constraints and removing axes labels where appropriate.
We blind parameter constraints until the following consistency tests are passed: (1) null tests, (2) comparison of a minimum-variance combination of band powers to the full multifrequency data vector, (3) conditional spectrum tests split by frequency, (4) conditional spectrum tests split by spectrum type assuming \lcdm{}, and (5) comparison of cosmological parameter constraints in \lcdm{} between subsets and the full data set.
Note that the last two tests are model dependent; in principle, failures of these tests do not prevent cosmological inference, but invite further analysis within the chosen model.
In addition to these quantitative preconditions, we test the robustness of our cosmological results under variations of the likelihood and commit to investigating any significant impact on key results.

\section{Parameter Fitting, Modelling, and External Data}
\label{sec:like}

We use the Markov Chain Monte Carlo (MCMC) package \textsc{CosmoMC} \citep{lewis02b}\footnote{\url{https://cosmologist.info/cosmomc/}} to obtain cosmological parameter constraints.
We compute theoretical CMB spectra using \textsc{camb} \citep{lewis00}\footnote{\url{https://camb.info/}} and \cosmopower{} \citep{spuriomancini22}.\footnote{\url{https://github.com/alessiospuriomancini/cosmopower/}}
We parametrize the \lcdm{} model using: the physical density of cold dark matter, $\Omega_{\mathrm{c}} h^2$, and baryons, $\Omega_{\mathrm{b}} h^2$, the optical depth to reionization $\tau$, the amplitude $A_{\mathrm{s}}$ and spectral index $n_{\mathrm{s}}$ of primordial density perturbations (with $A_{\mathrm{s}}$ defined at a pivot scale of $0.05\,\mathrm{Mpc^{-1}}$), and a parameter that approximates the sound horizon at recombination, $\theta_{\rm MC}$ \citep{hu96b}.

When not combining with \planck{} data, we include a \planck{}-based Gaussian prior on the optical depth to reionization of $\tau = 0.0540 \pm 0.0074$.
This parameter is primarily constrained by a bump at $\ell < 10$ in \EETE{}.
Omitting this prior leads to a degeneracy between $A_{\mathrm{s}}$ and $\tau$ as the amplitude of the power spectra over the angular multipole range probed by our data depends on $A_s$ and $\tau$ mostly through the combination $A_{\mathrm{s}} e^{-2\tau}$.

Similar to \citetalias{dutcher21}, we verify that the likelihood is unbiased using $100$ sets of simulated band powers generated using the data covariance matrix.
We obtain the best-fit model for each realization using the likelihood code.
We find that the average value for each cosmological parameter across the set of simulations lies within $<1.5$ standard errors (i.e. the standard deviation of the ensemble divided by $\sqrt{100}$) of the input value.
The likelihood code is made publicly available on the SPT website.\footnote{\url{https://pole.uchicago.edu/public/data/balkenhol22/}}

\subsection{\cosmopower{}}
\label{sec:cosmopower}

\citet{spuriomancini22} present \cosmopower{}, a neural-network-based CMB power spectrum emulator.
Akin to other emulators \citep[e.g.][]{fendt07a}, once trained, \cosmopower{} provides CMB power spectra in a fraction of the time it takes to evaluate Boltzmann solvers such as \textsc{CAMB} \citep{lewis00} or \textsc{CLASS} \citep{blas11}.
We train \cosmopower{} on a set of power spectra obtained using \textsc{CAMB} at high accuracy settings\footnote{We chose settings similar to the high accuracy settings \citet{hill22} use to update ACT DR4 results (c.f. Appendix A therein); we generate \textsc{CAMB} training spectra with 
\begin{itemize}
\setlength\itemsep{0em}
\item \texttt{k\_eta\_max = 144000},
\item \texttt{AccuracyBoost = 2.0},
\item \texttt{lSampleBoost = 2.0},
\item \texttt{lAccuracyBoost = 2.0}.
\end{itemize}}
for the \lcdm{}, $\mathrm{\Lambda CDM}+\Neff$, and $\mathrm{\Lambda CDM}+A_{\mathrm{L}}$ models.
The constraints obtained by \cosmopower{} and \textsc{CAMB} (run at default accuracy) are within $<0.1\,\sigma$ of each other for all models.
This also highlights that for the analysis of SPT-3G 2018 data, the default accuracy settings used in \textsc{CAMB} are sufficient.
The trained \cosmopower{} models are made publicly available on the SPT website.\footnote{\url{https://pole.uchicago.edu/public/data/balkenhol22/}}

\subsection{Foreground Model and Nuisance Parameters}
\label{sec:fg_model}

We introduce several foreground and nuisance parameters into our likelihood.
We account for the instrumental beam and calibration, aberration due to the relative motion with respect to the CMB rest frame \citep{jeong14}, and super-sample lensing \citep{manzotti14} in the same way as \citetalias{dutcher21}.
The polarized foreground model is minorly updated from \citetalias{dutcher21}, and we describe it briefly below.
Because we include the \TT{} spectrum in this work, we must model the much more complex temperature foregrounds, and we describe this modeling in detail below.
The baseline priors are summarized in Table \ref{tab:priors} in Appendix \ref{app:priors}.

\subsubsection{Temperature Foregrounds}
\label{sec:fg_model_TT}

For the SPT-3G 2018 data with a flux cut for point sources of $50\,\mathrm{mJy}$ at $150\,\mathrm{GHz}$, extragalactic foregrounds dominate over the CMB at $\ell \geq 2650$, $\ell \geq 3000$, and $\ell \geq 2450$ at $95$, $150$, and $220\,\mathrm{GHz}$, respectively.
We construct a foreground model largely based on the existing likelihoods of \citet{reichardt20, george15}, and \citet{dunkley13}.
We perform a re-analysis of \citet{reichardt20} data using the foreground model described below to derive constraints on nuisance parameters.
Where appropriate, we account for the different effective band centers of the data and the lower flux cut of \citet{reichardt20} using the population model of \citet{deZotti05}.
We conservatively widen the constraints from \citet{reichardt20} data on amplitude parameters and spectral indices by factors of four and two, respectively, before adopting them as priors in the cosmological analysis of SPT-3G data.
We perform an analysis of \planck{} data on the SPT-3G survey patch to set priors on the galactic cirrus contribution.

We model the contribution of the galactic cirrus as a modified black-body with temperature $T_d = 19.6\,\mathrm{K}$ and spectral index $\beta^{\mathrm{cirrus}}$ with a cross-frequency power spectrum of
\begin{equation}
\begin{split}
\label{eq:gal_cirrus}
D^{\mathrm{cirrus}}_{\ell, \nu\times\mu} = &A^{\mathrm{cirrus}}_{80} \frac{g(\nu)g(\mu)}{g(\nu^{\mathrm{cirrus}}_0)^2} \left(\frac{\nu \mu}{\nu^{\mathrm{cirrus}}_0 \nu^{\mathrm{cirrus}}_0}\right)^{\beta^{\mathrm{cirrus}}}\\
&\times \left( \frac{\ell}{80} \right)^{\alpha^{\mathrm{cirrus}}+2},
\end{split}
\end{equation}
where $\nu^{\mathrm{cirrus}}_0=150\,\mathrm{GHz}$ is the reference frequency, $A^{\mathrm{cirrus}}_{80}$ is the amplitude parameter, $\alpha^{\mathrm{cirrus}}$ the power law index, and $g=B_\nu(T_d)(\partial B_\nu(T)/\partial T )^{-1}|_{T_{\mathrm{CMB}}}$ with the Planck function $B_\nu(T)$ and CMB temperature taken from \citet{fixsen09b}.
The spectral index, amplitude parameter, and power law index are free parameters in this model.

We account for Poisson-distributed unresolved radio galaxies and dusty star-forming galaxies (DSFG) with a combined contribution to each cross-frequency spectrum of
\begin{equation}
\label{eq:poisson}
D^{\TT{}, \mathrm{Poisson}}_{\ell, \nu\times\mu} = D^{\TT{}, \mathrm{Poisson}}_{3000, \nu\times\mu} \left( \frac{\ell}{3000} \right)^{2},
\end{equation}
where we vary the six amplitude parameters $D^{\TT{}, \mathrm{Poisson}}_{3000, \nu\times\mu}$ in the likelihood.

Following \citet{george15} and \citet{dunkley13}, we model the clustering term of the cosmic infrared background (CIB) using a modified black-body spectrum at $25\,\mathrm{K}$ with spectral index $\beta^{\mathrm{CIB-cl.}}$.\footnote{Note that while the choice of CIB temperature is different from \citet{addison12a}, this has a negligible effect given that the SPT band passes are located in the Rayleigh-Jeans region of the spectrum \citep{george15, reichardt20}.}
Like \citet{george15} and \citet{dunkley13} we use a power law for the angular dependence of this foreground contaminant:
\begin{align}
\begin{split}
\label{eq:cib_clustering}
D^{\mathrm{CIB-cl.}}_{\ell, \nu\times\mu} = &A^{\mathrm{CIB-cl.}}_{80} \frac{g(\nu)g(\mu)}{g(\nu^{\mathrm{CIB-cl.}}_0)^2} \\
&\times \left(\frac{\nu \mu}{\nu^{\mathrm{CIB-cl.}}_0 \nu^{\mathrm{CIB-cl.}}_0}\right)^{\beta^{\mathrm{CIB-cl.}}} \left( \frac{\ell}{80} \right)^{0.8},
\end{split}
\end{align}
where the amplitude $A^{\mathrm{CIB-cl.}}_{80}$ and spectral index $\beta^{\mathrm{CIB-cl.}}$ are free parameters, $\nu^{\mathrm{CIB-cl.}}_0=150\,\mathrm{GHz}$ is the reference frequency, and the value of the power-law index is motivated by \citet{addison12a}.

Following \citet{reichardt20}, we account for the thermal Sunyaev–Zel’dovich (tSZ) effect by rescaling the power spectrum of \citet{shaw10} normalized at $\ell=3000$, $D^{\mathrm{tSZ, template}}_{\ell}$, at a reference frequency of $\nu^{\mathrm{tSZ}}_0=143\,\mathrm{GHz}$ via
\begin{equation}
\label{eq:tSZ}
D^{\mathrm{tSZ}}_{\ell, \nu\times\mu} = A^{\mathrm{tSZ}} \frac{f(\nu)f(\mu)}{f(\nu^{\mathrm{tSZ}}_0)^2} D^{\mathrm{tSZ, template}}_{\ell},
\end{equation}
where $f(x) = x \coth{(x/2)} - 4$ with $x=h\nu/k_B T_{\mathrm{CMB}}$ and we vary the amplitude parameter $A^{\mathrm{tSZ}}$ in the likelihood.

We model the correlation between the tSZ and CIB signals following \citet{george15} as
\begin{align}
\begin{split}
\label{eq:tSZ_CIB}
D^{\mathrm{tSZ-CIB}}_{\ell, \nu\times\mu} = &- \xi \left(\sqrt{D^{\mathrm{tSZ}}_{\ell, \nu\times\nu} D^{\mathrm{CIB-cl.}}_{\ell, \nu\times\nu}} \right.\\
&\left. + \sqrt{D^{\mathrm{tSZ}}_{\ell, \mu\times\mu} D^{\mathrm{CIB-cl.}}_{\ell, \mu\times\mu}} \right),
\end{split}
\end{align}
where $\xi$ is the correlation parameter, which we vary in the likelihood.
We define the sign here, such that $\xi>0$ corresponds to a reduction in power at $150\,\mathrm{GHz}$.

Finally, we account for the kinematic Sunyaev–Zel’dovich (kSZ) effect similar to \citet{reichardt20} by rescaling a combined template for the homogeneous \citep{shaw12} and patchy \citep{zahn12} kSZ effects normalized at $\ell=3000$, $D^{\mathrm{kSZ, template}}_{\ell}$, via
\begin{equation}
\label{eq:kSZ}
D^{\mathrm{kSZ}}_{\ell} = A^{\mathrm{kSZ}} D^{\mathrm{kSZ, template}}_{\ell},
\end{equation}
where we vary the amplitude parameter $A^{\mathrm{kSZ}}$ in the likelihood.

\subsubsection{Polarization Foregrounds}
\label{sec:fg_model_pol}

We adopt the polarization foreground model of \citetalias{dutcher21}.
We account for Poisson sources in the \EE{} power spectrum and polarized galactic dust in the \EE{} and \TE{} data.
The priors for the former contaminant are unaltered from \citetalias{dutcher21}, while we amend priors on polarized galactic dust using the updated analysis of \planck{} data within our survey region (see Appendix \ref{app:changes} for details).

\subsection{External Data Sets}
\label{sec:data}

We use \planck{} data in combination with SPT-3G 2018 data to derive cosmological constraints.
\planck{} and SPT-3G data complement one another by providing high-precision measurements of the CMB power spectra on large and small angular scales, respectively.
Specifically, the SPT-3G data are more precise than \planck{} for \TT{} at $\ell > 2000$, for \TE{} at $\ell > 1400$, and for \EE{} at $\ell > 1000$.
We use the \textsc{base\_plikHM\_TTTEEE\_lowl\_lowE} \planck{} data set \citep{planck18-5}.

We also report joint results for SPT-3G 2018 and \WMAP{} data for key scenarios, to be as independent of \planck{} data as possible.
We use the year nine data set \citep{bennett13} with \TT{} data at $2 < \ell < 1200$, and \TE{} and \EE{} data at $24 < \ell < 800$.
We exclude polarization data at $\ell < 24$, due to the possibility of dust contamination \citep{planck13-15}, and include our baseline prior on $\tau$ to constrain the optical depth to reionization instead.
This setup is the same that \citet{aiola20} used for joint ACT DR4 and \WMAP{} constraints.

We ignore correlations between SPT-3G and satellite data.
\planck{} and \WMAP{} data cover a large amount of sky not observed by SPT.
Moreover, the SPT-3G data are weighted towards higher $\ell$.

\section{Internal Consistency and Robustness of Results}
\label{sec:consistency}

In this section, we perform null tests, consistency tests on the final band powers, parameter-level consistency tests, and an assessment of the robustness of cosmological constraints.
For each test category, we compute a set of probability-to-exceed (PTE) values, which we require to lie within some predetermined limits.
We require the PTE values to lie above the threshold $5\%/N$ for null tests and within the symmetric interval $[(2.5/N)\%, (100-2.5/N)\%]$ for all other tests, where $N$ is the number of independent tests, i.e. using the Bonferroni correction for the look-elsewhere effect \citep{dunn61}.
We determine $N$ for each test category individually within the relevant section and conservatively do not correct for the look-elsewhere effect across different test categories.
As noted in \S\ref{sec:blinding}, this work was done prior to unblinding parameter constraints.

\subsection{Null Tests}
\label{sec:null_tests}

\begin{table*}[hbt!]
\centering
\setlength{\tabcolsep}{2pt}
\begin{tabular}{l C{2cm} C{2cm} C{2cm} C{2cm} C{2cm} C{2cm} C{2cm} C{2cm}}
\hline\hline
\vphantom{\makecell{A\\A}}
& Azimuth
& First/Second
& Left/Right
& Moon
& Saturation
& Wafer\\
\hline
\multicolumn{7}{l}{95\,GHz}\\
\hphantom{XX}$\TT{}$ & 0.116 & 0.614 & 0.630 & 0.991 & 0.882 & 0.492\\
\hphantom{XX}$\TE{}$ & 0.294 & 0.067 & 0.028 & 0.938 & 0.234 & 0.620\\
\hphantom{XX}$\EE{}$ & 0.765 & 0.398 & 0.015 & 0.866 & 0.340 & 0.037\\
\hphantom{XX}$\TTTEEE{}$ & 0.284 & 0.210 & 0.012 & 0.999 & 0.508 & 0.184\\
\hline
\multicolumn{7}{l}{150\,GHz}\\
\hphantom{XX}$\TT{}$ & 0.075 & 0.549 & 0.861 & 0.305 & 0.884 & 0.485\\
\hphantom{XX}$\TE{}$ & 0.879 & 0.539 & 0.859 & 0.894 & 0.238 & 0.465\\
\hphantom{XX}$\EE{}$ & 0.002 & 0.970 & 0.432 & 0.486 & 0.268 & 0.005\\
\hphantom{XX}$\TTTEEE{}$ & 0.012 & 0.882 & 0.889 & 0.667 & 0.460 & 0.045\\
\hline
\multicolumn{7}{l}{220\,GHz}\\
\hphantom{XX}$\TT{}$ & 0.310 & 0.548 & 0.635 & 0.635 & 0.128 & 0.077\\
\hphantom{XX}$\TE{}$ & 0.420 & 0.929 & 0.169 & 0.834 & 0.784 & 0.510\\
\hphantom{XX}$\EE{}$ & 0.991 & 0.735 & 0.222 & 0.835 & 0.875 & 0.501\\
\hphantom{XX}$\TTTEEE{}$ & 0.751 & 0.914 & 0.243 & 0.931 & 0.635 & 0.227\\
\hline
\end{tabular}
\caption{
\label{tab:null_tests}
Individual null test PTE values for $95$, $150$, and $220\,\mathrm{GHz}$ and \TT{}, \TE{}, and \EE{} spectra. Additionally, we show the combined \TTTEEE{} null test PTE values. All PTE values lie above the required threshold of $0.05 / (9 \times 6) \approx 0.001$.
}
\end{table*}

We test that the data are free of significant systematic effects through six types of null tests.
Following \citetalias{dutcher21}, we analyze the following data splits (to test for the corresponding category of systematic errors): azimuth (ground pick-up), first-second (chronological effects), left-right (scan-direction dependent effects), moon up - moon down (beam sidelobe pickup), saturation (decreased array responsivity), and detector module or ``wafer'' (non-uniform detector properties).
The data are ranked or divided into groups based on a given possible systematic and we take the difference of these map bundles to form null maps.
We then calculate the null spectra as the average of null map cross-spectra for each test and use their distribution to compute uncertainties.
We verify that the average of these spectra is consistent with the expectation for a given test using a $\chi^2$ statistic.

We update the null test framework employed by \citetalias{dutcher21} as follows.
First, we scale null spectra by $\ell(\ell+1)/2\pi$ and apply the debiasing kernel of the corresponding auto-frequency spectrum to the null spectra.
This change corresponds to a linear transformation and does not change the pass state of tests while making it easier to interpret the amplitude of null spectra.

Second, we cast the \TE{} and \EE{} null spectra in nine bins of width $\Delta\ell=300$ spanning the angular multipole range $300 < \ell < 3000$, whereas for \TT{} we use ten bins of width $\Delta\ell=250$ across $750 < \ell < 3000$.
This change makes the tests more sensitive to plateaus in power.
Furthermore, this allows us to ignore bin-to-bin correlations induced by the flat-sky projection step, which only drop to $\leq 20\%$ for bins separated by $\Delta \ell \geq 100$.

Third, we add $1\%$ of uncorrelated sample variance to the covariance of the \TT{} null spectra.
SPT-3G produces a high signal-to-noise measurement of the \TT{} power spectrum.
Minor low-level systematic effects may appear above the noise level, while having a negligible effect on cosmological results due to the high sample variance of the \TT{} spectrum across the $\sim\,1500\,\mathrm{deg^2}$ field.
We verify this by artificially displacing the final \TT{} data band powers by vectors mimicking systematic effects and rerunning the temperature likelihood.
We asses the potential impact of two potential systematic effects:
\begin{itemize}
\item We asses the impact of unmodeled time constants by injecting a left-right expectation spectrum large enough to produce a null test failure.
\item We asses the impact of an overall miscalibration by increasing the amplitude of \TT{} band powers by the square root of $1\%$ of their total covariance.
\end{itemize}
In both cases, we find that the best-fit parameters in \lcdm{} shift by $<0.2\,\sigma^{\TT{}}$, where $\sigma^{\TT{}}$ represents the size of parameter errors when using only \TT{} data.

Fourth, we model the effect of detector time-constants in the \TT{} scan-direction expectation spectrum.
The maps presented in \citetalias{dutcher21} are not corrected for time-constants, which we see in the scan-direction test.
We model this null spectrum as a constant offset between left- and right-going scans of $2vt$, where we assume a uniform on-sky scan speed of $v=0.7\,\mathrm{deg\,s^{-1}}$ across the survey field and $\tau=4.6\,\mathrm{ms}$ is the median time constant.
This effect does not appear above the noise level in the \TE{} and \EE{} data.
Detector time-constants act as an effective beam.
The maps used for the beam measurement in \S\RNum{4} E of \citetalias{dutcher21} include this effect and therefore when we remove the instrumental beam during the debiasing procedure, we also remove the signature of detector time-constants from the data band powers.
The expectation spectrum for all other \TT{} null tests is approximated as zero.

In addition to the individual \TT{}, \TE{}, and \EE{} null tests, we also report results for all three spectra (\TTTEEE{}) at a single frequency.
We forego quantifying the correlation between the combined and individual tests and exclude this combined test in setting the PTE threshold.
We assume that the remaining tests are independent from one another, such that across three frequencies and three spectrum types and six test categories, there are $N=3\times3\times6=54$ independent tests.
We require all PTE values to lie above $0.05 / 54 \approx 0.001$.
We do not repeat the meta-analyses (i.e. the per-row and full-table tests) carried out by \citetalias{dutcher21} since the addition of sample-variance to the \TT{} null spectra means the PTE values are not expected to be uniformly distributed.
For this reason, we do not flag and investigate high PTE values in the \TT{} and \TT{}/\TE{}/\EE{} tests.
Due to the updates detailed above we expect the PTE values of the \TE{} and \EE{} null tests to change from \citetalias{dutcher21}.

We report the null test PTE values in Table \ref{tab:null_tests}.
All of the PTE values lie above the set threshold.
Across the 72 tests the lowest PTE value is $0.002$ (\EE{} 150\,GHz Azimuth test).
There is no significant mean change to the PTE values of the \EE{} and \TE{} reported in \citetalias{dutcher21}.
The largest individual change is an increase to the PTE value of the \TE{} 150\,GHz Azimuth test by $0.683$.
We have confirmed that all PTE values also lie above the required threshold when adopting a finer bin width of $\Delta\ell=125$ for \TT{} and $\Delta\ell=100$ for \EETE{} null spectra.\footnote{The different bin widths are due to the different $\ell$ ranges covered by temperature and polarization data.}
We conclude that the data are free of significant systematic errors and proceed with the analysis.

\subsection{Power Spectrum Tests}
\label{sec:ps_consistency}

In this section, we perform a series of power-spectrum level tests to assess the internal consistency of the SPT-3G 2018 \TTTEEE{} data set.
We begin by combining the six cross-frequency band powers, $\hat{D}$, for each spectrum type into a minimum-variance combination, $\hat{D}^{MV}$, that represents our best, foreground-free measurement of the CMB anisotropies.
Following \citet{planck15-11} and \citet{mocanu19}
\begin{equation}
    \label{eq:MV}
    \hat{D}^{MV} = \left( X^T \mathcal{C}^{-1} X \right)^{-1} X^T \mathcal{C}^{-1} \hat{D},
\end{equation}
where $\mathcal{C}$ is the band power covariance matrix and $X$ is the design matrix, which is populated with ones and zeros and connects the six cross-frequency estimates of the same CMB signal per multipole bin in $\hat{D}$ to the corresponding single element in $\hat{D}^{MV}$ \citep{planck15-11}.
We subtract the best-fit foreground model from the data prior to the above procedure, though this only matters for the \TT{} spectra since the foreground contamination in polarization is negligible.

For our first test, we compare the minimum-variance spectrum to the full set of multifrequency band powers and require that the PTE values lie within $[2.5\%, 97.5\%]$ for each spectrum-type and the full combination of \TTTEEE{} spectra.
This test ensures that the data are consistent with measuring the same underlying signal and free from any significant unmodelled foreground contamination.
We use the test-statistic
\begin{equation}
    \label{eq:6:MV_test}
    \chi^2 = \left( X \hat{D}^{MV} - \hat{D} \right)^{T} \mathcal{C}^{-1} \left( X \hat{D}^{MV} - \hat{D} \right).
\end{equation}
We obtain $\chi^2=668$ for $605$ degrees of freedom.\footnote{We follow \citetalias{dutcher21} and use the number of multifrequency band powers minus the number of minimum-variance band powers as the number of degrees of freedom.}
This corresponds to a PTE value of $4\%$ for \TTTEEE{}.
For \TT{}, \TE{}, and \EE{} spectra individually, we find PTE values of $22\%$, $12\%$, and $16\%$, respectively.
The PTE value of the combined test is driven low by the $220\,\mathrm{GHz}$ data in temperature and polarization.
However, all PTE values lie within the 95th percentile and we report no sign of significant internal inconsistency.

\begin{figure*}[ht!]
  \centering
  \includegraphics[width=\linewidth]{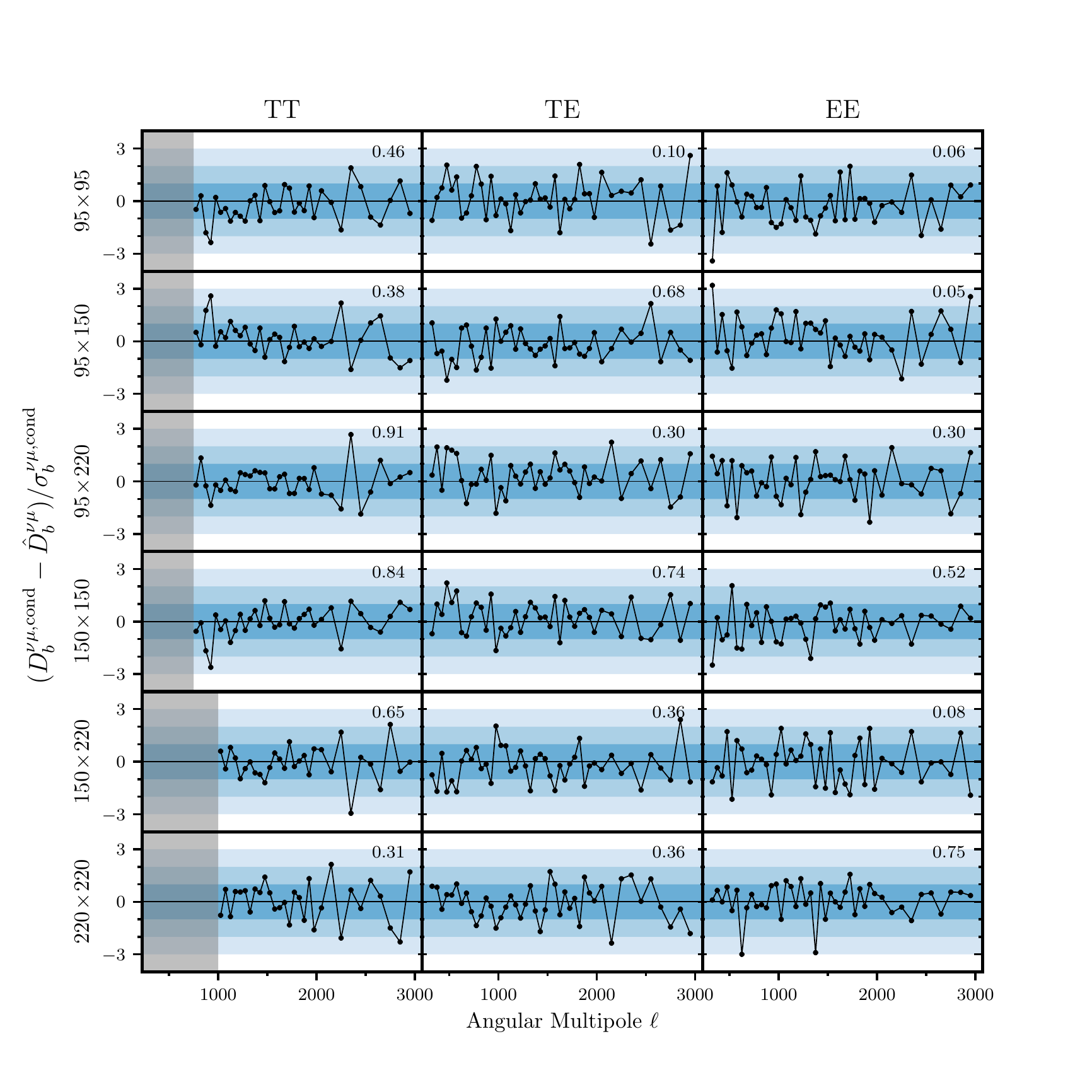}
  \caption{\label{fig:freq_cond}
  Relative conditional residuals, $(D_b^{\nu\mu, \mathrm{cond}} -\hat{D}_b^{\nu\mu})/\sigma_b^{\nu\mu, \mathrm{cond}}$, i.e. the difference between conditional predictions for a given set of multifrequency band powers and the measured data, divided by the square-root of the diagonal of the conditional covariance.
  The blue shaded region corresponds to the $3\,\sigma$ range and the grey shaded area in the first column indicates the \TT{} angular multipole lower limit.
  The conditional residuals are consistent with zero, as evidenced by the PTE values indicated in the upper right corner of each panel.
  This speaks to the inter-frequency consistency of the SPT-3G 2018 \TTTEEE{} data set.
  }
\end{figure*}

\begin{figure*}[ht!]
  \centering
  \includegraphics[width=\linewidth]{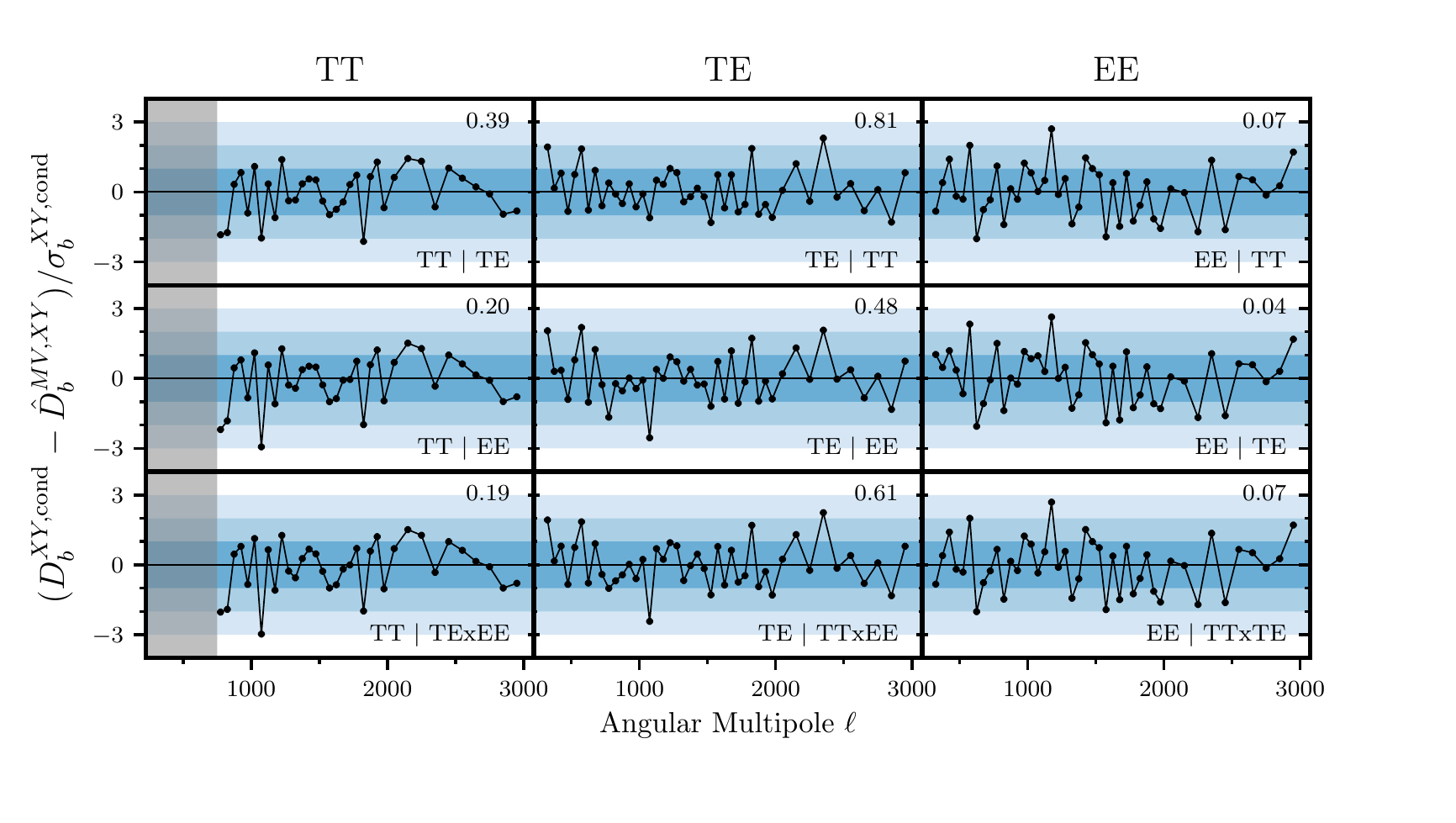}
  \caption{\label{fig:fld_cond}
  Relative conditional residuals, $(D_b^{XY, \mathrm{cond}} -\hat{D}_b^{MV,XY})/\sigma_b^{XY, \mathrm{cond}}$ with $XY\in \{ \TT, \TE, \EE \}$, i.e. the difference between conditional predictions for a given set of minimum-variance band powers and the measured data, divided by the square-root of the diagonal of the conditional covariance.
  The blue shaded region corresponds to the $3\,\sigma$ range and the grey shaded area in the first column indicates the \TT{} angular multipole range.
  The spectra used in the conditional prediction are specified in the bottom right corner of each panel and the PTE values are indicated in the top right corner of each panel.
  We find good agreement between the different spectra of the SPT-3G 2018 \TTTEEE{} data set.
  }
\end{figure*}

Second, we perform a conditional spectrum test to probe the interfrequency agreement within each spectrum type.
This test is largely agnostic to the cosmological model, though it assumes that the foreground model describes the data well.
We compare each set of multifrequency band powers, $\hat{D}^{\nu\mu}$, where $\nu, \mu$ denote the frequency combination, to the ensemble of other band powers of the same spectrum type.
Following \citet{planck18-5}, we split the data band powers into $\hat{D} = \left[\hat{D}^{\nu\mu}, \hat{D}^{\mathrm{others}}\right]$, where ``others'' indicates the part of the data we use for the prediction of the remainder.
We decompose the best-fit spectrum, $D$, and the covariance, $\mathcal{C}$, in the same way.
The conditional prediction and the associated covariance are
\begin{align}
\begin{split}
    D^{\nu\mu, \mathrm{cond}} &= D^{\nu\mu} + \mathcal{C}^{\nu\mu \times \mathrm{others}} \left( \mathcal{C}^{\mathrm{others \times others}} \right)^{-1}\\
    &\hphantom{= D^{\nu\mu} + } \left( \hat{D}^{\mathrm{others}} - D^{\mathrm{others}} \right),\\
    \mathcal{C}^{\nu\mu \times \nu\mu, \mathrm{cond}} &= \mathcal{C}^{\nu\mu \times \nu\mu} - \mathcal{C}^{\nu\mu \times \mathrm{others}}\\
    & \hphantom{= \mathcal{C}^{\nu\mu \times \nu\mu} - } \left( \mathcal{C}^{\mathrm{others \times others}} \right)^{-1} \mathcal{C}^{\mathrm{others} \times \nu\mu}.
\end{split}
\end{align}
We compare this prediction to the measured data band powers using a $\chi^2$ statistic and require all PTE values to lie within the interval $[(2.5/N)\%, (100-2.5/N)\%]$, where $N$ is the number of independent tests.
Given that there are six cross-frequency combinations and three spectrum types, there are $18$ tests in total.
However, the number of independent tests is lower.
We conservatively set $N=5$; due to the absence of correlated noise in the polarization data, the auto-frequency \EE{} tests are independent and we discount the remaining \EE{} tests and assume that the \TE{} and \TT{} tests only add one independent test each.
We list the PTE values and plot the results for the conditional residuals in Figure \ref{fig:freq_cond}.
We find that all PTE values lie within the required interval; the conditional spectra are in good agreement with the measured data.
This agreement is noteworthy, as across the different spectra we have data that are highly correlated (\TT{} on intermediate scales) and uncorrelated beyond the common CMB sample variance (\EE{} spectra).

Next, we apply the conditional test framework across the different spectrum types and probe the consistency between the \TT{}, \TE{}, and \EE{} data.
In contrast to the per-frequency conditional test, this test is dependent on the cosmological model and we carry it out assuming \lcdm{}.
As in \citet{planck18-5}, this test is performed using the minimum-variance band powers.
For each spectrum, we compare the data minimum-variance combination to the conditional prediction given each other spectrum individually and jointly. 
We require all PTE values to lie within the interval $[(2.5/N)\%, (100-2.5/N)\%]$, where $N$ is the number of independent tests.
Given the mild correlation between the temperature and polarization anisotropies, we conservatively set $N=2$.
We show the conditional residuals in Figure \ref{fig:fld_cond} and list the PTE values therein.
We find no statistically significant outliers when comparing the conditional predictions and the measured data; all PTE values are in the required interval.
The series of tests we have carried out provide a stringent assessment of the consistency of the SPT-3G 2018 \TTTEEE{} band powers across frequencies and spectra; we conclude that the data are free of any significant internal tension at the power-spectrum level.

Though the tests above already complete our passing criteria to proceed with the analysis, we additionally investigate the difference spectra in Appendix \ref{app:diff_specs}.
This allows us to build further expertise with the data.
We observe no significant features, such as slopes, constant offsets, or signal leakage.

\begin{figure*}[ht!]
  \centering
  \includegraphics[width=0.8\linewidth]{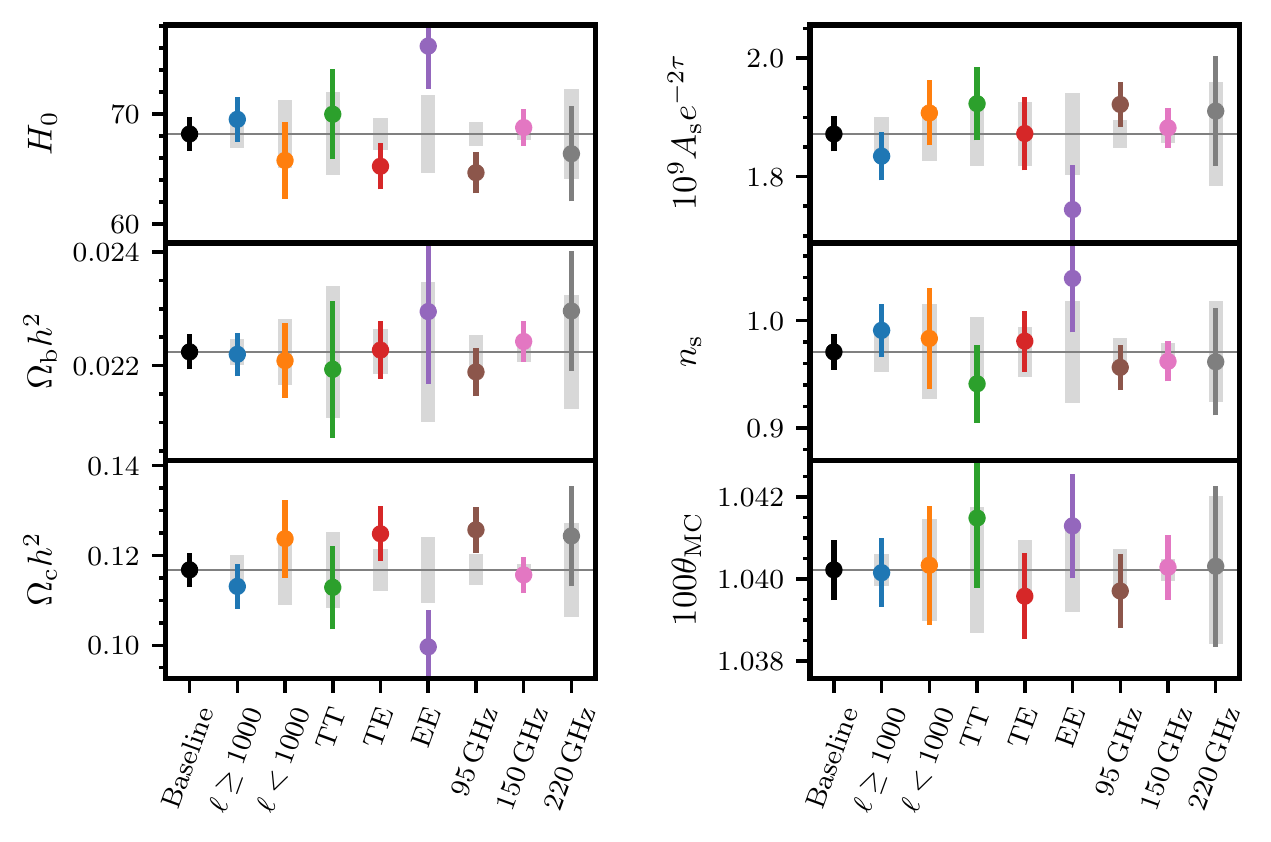}
  \caption{\label{fig:param_fluct}
  Parameter constraints from the full SPT-3G 2018 \TTTEEE{} data set (black points) and select subsets (coloured points as indicated) in \lcdm{}.
  The grey boxes indicate the expected $1\,\sigma$ fluctuations between each subset and the full data set, taking the shared data into account.
  The observed shifts between subsets and the full data are consistent with statistical fluctuations.
  During the blind stage of this analysis, the parameter values along the vertical axes were not shown.
  }
\end{figure*}

\subsection{Parameter-Level Tests}
\label{sec:par_consistency}

We now turn to the internal consistency of the SPT-3G 2018 \TTTEEE{} data set at the parameter level.
This test is explicitly model dependent and is performed in \lcdm{} using the following parameters: $\Omega_{\mathrm{b}} h^2$, $\Omega_{\mathrm{c}} h^2$, $\theta_{\mathrm{MC}}$, $n_{\mathrm{s}}$, and $10^9 A_{\mathrm{s}}(k=0.1\,\mathrm{Mpc}^{-1}) e^{-2\tau}$.
Here, $A_{\mathrm{s}}(k=0.1\,\mathrm{Mpc}^{-1})$ is the amplitude of the primordial power spectrum at $k=0.1\,\mathrm{Mpc}^{-1}$.
This definition provides a better match to the scales constrained by the SPT data compared to the conventional reference point of $k=0.05\,\mathrm{Mpc}^{-1}$ and improves the numerical stability of the test by reducing the correlation between the combined amplitude parameter and $n_{\mathrm{s}}$.
We use the conventional reference point for $A_{\mathrm{s}}$ when reporting cosmological results in \S\ref{sec:cosmo}.

We investigate parameter constraints from the following subsets of the data: \TT{}, \TE{}, and \EE{} spectra individually, the three sets of auto-frequency spectra ($95 \times 95\,\mathrm{GHz}$, $150 \times 150\,\mathrm{GHz}$, and $220 \times 220\,\mathrm{GHz}$), large angular scales ($\ell < 1000$), and small angular scales ($\ell \geq 1000$).
We follow \citet{gratton19} and quantify the significance of the shift of mean parameter values from the full data set to a given subset, $\Delta p$, using the parameter-level $\chi^2$:
\begin{equation}
\chi^2 = \Delta p^T \mathcal{C}_p^{-1} \Delta p,
\end{equation}
where $\mathcal{C}_p$ is the difference of the parameter covariances of the full data set and a given subset.
This formalism takes the correlation between parameter constraints from the full data set and any given subset into account.
As with the other tests, we require all PTE values to lie within $[(2.5/N)\%, (100-2.5/N)\%]$, where $N$ is the number of independent tests.
The large and small angular scale tests are independent from one another and we conservatively assume that the remaining six subsets only count as one independent test setting $N=3$.

\begin{table}[ht!]
\def\arraystretch{1.5}
\small
\setlength{\tabcolsep}{10pt}
\centering
\begin{tabular}{l c  c}
\hline\hline
 Subset & $\chi^2$ & PTE\\
 \hline
$\ell \le 1000$ & 4.8 & 44.7\% \\
$\ell > 1000$ & 4.9 & 43.4\% \\
\TT{} & 10.3 & 6.7\% \\
\TE{} & 4.9 & 43.1\% \\
\EE{} & 14.8 & 1.1\% \\
$95\,\mathrm{GHz}$ & 9.8 & 8.0\% \\
$150\,\mathrm{GHz}$ & 3.5 & 61.7\% \\
$220\,\mathrm{GHz}$ & 1.9 & 86.5\% \\
\hline
\end{tabular}
\caption[
Parameter-level $\chi^2$ difference and PTE between subsets of the data and the full dataset.
]{
Parameter-level $\chi^2$ and PTE values between subsets of the data and the full data set.
Note that there are five degrees of freedom as we perform the comparison across $[\Omega_b h^2, \Omega_c h^2, \theta_{MC}, 10^9 A_s(k=0.1\,\mathrm{Mpc}^{-1}) e^{-2\tau}, n_s]$, due to the common $\tau$ prior.
Here, we use $A_s(k=0.1\,\mathrm{Mpc}^{-1})$, the amplitude of the primordial power spectrum at $k=0.1\,\mathrm{Mpc}^{-1}$, to improve the numerical stability of the test.
All PTE values lie within the required interval of $[(2.5/3)\%, (100-2.5/3)\%]$ and we conclude that the parameter shifts are compatible with statistical fluctuations.
}
\label{tab:datasplit_chisq}
\end{table}

We plot parameter fluctuations for the standard \lcdm{} parameters in Figure \ref{fig:param_fluct} and list the subset $\chi^2$ and associated PTE values in Table \ref{tab:datasplit_chisq}.
We note that the \EE{} parameter constraints deviate the most from the full data set and have the lowest PTE value of any of the subsets.
However, this PTE value is still above our preset criterion and we therefore consider the parameter shifts compatible with statistical fluctuations.
We conclude that the data are internally consistent at the parameter level and proceed to unblind parameter constraints.

\begin{figure*}[ht!]
  \centering
  \includegraphics[width=\linewidth]{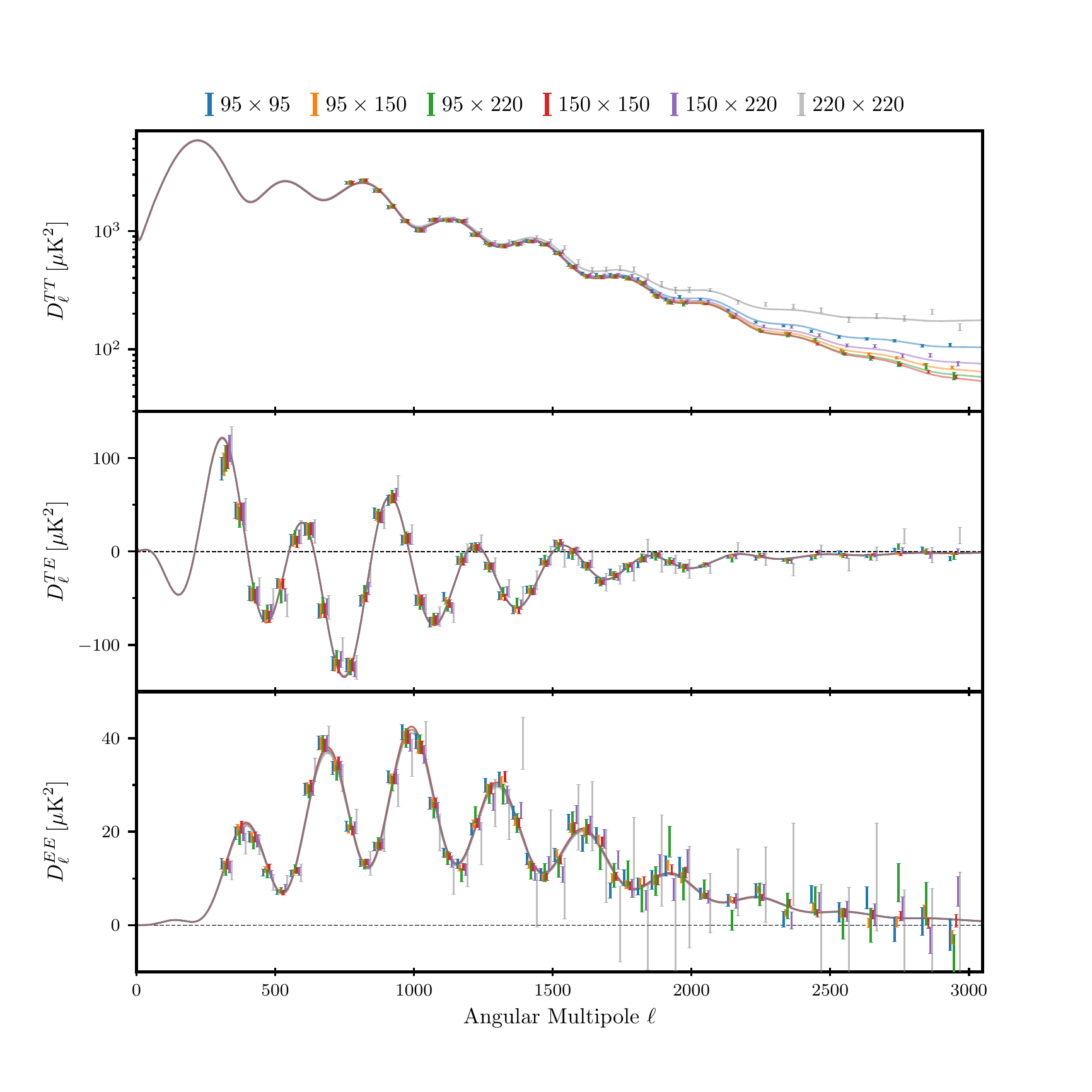}
  \caption{\label{fig:PS_all_freqs}
  SPT-3G 2018 multifrequency \TTTEEE{} band powers in colors as indicated in the legend, along with the best-fit \lcdm{} model to the SPT data including foregrounds (solid lines of matching color).
  The SPT-3G data provide a precision measurement of the CMB temperature and polarization anisotropies on intermediate and small angular scales.
  }
\end{figure*}

\begin{figure*}[ht!]
  \centering
  \includegraphics[width=\linewidth]{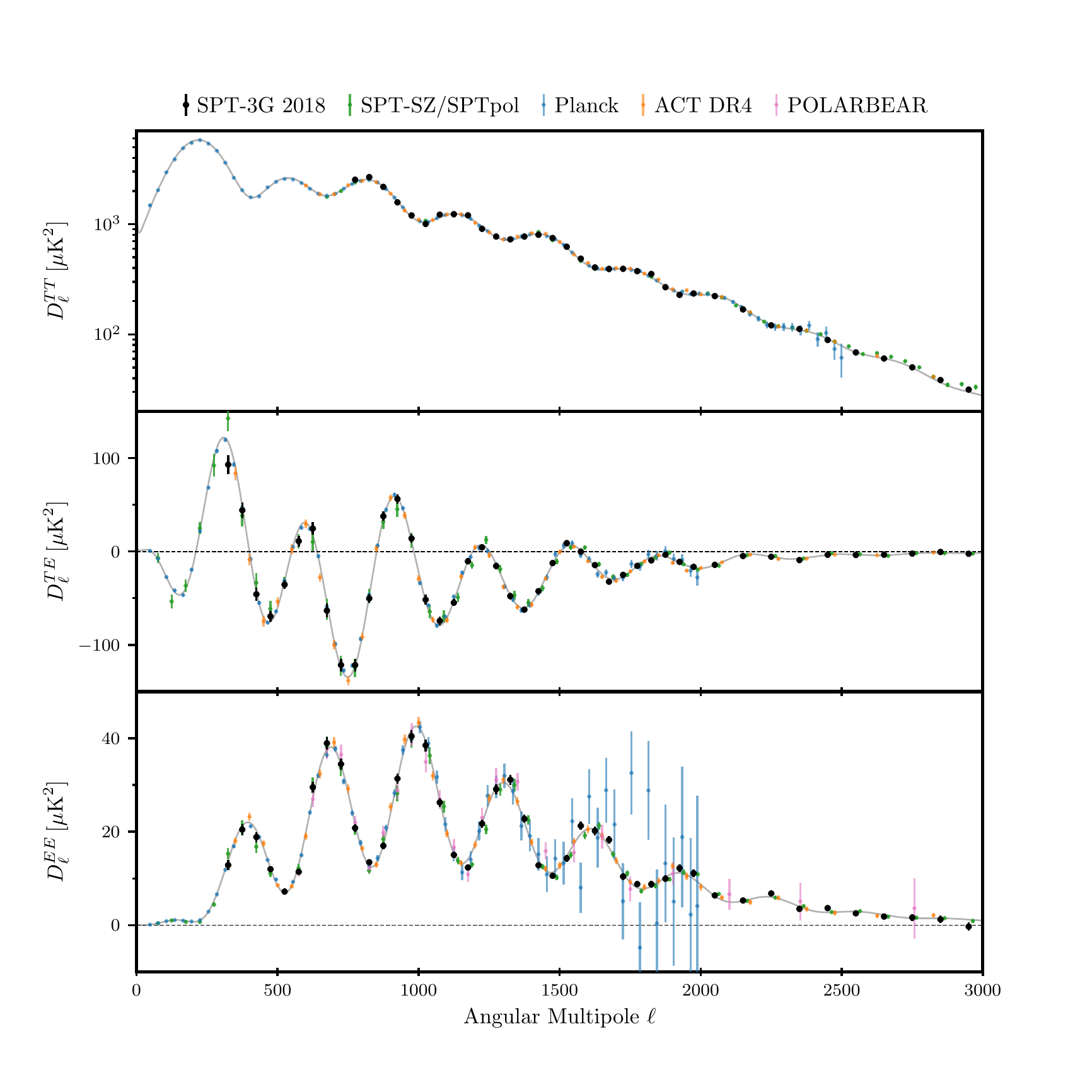}
  \caption{\label{fig:PS_MV}
  SPT-3G 2018 minimum-variance \TTTEEE{} band powers (black) along with a selection of contemporary power spectrum measurements: \planck{} (blue) \citep{planck18-5}, SPT-SZ (green, top panel only) \citep{mocanu19}, SPTpol (green, bottom two panels only, horizontally offset for clarity) \citep{henning18}, ACT DR4 (orange) \citep{choi20}, \textsc{POLARBEAR} (pink, bottom panel only) \citep{polarbear20}.
  The SPT-3G 2018 best-fit CMB power spectrum is indicated in gray.
  The ensemble of CMB data is visually consistent and yields a high signal-to-noise measurement of the power spectrum.
  }
\end{figure*}

\subsection{Robustness of Cosmological Constraints}
\label{sec:like_robust}

We verify the robustness of our cosmological results with respect to variations of the likelihood presented in \S\ref{sec:like}.
We test the following cases in \lcdm{}: removing the priors on each set of amplitude parameters for a given foreground source; removing the priors on all temperature amplitude parameters simultaneously; widening the CIB spectral index prior by a factor of two; introducing the CIB power law index as a free parameter either with a wide uniform prior or adopting the result of \citet{addison12a} as a prior; introducing CIB decorrelation parameters $\zeta^{\nu}$ for each frequency band with uniform priors between zero and unity that multiply Equation \ref{eq:cib_clustering} by $\sqrt{\zeta^{\nu}\zeta^{\mu}}$; ignoring the tSZ-CIB correlation; ignoring galactic cirrus; ignoring or quadrupling the beam covariance; adopting the $\tau$ constraint found by \citet{natale20} as a prior. 
In addition to these tests for constraints from the full \TTTEEE{} data set, we also investigate the effect of foreground model variations on constraints from \TT{} alone.
We find no significant change to cosmological constraints for any of the cases tested; all parameter shifts are $<0.3\,\sigma$, where $\sigma$ indicates the width of the respective \TTTEEE{} or \TT{} constraint using baseline priors.\footnote{We also test the case of removing all priors on foreground amplitude parameters when analyzing \TT{} data alone in \lcdm{}+$A_L$ and \lcdm{}+$\Neff{}$ and report no significant change to cosmological constraints.}
We conclude that none of the likelihood variations above have a significant impact on cosmological constraints.
Together with the consistency tests at the band power level in \S\ref{sec:ps_consistency}, this indicates that our results are robust with respect to a mismodelling of the foreground contamination.

\begin{table}[ht!]
\footnotesize
\setlength{\tabcolsep}{5pt}
\def\arraystretch{1.0}
\centering
\begin{tabular}{c |  D{.}{.}{4.2}  D{.}{.}{2.2} | D{.}{.}{4.2}  D{.}{.}{2.2} |  D{.}{.}{2.2}  D{.}{.}{1.2}}
\hline\hline
\rule{0pt}{3ex} $\ell$ Range & \multicolumn{1}{r}{$D_b^{\TT}$} & \multicolumn{1}{r}{$\sigma^{\TT}$} & \multicolumn{1}{r}{$D_b^{\TE}$} & \multicolumn{1}{r}{$\sigma^{\TE}$} & \multicolumn{1}{r}{$D_b^{\EE}$} & \multicolumn{1}{r}{$\sigma^{\EE}$} \\[2pt]
\hline
300 -- 350 & - & - & 92.96 & 10.32 & 12.87 & 1.02 \\
350 -- 400 & - & - & 44.06 & 8.46 & 20.46 & 1.23 \\
400 -- 450 & - & - & -45.80 & 7.15 & 18.85 & 1.08 \\
450 -- 500 & - & - & -69.45 & 5.99 & 11.99 & 0.64 \\
500 -- 550 & - & - & -35.48 & 4.67 & 7.19 & 0.39 \\
550 -- 600 & - & - & 11.07 & 5.70 & 11.42 & 0.61 \\
600 -- 650 & - & - & 24.52 & 6.71 & 29.50 & 1.14 \\
650 -- 700 & - & - & -63.28 & 7.39 & 38.95 & 1.33 \\
700 -- 750 & - & - & -121.54 & 6.85 & 34.48 & 1.24 \\
750 -- 800 & 2531.89 & 82.90 & -121.56 & 6.65 & 20.80 & 0.88 \\
800 -- 850 & 2674.59 & 78.11 & -50.31 & 4.71 & 13.47 & 0.55 \\
850 -- 900 & 2179.55 & 72.87 & 37.67 & 5.07 & 17.01 & 0.70 \\
900 -- 950 & 1578.46 & 52.45 & 56.22 & 4.89 & 31.37 & 1.05 \\
950 -- 1000 & 1201.33 & 38.99 & 13.95 & 4.83 & 40.44 & 1.33 \\
1000 -- 1050 & 1003.98 & 33.71 & -51.61 & 5.19 & 38.49 & 1.30 \\
1050 -- 1100 & 1219.01 & 35.13 & -74.30 & 4.69 & 26.27 & 0.96 \\
1100 -- 1150 & 1231.40 & 36.35 & -54.77 & 3.82 & 15.05 & 0.64 \\
1150 -- 1200 & 1202.46 & 36.99 & -10.53 & 3.28 & 12.34 & 0.59 \\
1200 -- 1250 & 907.07 & 28.30 & 4.39 & 3.30 & 21.73 & 0.85 \\
1250 -- 1300 & 771.75 & 22.69 & -15.57 & 3.36 & 29.12 & 1.07 \\
1300 -- 1350 & 727.84 & 21.05 & -47.79 & 3.42 & 31.14 & 1.08 \\
1350 -- 1400 & 771.56 & 24.02 & -62.26 & 3.43 & 22.76 & 0.87 \\
1400 -- 1450 & 800.59 & 23.88 & -42.49 & 3.04 & 12.82 & 0.65 \\
1450 -- 1500 & 748.60 & 21.56 & -12.44 & 2.70 & 10.57 & 0.62 \\
1500 -- 1550 & 623.76 & 18.81 & 8.95 & 2.49 & 14.31 & 0.71 \\
1550 -- 1600 & 485.77 & 13.93 & -0.16 & 2.53 & 21.27 & 0.86 \\
1600 -- 1650 & 404.60 & 12.95 & -14.62 & 2.46 & 20.19 & 0.91 \\
1650 -- 1700 & 392.84 & 11.13 & -32.37 & 2.25 & 18.27 & 0.81 \\
1700 -- 1750 & 393.10 & 12.46 & -25.07 & 2.20 & 10.40 & 0.71 \\
1750 -- 1800 & 374.26 & 11.31 & -15.43 & 2.05 & 8.78 & 0.65 \\
1800 -- 1850 & 353.00 & 10.17 & -9.56 & 1.93 & 8.78 & 0.70 \\
1850 -- 1900 & 267.74 & 9.01 & -3.44 & 1.89 & 9.95 & 0.77 \\
1900 -- 1950 & 227.93 & 7.76 & -11.16 & 1.86 & 12.21 & 0.83 \\
1950 -- 2000 & 234.80 & 7.47 & -16.46 & 1.83 & 11.11 & 0.82 \\
2000 -- 2100 & 222.41 & 3.97 & -14.31 & 0.93 & 6.37 & 0.42 \\
2100 -- 2200 & 168.32 & 3.53 & -4.86 & 0.87 & 5.28 & 0.44 \\
2200 -- 2300 & 120.67 & 2.64 & -5.61 & 0.82 & 6.79 & 0.49 \\
2300 -- 2400 & 111.78 & 2.44 & -9.24 & 0.80 & 3.49 & 0.51 \\
2400 -- 2500 & 88.87 & 2.16 & -3.60 & 0.77 & 3.65 & 0.54 \\
2500 -- 2600 & 68.44 & 1.92 & -3.78 & 0.75 & 2.54 & 0.59 \\
2600 -- 2700 & 60.31 & 1.78 & -3.49 & 0.76 & 1.85 & 0.64 \\
2700 -- 2800 & 50.13 & 1.69 & -2.32 & 0.78 & 1.63 & 0.71 \\
2800 -- 2900 & 38.42 & 1.55 & -0.52 & 0.79 & 1.23 & 0.80 \\
2900 -- 3000 & 31.51 & 1.51 & -2.48 & 0.82 & -0.29 & 0.90 \\
\hline
\end{tabular}
\caption[Minimum-variance SPT-3G 2018 \TTTEEE{} band powers]{
The SPT-3G 2018 \TTTEEE{} minimum-variance band powers $D_b$ and their associated uncertainties $\sigma_B$ for each angular multipole bin.
The band powers and errors are quoted in units of $\mu$K$^2$.}
\label{tab:MV_bandpowers_table}
\end{table}

\section{The SPT-3G 2018 Power Spectra}
\label{sec:ps}

We report the SPT-3G 2018 \TTTEEE{} multifrequency band powers in Appendix \ref{app:bandpowers} and plot the power spectrum measurement in Figure \ref{fig:PS_all_freqs}.
The SPT-3G 2018 \TT{} power spectra are sample-variance-dominated across the entire multipole range.
The \EE{} and \TE{} band powers are sample-variance-dominated for $\ell < 1275$ and $\ell < 1425$, respectively.

\begin{figure}[ht!]
  \centering
  \includegraphics[width=\linewidth]{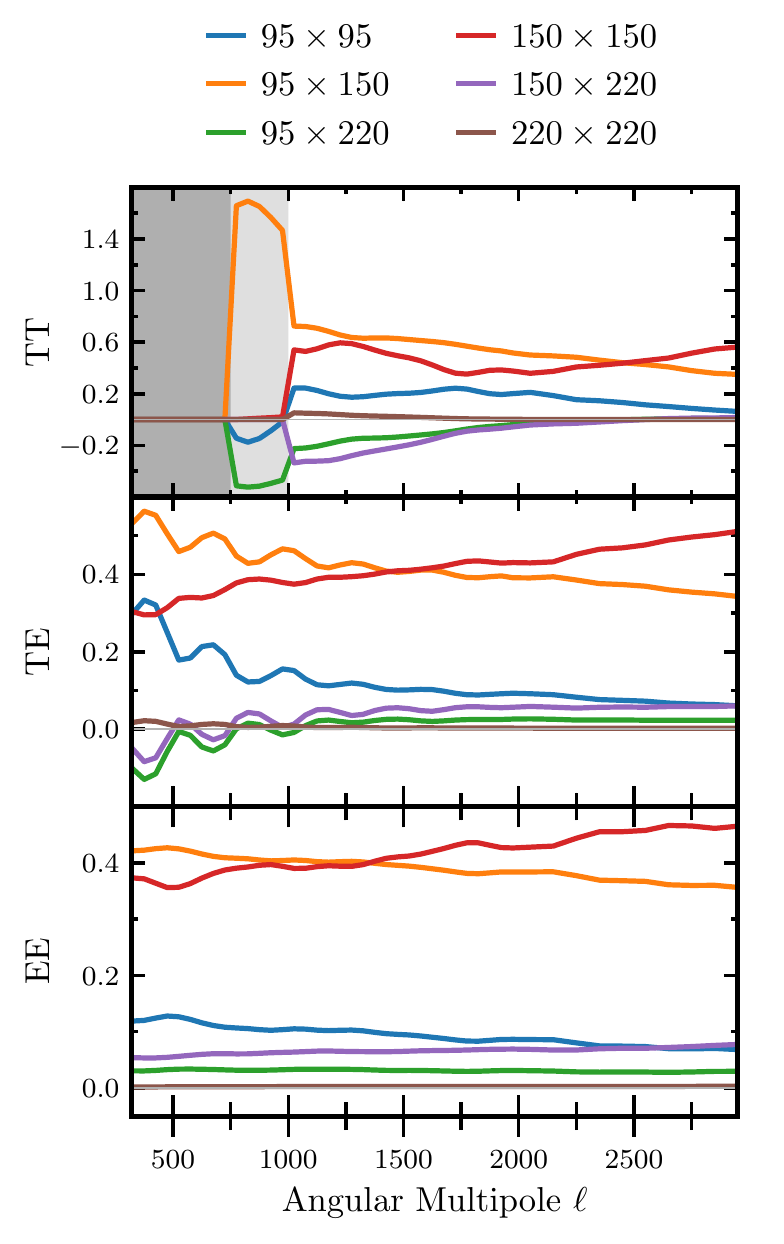}
  \caption{\label{fig:mixing_matrix}
  Relative weight of each multifrequency spectrum entering the minimum-variance combination (diagonal elements of the mixing matrix).
  The gray shaded areas indicate the different $\ell_{\mathrm{min}}$ cuts of the \TT{} spectra.
  Overall, the $95\times 150\,\mathrm{GHz}$ and $150\times 150\,\mathrm{GHz}$ spectra contribute the most weight.
  For \TT{} data, all spectra bar the $220\times 220\,\mathrm{GHz}$ band powers are non-negligible at intermediate $\ell$ and the $95\times 95\,\mathrm{GHz}$ \TE{} data are important on large angular scales.
  }
\end{figure}

\begin{figure*}[ht!]
  \centering
  \includegraphics[width=\linewidth]{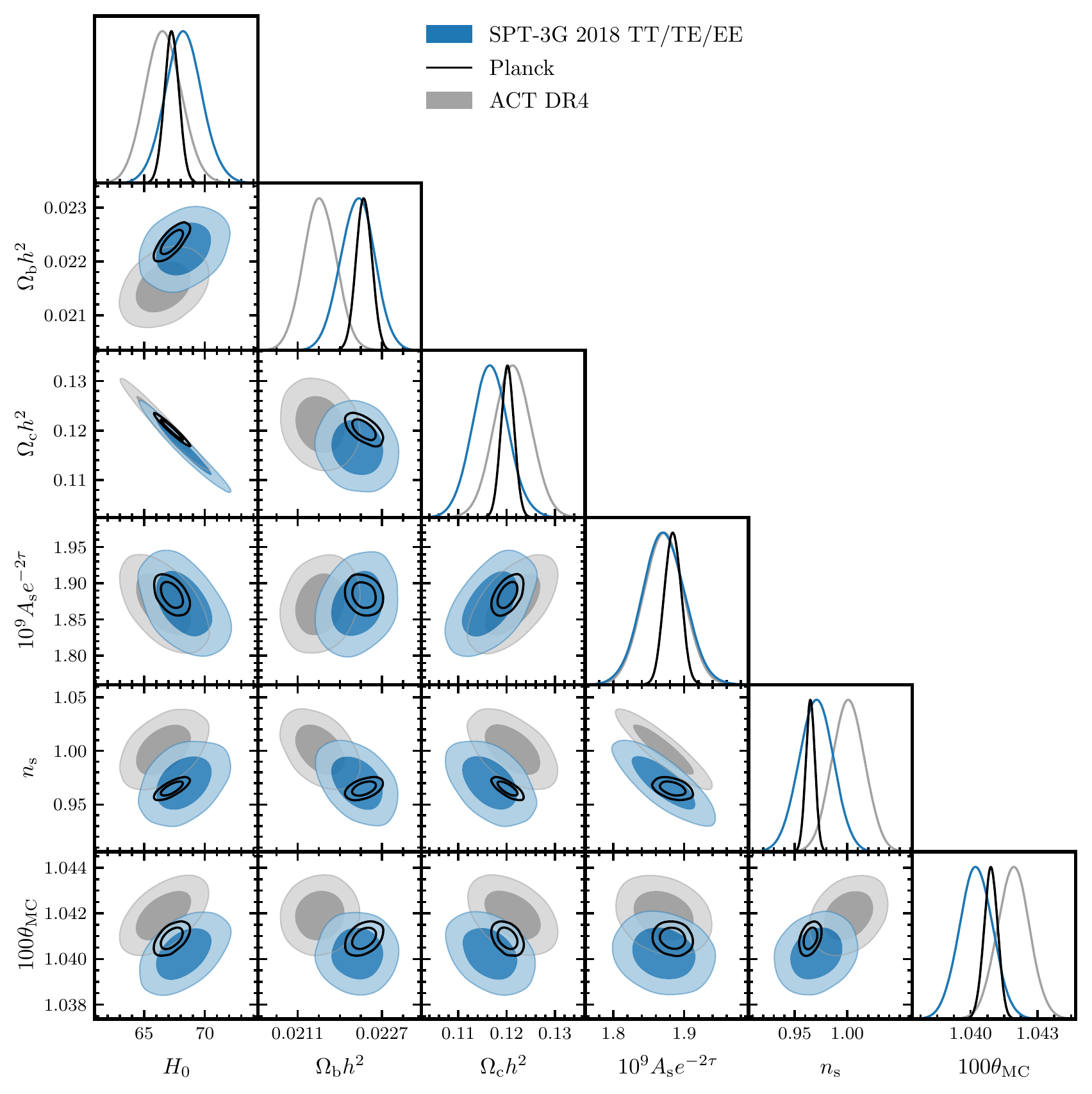}
  \caption{\label{fig:lcdm_triangle}
  Marginalized one- and two- dimensional posterior distributions for the SPT-3G 2018 \TTTEEE{} data set (blue contours), \planck{} (black line contours), and ACT DR4 (gray contours) in \lcdm{}.
  The constraints derived from SPT-3G data are in excellent agreement with the \planck{} constraints, including for \Hubble{}.
  The SPT-3G and ACT data have similar constraining power and the differences in their constraints are compatible with statistical fluctuations.
  }
\end{figure*}

\begin{figure*}[ht!]
  \centering
  \includegraphics[width=\linewidth]{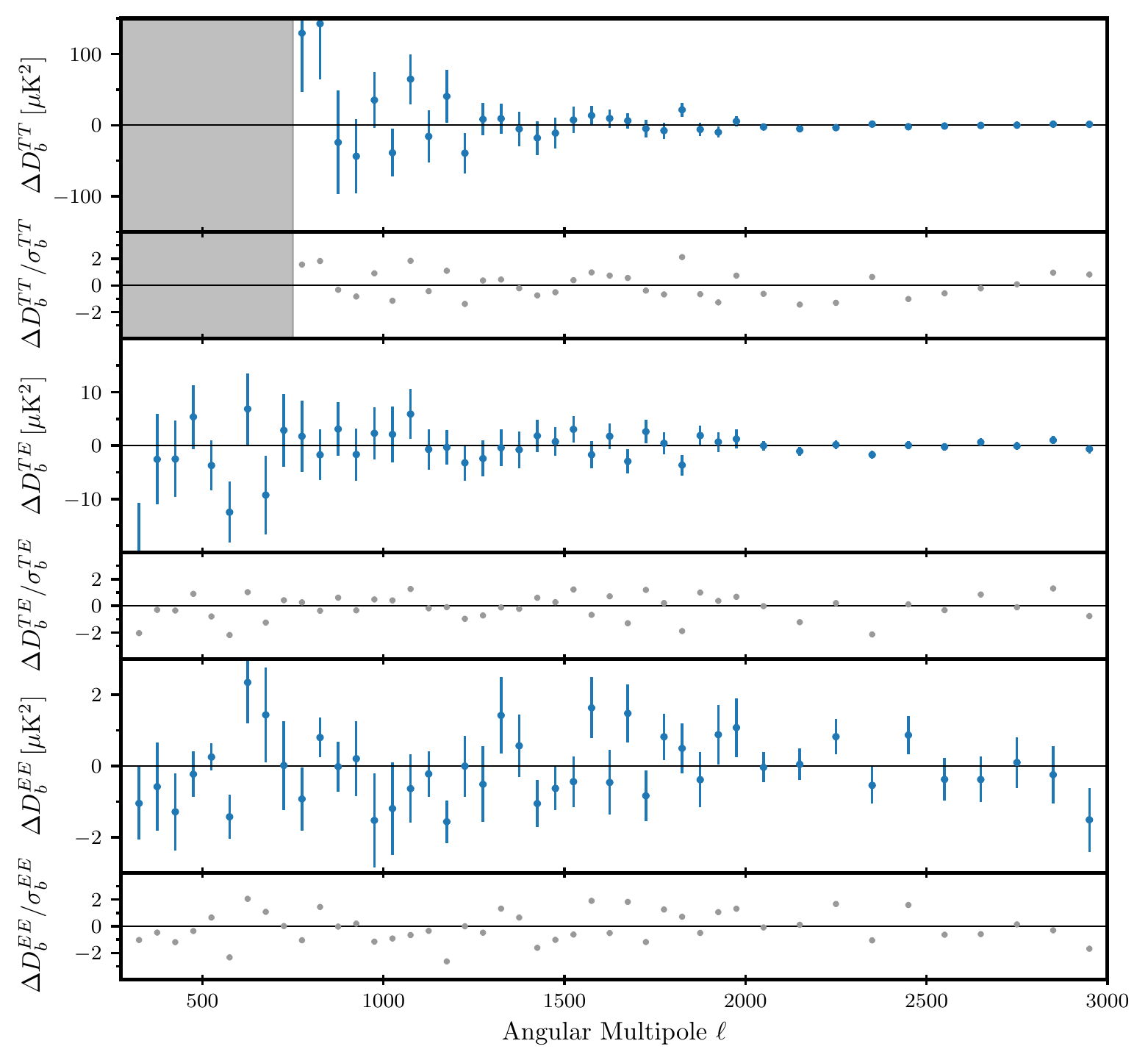}
  \caption{\label{fig:lcdm_resids}
  Residuals of the SPT-3G 2018 \TTTEEE{} minimum-variance data band powers to the best-fit \lcdm{} model.
  Note that the SPT-3G band powers are correlated by up to $40\%$ for neighboring bins.
  The standard model fits the data well and we report $\chi^2=763$ for $723$ degrees of freedom.
  Residuals for the full array of multifrequency band powers are shown in Appendix \ref{app:full_resids}.
  }
\end{figure*}

We report the minimum-variance band powers formed in \S\ref{sec:ps_consistency} in Table \ref{tab:MV_bandpowers_table} and plot them together with other select power spectrum measurements in Figure \ref{fig:PS_MV}.
Note that the minimum-variance band powers are only intended for plotting purposes and the likelihood uses the full set of multifrequency spectra.
The uncertainty of the minimum-variance combination is reduced by $3\%$, $2-19\%$, and $4-31\%$ compared to the $150 \times 150\,\mathrm{GHz}$ \TT{}, \TE{}, and \EE{} band powers, respectively.
This improvement is constant across scales for the sample-variance-limited \TT{} spectra and increases at higher $\ell$ for the noise-limited polarization spectra.

We can assess the relative weight of each multifrequency spectrum entering the minimum-variance contribution using the diagonals of the mixing matrix, $\left( X^T \mathcal{C}^{-1} X \right)^{-1} X^T \mathcal{C}^{-1}$, which are shown in Figure \ref{fig:mixing_matrix}.
Note that the absolute amplitudes of these elements correspond to the relative weights; the signs depend on the correlation structure and ensure that the sum of all elements is unity.
We find that the $95\times 150\,\mathrm{GHz}$ and $150\times 150\,\mathrm{GHz}$ spectra generally dominate the minimum-variance combination.
For \TT{}, these spectra combine to contribute $60\%$ of the total weight at $\ell=1000$, which increases to $91\%$ at $\ell=3000$.
There is an abrupt change at $\ell=1000$, i.e. when all multifrequency spectra are considered, while at larger angular scales the $95\times 150\,\mathrm{GHz}$ frequency combination alone dominates the minimum-variance contribution.
This is because (1) the $95\times 150\,\mathrm{GHz}$ and $150\times 150\,\mathrm{GHz}$ spectra are highly correlated on large angular scales while the former has a lower noise level and (2) the high degree of correlation between $150\,\mathrm{GHz}$ and $220\,\mathrm{GHz}$ noise leads to a more complex interplay between data from all three frequency channels in the minimum-variance combination when the $150\times 220\,\mathrm{GHz}$ and $220\times 220\,\mathrm{GHz}$ spectra are available.
For \EE{} and \TE{}, the $95\times 150\,\mathrm{GHz}$ and $150\times 150\,\mathrm{GHz}$ data contribute $65\%$ and $79\%$ at $\ell=300$ and $85\%$ and $82\%$ at $\ell=3000$, respectively.
Though the $95\times 150\,\mathrm{GHz}$ and $150\times 150\,\mathrm{GHz}$ data have a high combined weight, a wide frequency coverage is essential to control the foreground contamination and provides sensitivity to systematics.

\section{Cosmological Constraints}
\label{sec:cosmo}

\subsection{\lcdm{}}
\label{sec:lcdm}

We report constraints on cosmological parameters in \lcdm{} from SPT-3G 2018 \TTTEEE{} in Table \ref{tab:lcdm_param_table} and show one- and two-dimensional marginalized posterior distributions in Figure \ref{fig:lcdm_triangle}.
The best-fit values for nuisance parameters all lie within $1.2\,\sigma$ of the central value of their respective prior and are given in Appendix \ref{app:priors}.
We show residuals between the minimum-variance data band powers and the best-fit model in Figure \ref{fig:lcdm_resids} and plot the residuals for all multifrequency spectra in Appendix \ref{app:full_resids}.

\begin{figure}[ht!]
  \centering
  \includegraphics[width=\linewidth]{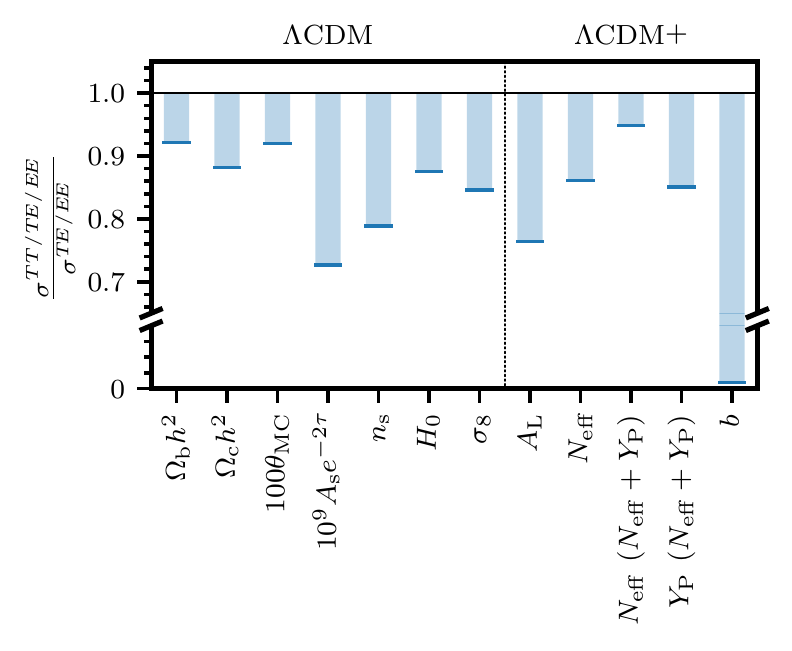}
  \caption{\label{fig:TT_improv}
  Ratio of the widths of marginalized posteriors from SPT-3G 2018 \TTTEEE{} and \EETE{} for select \lcdm{} parameters (left half) and extension parameters (right half).
  The addition of \TT{} data leads to improvements on core \lcdm{} parameters between $8-27\%$ and the $\Hubble{}$ and $\sigma_8$ posteriors tighten by $12\%$ and $15\%$, respectively.
  For \lcdm{}$+A_{\mathrm{L}}$, \lcdm{}$+\Neff{}$, and \lcdm{}$+\Neff{}+\Yp{}$ we report improvements for extension parameters between $5-24\%$.
  In the case of primordial magnetic fields, \lcdm{}$+b$, \EETE{} data alone suffers from a degeneracy between $n_\mathrm{s}$ and $b$ and only the addition of \TT{} data allows for a meaningful constraint.
  The vertical axis is split and the improvement on $b$ shown only for visualization purposes.}
\end{figure}

We find that the \lcdm{} model provides a good fit to the data.
We report $\chi^2=763.0$ across the $728$ band powers of the full data set.
We ignore the effect of nuisance parameters and translate this $\chi^2$ value to a PTE value of $15\%$.
This agreement also applies to the three spectrum types individually.
For \TT{}, \TE{}, and \EE{} data we report $\chi^2$ (PTE) values of $194.4\,(60\%)$, $273.4\, (33\%)$, and $285.5\, (17\%)$, respectively.\footnote{While the foreground model helps improve the fit to the temperature data substantially, determining the effective number of degrees of freedom is not straightforward. If we conservatively account for $15$ additional parameters, covering all baseline nuisance parameters, bar $\bar{\kappa}$, the polarization foreground parameters, and the calibration parameters (following \citetalias{dutcher21}), we find a PTE value of $8\%$ for the full data set and $30\%$ for \TT{}. These values still indicate that \lcdm{} provides a good fit to the data.}
All PTE values lie in the central 95th percentile, indicating the data are well fit by the standard model of cosmology.

The addition of temperature data to the \EETE{} spectra noticeably improves constraints on all cosmological parameters as shown in Figure \ref{fig:TT_improv}.
The posteriors of $\Omega_{\mathrm{b}} h^2$, $\Omega_{\mathrm{c}} h^2$, $\theta_{\mathrm{MC}}$, $10^9 A_{\mathrm{s}} e^{-2\tau}$, and $n_{\mathrm{s}}$ tighten by $8\%$, $12\%$, $8\%$, $27\%$, and $21\%$, respectively.
The uncertainty on the $H_0$ constraint shrinks by $12\%$.
We use the determinant of the parameter covariance as a metric for the allowed multi-dimensional volume, finding a reduction of the five-dimensional allowed parameter volume by a factor of $2.7$.

\begin{figure}[ht!]
  \centering
  \includegraphics[width=0.9\linewidth]{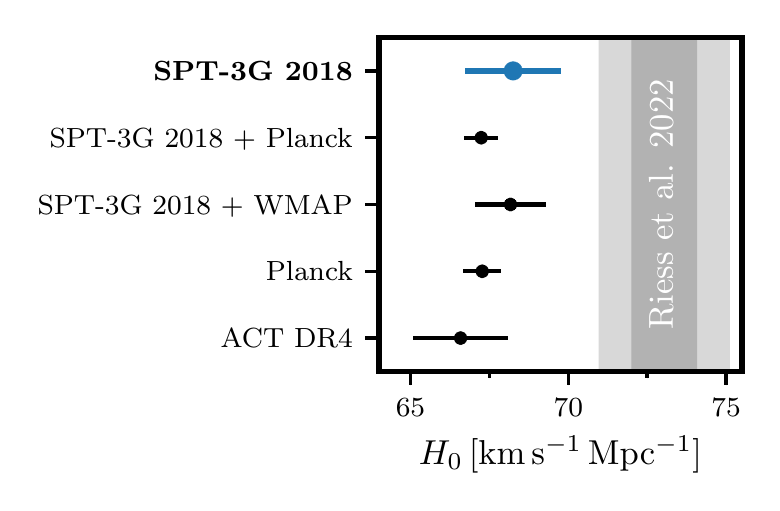}
  \caption{\label{fig:H0}
  Compilation of $\Hubble$ constraints from combinations of different CMB data sets assuming \lcdm{}: SPT-3G 2018, \planck{} \citep{planck18-6}, \WMAP{} \citep{bennett13}, ACT DR4 \citep{aiola20}.
  The vertical gray band indicates the $2\,\sigma$ constraint from the most precise supernovae and distance ladder analysis \citep{riess22}.
  SPT-3G 2018 data allow for a precision constraint on $H_0$ effectively independent from \planck{} data that deepens the Hubble tension.
  }
\end{figure}

Constraints on the expansion rate today based on CMB data and supernovae and distance-ladder analyses are discrepant at the $4-5\,\sigma$ level \citep{dutcher21, planck18-6, aiola20, balkenhol21, riess22}.
With SPT-3G 2018 \TTTEEE{} data we constrain the Hubble constant to
\begin{equation}
H_0 = 68.3 \pm 1.5\,\kmsmpc{}.
\end{equation}
This value is in excellent agreement with the most recent results from \planck{} \citep{planck18-6} and ACT \citep{aiola20}.
Conversely, our result lies $2.6\,\sigma$ below the most precise local determination of the Hubble constant, the Cepheid-calibrated supernovae distance-ladder analysis of \citet{riess22}.
The SPT-3G 2018 \TTTEEE{} data set is effectively independent of \planck{} and ACT data so this result deepens the Hubble tension.
Our \Hubble{} constraint lies $0.6\,\sigma$ below the distance-ladder analysis using the tip-of-the-red-giant-branch approach by \citet{freedman19}.
Moreover, it is $2.1\,\sigma$ and $1.0\,\sigma$ below the result of \citet{wong19} and \citet{birrer20} using strong-lensing time delays.

\begin{figure}[ht!]
  \centering
  \begin{subfigure}{}
    \includegraphics[width=\linewidth]{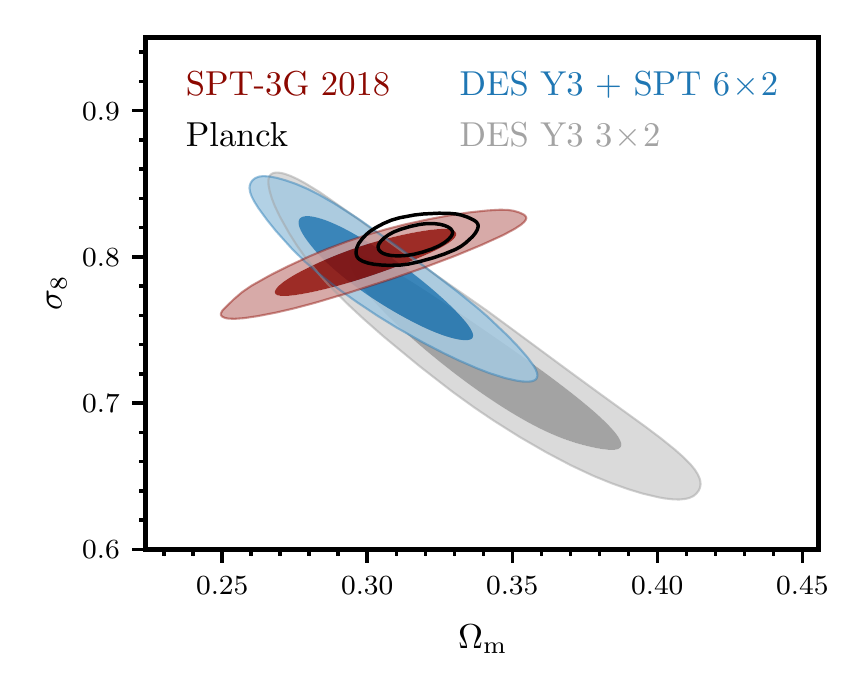}
  \end{subfigure}\hfill
  \begin{subfigure}{}
    \includegraphics[width=0.9\linewidth]{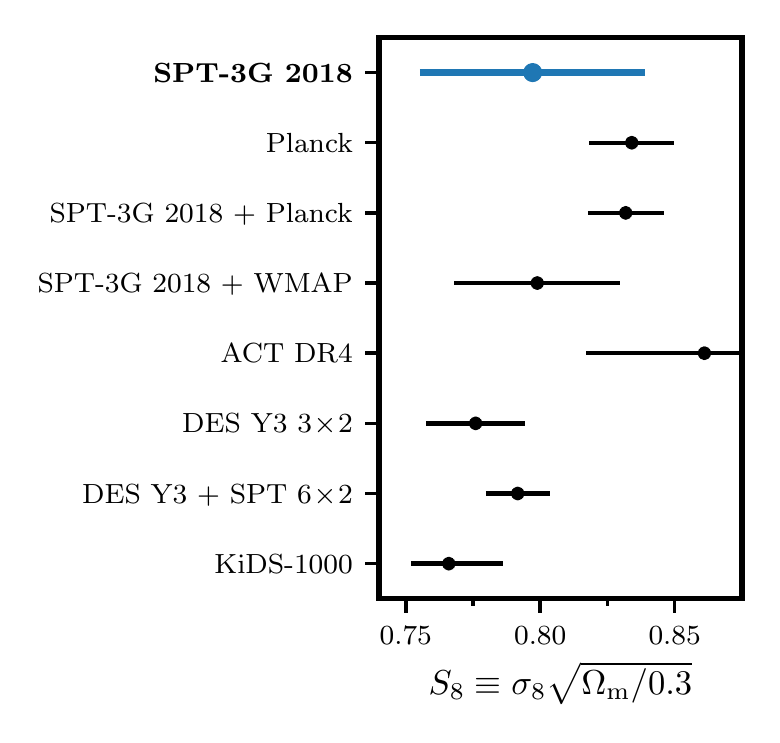}
  \end{subfigure}
  \caption{  \label{fig:S8}
\emph{Top panel:} Constraints in the $\sigma_8$ vs. $\Omega_{\mathrm{m}}$ plane from SPT-3G 2018 (red), \planck{} (black line), a joint analysis of DES Y3 galaxy position and lensing data and SPT and \planck{} CMB lensing data ($6\!\times\!2$, blue) \citep{abbott22b}, and DES Y3 joint galaxy density and weak lensing data ($3\!\times\!2$, gray) \citep{abbott22a}.
The combined structure growth parameter, $S_8 \equiv \sigma_8 \sqrt{\Omega_{\mathrm{m}}/0.3}$, varies perpendicular to the degeneracy direction of the DES data.\\
\emph{Bottom panel:} A compilation of $S_8$ constraints using different cosmological data sets: SPT-3G 2018, \planck{} \citep{planck18-6}, \WMAP{} \citep{bennett13}, ACT DR4 \citep{aiola20}, DES Y3 \citep{abbott22a}, DES Y3 + SPT \citep{abbott22b}, and KiDS-1000 \citep{heymans20}.
Note that all constraints are produced assuming \lcdm{}.
The central value of the SPT-3G constraint lies between those of low-redshift analyses and \planck{}.
}
\end{figure}

Next, we look at structure growth as parametrized by the amplitude of matter fluctuations within a sphere with comoving volume of $8\,\mathrm{Mpc^{-1}}$, $\sigma_8$, and the combined structure growth parameter $S_8 \equiv \sigma_8 \sqrt{\Omega_m/0.3}$.
The \planck{} constraint on $S_8$ using primary CMB data lies approximately $3\,\sigma$ above the results of joint galaxy clustering and weak lensing analyses \citep{planck18-6, heymans20, abbott22a} as shown in the bottom panel of Figure \ref{fig:S8}.
For SPT-3G 2018 \TTTEEE{} we report:
\begin{equation}
\begin{aligned}
\sigma_8 &= 0.797 \pm 0.015,\\
S_8 &= 0.797 \pm 0.042.
\end{aligned}
\end{equation}
This result lies between $S_8$ constraints from \planck{} data and low redshift data as shown in the top panel of Figure \ref{fig:S8}; our central value is  $0.8\,\sigma$ below the \planck{} constraint \citep{planck18-6} and $0.5\,\sigma$ and $0.7\,\sigma$ higher than the DES-Y3 \citep{abbott22a} and KiDS-1000 \citep{heymans20} results, respectively.
Adjusting our definition of $S_8$ appropriately, we find agreement at $0.9\,\sigma$ with the SZ-cluster analysis of \citet{bocquet19}.

We find the scalar spectral index of primordial fluctuations to be $n_s = 0.970 \pm 0.016$, which corresponds to a $1.8\,\sigma$ preference for $n_{\mathrm{s}} < 1$.
We note that when excising our measurement of the third acoustic peak of the temperature power spectrum, i.e. \TT{} data at $\ell < 1000$, we find $n_{\mathrm{s}} = 0.994 \pm 0.018$.
The corresponding five-dimensional parameter shift from the baseline result is a $2.2\,\sigma$ event, where $\sigma$ denotes the number of standard deviations equivalent to the associated PTE for a Gaussian distribution.
This is compatible with a statistical fluctuation and we therefore expect that the addition of more data to the subset, i.e. our baseline configuration with \TT{} data at $\ell < 1000$, yields constraints closer to the underlying mean.
This matches what we observe when comparing to the tight constraints of \planck{} and \WMAP{} \citep{planck18-6, hinshaw13}, which are enabled by the broad coverage of scales in $\log\ell$ space of satellite data; adding \TT{} data at $\ell < 1000$ to the $\TT{}\,\ell>1000/\TE{}/\EE{}$ subset shifts our $n_{\mathrm{s}}$ result towards these tight constraints.

For a less model-dependent check on our \TT{} measurement at $750 < \ell < 1000$ we compare our minimum-variance band powers to the \planck{} full-sky power spectrum.
Given that both data sets are sample-variance-dominated on these angular scales, we assume that the SPT data are a subset of the \planck{} data; we use the difference of the SPT and \planck{} band power covariance matrices as the covariance of the difference between the two \TT{} data sets.
We report a PTE value of $9\%$.
This indicates that the two power spectrum measurements are in good agreement and we conclude that the effect the SPT-3G \TT{} data at $\ell < 1000$ has on $n_{\mathrm{s}}$ is not statistically anomalous.

\begin{table*}[ht!]
\def\arraystretch{1.2}
\footnotesize
\setlength{\tabcolsep}{8pt}
\centering
\begin{tabular}{l @{\hskip 20pt} D{+}{\,\pm\,}{7.7} D{+}{\,\pm\,}{7.7} D{+}{\,\pm\,}{7.7} D{+}{\,\pm\,}{7.7}}
\hline\hline
\vphantom{\makecell{A\\A\\A}} & \multicolumn{1}{c}{\makecell{SPT-3G 2018}} & \multicolumn{1}{c}{\makecell{SPT-3G 2018\\ + \planck{}}} & \multicolumn{1}{c}{\makecell{SPT-3G 2018\\ + \WMAP{}}} & \multicolumn{1}{c}{\makecell{\planck{}}}\\
\hline
$\Omega_{\mathrm{b}}h^2$ &
  0.02224+0.00032&
  0.02233+0.00013&
  0.02240+0.00020&
  0.02236+0.00015\\
$\Omega_{\mathrm{c}}h^2$ &
  0.1166+0.0038&
  0.1201+0.0012&
  0.1171+0.0027&
  0.1202+0.0014\\
$100\theta_{\mathrm{MC}}$ &
  1.04025+0.00074&
  1.04075+0.00028&
  1.04016+0.00067&
  1.04090+0.00031\\
$10^9 A_{\mathrm{s}} e^{-2\tau}$ &
  1.871+0.030&
  1.884+0.010&
  1.867+0.016&
  1.884+0.012\\
$n_{\mathrm{s}}$ &
  0.970+0.016&
  0.9649+0.0041&
  0.9671+0.0063&
  0.9649+0.0044\\
\hline
$H_{\mathrm{0}}\,[\kmsmpc{}]$ &
  68.3+1.5&
  67.24+0.54&
  68.2+1.1&
  67.27+0.60\\
$\sigma_8$ &
  0.797+0.015&
  0.8099+0.0067&
  0.796+0.012&
  0.8120+0.0073\\
$S_8 \equiv \sigma_8 \sqrt{\Omega_{\mathrm{m}}/0.3}$ &
  0.797+0.042&
  0.832+0.014&
  0.799+0.031&
  0.834+0.016\\
$\Omega_{\mathrm{\Lambda}}$ &
  0.700+0.021&
  0.6835+0.0075&
  0.698+0.015&
  0.6834+0.0084\\
${\mathrm{Age}}/{\mathrm{Gyr}}$ &
  13.815+0.047&
  13.807+0.021&
  13.804+0.037&
  13.800+0.024\\
\hline
\end{tabular}
\caption[
LCDM table
]{
Marginalized constraints and 68\% uncertainties on $\Lambda$CDM parameters from SPT-3G 2018 \TTTEEE{}, along with joint constraints from SPT-3G 2018 \TTTEEE{} + \planck{}, SPT-3G 2018 \TTTEEE{} + \WMAP{}, and results from \planck{} alone \citep{planck18-6, bennett13}. We show constraints on the baseline $\Lambda$CDM parameters in the top half of the table, combining the optical depth to reionization and amplitude of primordial fluctuations into $10^9 A_{\mathrm{s}} e^{-2\tau}$. The bottom half shows select derived parameters. Note that we do not use \WMAP{} polarization data at $\ell < 24$ and SPT-3G data alone do not constrain the optical depth to reionization $\tau$; instead, we use a \planck-based Gaussian prior of $\tau = 0.0540 \pm 0.0074$.}
\label{tab:lcdm_param_table}
\end{table*}

We find excellent agreement between cosmological constraints from SPT-3G 2018 \TTTEEE{} and \planck{} data.
For individual \lcdm{} parameters, all differences are $<1\,\sigma$.
Comparing all five parameters constrained by the SPT data, we find $\chi^2=2.6$, corresponding to a PTE value of $76\%$.
This indicates a high level of agreement between the two data sets.
This is particularly striking given that SPT-3G and \planck{} constraints are effectively independent of one another, given the large amount of sky observed by \planck{} that is not observed by SPT and the different $\ell$ weighting of the data as well as the different weightings of the \TT{}, \TE{}, and \EE{} spectra.
Though we use \planck{} data to calibrate our power spectrum measurement, we marginalize over the temperature calibration and polarization efficiency in the likelihood analysis.
Furthermore, as per \S\ref{sec:like_robust} we find that our cosmological results are robust when replacing the \planck{}-based prior on the optical depth to reionization with the result of \citet{natale20}.
The agreement between SPT-3G and \planck{} data is not only a strong argument for the consistency and robustness of both experiments' cosmological results, but implies consistency of the \lcdm{} model across angular scales and temperature and polarization spectra.

We find acceptable agreement between constraints from SPT-3G 2018 \TTTEEE{} and ACT DR4.
Across the five \lcdm{} parameters constrained by the ground-based experiments, we find $\chi^2=10.4$, which translates to a PTE value of $6\%$.
Interestingly, the largest difference is in $\theta_{\mathrm{MC}}$, which controls the positions of acoustic peaks; CMB data constrain this parameter with great precision and SPT-3G 2018 \TTTEEE{} yields a $0.07\%$ measurement.
ACT data yield a value $2.0\,\sigma$ and $1.7\,\sigma$ larger than SPT-3G and \planck{} data, respectively.
\citet{aiola20} note an offset in the cosmological parameter constraints on $n_{\mathrm{s}}$ and $\Omega_{\mathrm{b}}h^2$ when comparing \planck{} and ACT results (also visible in Fig. \ref{fig:lcdm_triangle}).
Due to the degeneracy of these parameters with $\theta_{\mathrm{MC}}$, the observed offset between ACT and SPT-3G constraints is likely related and from a similar origin.
Regardless, the multi-dimensional test indicates that the observed parameter shifts are compatible with statistical fluctuations.

We report joint constraints from SPT-3G 2018 \TTTEEE{} and \planck{} data in Table \ref{tab:lcdm_param_table} and find $H_0 = 67.24 \pm 0.54\,\kmsmpc{}$.
This is a refinement of the \planck{} constraint on \Hubble{} by $11\%$.
The precision measurement of the CMB anisotropies at small angular scales in temperature and polarization provided by SPT-3G shrinks the \planck{} posteriors by approximately $10\%$ for each \lcdm{} parameter.
Across the six-dimensional parameter space we report a reduction of the allowed volume by a factor of $1.7$; for comparison, only adding the SPT \EETE{} data to \planck{} leads to a reduction of the allowed parameter volume by a factor of $1.4$.
Due to the excellent agreement of SPT and \planck{} data, the shift to central values of parameter constraints compared to \planck{} alone is small.

The SPT-3G 2018 data are in good agreement with \WMAP{} and we report a PTE value for a five-dimensional parameter-space comparison of $95\%$.
Combining the SPT-3G and \WMAP{} data yields constraints largely independent of \planck{}, which we list in Table \ref{tab:lcdm_param_table}.
We report $\Hubble = 68.2 \pm 1.1\,\kmsmpc$, which lies $3.2\,\sigma$ below the distance-ladder analysis of \citet{riess22} and deepens the Hubble tension.
We report a constraint on the combined structure growth parameter of $S_8 = 0.799 \pm 0.031$, which is compatible with \planck{}, as well as DES Y3 and KiDS-1000 data and the SZ-cluster analysis of \citet{bocquet19} within $1\,\sigma$. \citep{planck18-6, heymans20, abbott22a}.
The addition of the low $\ell$ power spectrum measurement of \WMAP{} to SPT-3G data refines our $n_{\mathrm{s}}$ constraint by $62\%$.
We report $n_{\mathrm{s}} = 0.9671 \pm 0.0063$, which disfavors a scale-invariant Harrison-Zel'dovich spectrum at $5.2\,\sigma$.
For comparison, from \WMAP{} data alone we infer $n_s=0.967 \pm 0.012$, which is $2.8\,\sigma$ from unity; the addition of SPT data tightens the $n_s$ constraint derived from \WMAP{} data alone by $46\%$.

\subsection{Gravitational Lensing, $A_{\mathrm{L}}$}
\label{sec:alens}

The lensing of CMB photons emitted at the surface of last scattering by intervening large scale structure causes a characteristic distortion of the CMB anisotropies leading to changes in the power spectrum: a smoothing of acoustic peaks and a transfer of power to the damping tail.
Though the magnitude of this effect is derived from the values of cosmological parameters in the \lcdm{} model, marginalizing over the effect of lensing on the primary CMB power spectra assesses the compatibility of the data with the standard model \citep{seljak96b, lewis06, calabrese08}.
\citet{planck18-6} find a preference for increased lensing at $2.8\,\sigma$.

We marginalize over an artificial scaling of the lensing power spectrum that smears the primary CMB, $A_{\mathrm{L}}$, and report parameter constraints in Table \ref{tab:ext_param_table_w_planck}.
We find
\begin{equation}
A_{\mathrm{L}} = 0.87 \pm 0.11.
\end{equation}
which is compatible with the standard model prediction of unity at $1.3\,\sigma$.
Adding $A_{\mathrm{L}}$ does not lead to a statistically significant improvement to the goodness-of-fit compared to \lcdm{} ($\Delta \chi^2 = -1.3$).

\begin{table*}[ht!]
\def\arraystretch{1.2}
\footnotesize
\setlength{\tabcolsep}{1.5pt}
\centering
\begin{tabular}{l D{+}{\,\pm\,}{7.7} D{+}{\,\pm\,}{7.7} D{+}{\,\pm\,}{7.7} D{+}{\,\pm\,}{7.7} D{+}{\,\pm\,}{7.7} D{+}{\,\pm\,}{7.7}}
\hline\hline
& \multicolumn{2}{c}{\text{$A_{\rm L}$}}
& \multicolumn{2}{c}{\text{$N_{\rm eff}$}}
& \multicolumn{2}{c}{\text{$N_{\rm eff} + Y_{\mathrm{P}}$}} \\
\cmidrule(lr){2-3} \cmidrule(lr){4-5} \cmidrule(lr){6-7}
& \multicolumn{1}{c}{SPT-3G 2018} & \multicolumn{1}{c}{\makecell{SPT-3G 2018\\ + \planck{}}} & \multicolumn{1}{c}{SPT-3G 2018} & \multicolumn{1}{c}{\makecell{SPT-3G 2018\\ + \planck{}}} & \multicolumn{1}{c}{SPT-3G 2018} & \multicolumn{1}{c}{\makecell{SPT-3G 2018\\ + \planck{}}}\\
 \hline
$\Omega_{\mathrm{b}}h^2$ &
  0.02213+0.00033 &
  0.02243+0.00015 &
  0.02254+0.00046 &
  0.02229+0.00020 &
  0.02235+0.00050 &
  0.02228+0.00020 \\
$\Omega_{\mathrm{c}}h^2$ &
  0.1222+0.0060 &
  0.1190+0.0014 &
  0.1235+0.0089 &
  0.1194+0.0028 &
  0.139+0.018 &
  0.1208+0.0042 \\
$100\theta_{\mathrm{MC}}$ &
  1.03982+0.00081 &
  1.04087+0.00029 &
  1.03980+0.00092 &
  1.04083+0.00039 &
  1.0359+0.0030 &
  1.0404+0.0011 \\
$10^9 A_{\mathrm{s}} e^{-2\tau}$ &
  1.905+0.041 &
  1.879+0.011 &
  1.886+0.037 &
  1.881+0.016 &
  1.918+0.046 &
  1.884+0.017 \\
$n_{\mathrm{s}}$ &
  0.956+0.020 &
  0.9677+0.0043 &
  1.001+0.040 &
  0.9628+0.0084 &
  0.985+0.043 &
  0.9630+0.0080 \\
$A_{\mathrm{L}}$ &
  0.87+0.11 &
  1.078+0.054 &
 \multicolumn{1}{D{+}{\,-\,}{7.7}}{+} &
 \multicolumn{1}{D{+}{\,-\,}{7.7}}{+} &
 \multicolumn{1}{D{+}{\,-\,}{7.7}}{+} &
 \multicolumn{1}{D{+}{\,-\,}{7.7}}{+} \\
$N_{\mathrm{eff}}$ &
 \multicolumn{1}{D{+}{\,-\,}{7.7}}{+} &
 \multicolumn{1}{D{+}{\,-\,}{7.7}}{+} &
  3.55+0.58 &
  3.00+0.18 &
  4.7+1.3 &
  3.09+0.28 \\
$Y_{\mathrm{P}}$ &
 \multicolumn{1}{D{+}{\,-\,}{7.7}}{+} &
 \multicolumn{1}{D{+}{\,-\,}{7.7}}{+} &
 \multicolumn{1}{D{+}{\,-\,}{7.7}}{+} &
 \multicolumn{1}{D{+}{\,-\,}{7.7}}{+} &
  0.165+0.058 &
  0.238+0.016 \\
\hline
$H_{\mathrm{0}}\,[\kmsmpc{}]$ &
  66.1+2.3 &
  67.73+0.64 &
  71.7+4.3 &
  66.9+1.4 &
  77.5+7.2 &
  67.4+1.7\\
$\sigma_8$ &
  0.819+0.023 &
  0.8031+0.0085 &
  0.817+0.029 &
  0.807+0.010 &
  0.831+0.035 &
  0.810+0.012\\
$S_8 \equiv \sigma_8 \sqrt{\Omega_{\mathrm{m}}/0.3}$ &
  0.864+0.071 &
  0.816+0.018 &
  0.799+0.043 &
  0.831+0.015 &
  0.791+0.043 &
  0.832+0.015\\
$\Omega_{\mathrm{\Lambda}}$ &
  0.666+0.037 &
  0.6901+0.0087 &
  0.713+0.026 &
  0.6821+0.0098 &
  0.727+0.029 &
  0.6832+0.0098\\
${\mathrm{Age}}/{\mathrm{Gyr}}$ &
  13.861+0.058 &
  13.789+0.024 &
  13.36+0.54 &
  13.86+0.19 &
  12.59+0.89 &
  13.78+0.25\\
\hline
\end{tabular}
\caption[
$\Lambda$CDM+ parameter constraints.
]{
Constraints on $\Lambda$CDM model extensions $A_L$, $\Neff$, and $\Neff+\Yp$ from SPT-3G 2018 \TTTEEE{} alone and in combination with \planck{} data.
}
\label{tab:ext_param_table_w_planck}
\end{table*}

The SPT-3G 2018 \TT{} band powers provide a sample-variance-limited measurement of the third and higher order acoustic peaks, which helps constrain cosmological parameters in this model.
The $A_{\mathrm{L}}$ constraint improves by $24\%$ for \TTTEEE{} compared to \EETE{} as shown in Figure \ref{fig:TT_improv}.
Across all six dimensions, the allowed parameter volume shrinks by a factor of $3.1$.

In this model the SPT-3G and \planck{} constraints slightly diverge.
\planck{} data yield $A_{\mathrm{L}} = 1.180 \pm 0.065$, which is $2.5\,\sigma$ away from our result.
Nevertheless, comparing the two data sets across the full six-dimensional parameter space gives $\chi^2=10.2$, which translates to a PTE value of $12\%$ and indicates that the parameter shifts are consistent with statistical fluctuations.

We report joint constraints from SPT-3G 2018 and \planck{} data in Table \ref{tab:ext_param_table_w_planck}.
We find $A_{\mathrm{L}} = 1.078 \pm 0.054$, which is within $1.5\,\sigma$ of the standard model prediction.
Adding SPT-3G to \planck{} data lowers the significance of the $A_{\mathrm{L}}$ deviation from unity and constraints on other cosmological parameters shift closer to the \planck{} only \lcdm{} results.
The width of the $A_{\mathrm{L}}$ posterior shrinks by $18\%$ when adding SPT-3G to \planck{} data and the seven-dimensional allowed parameter volume decreases by a factor of $2.0$.

We revisit the investigation of lensing convergence on the SPT-3G survey patch from \citet{balkenhol21} using the complete SPT-3G 2018 \TTTEEE{} data set.
We analyze joint constraints from SPT-3G 2018 and \planck{} data in \lcdm{} foregoing the baseline Gaussian prior on $\kappa$.
We adjust the sign of the $\kappa$ definition in \S\ref{sec:like} to match \citet{motloch19} and the appendix of \citet{balkenhol21}.
We find
\begin{equation}
10^3 \kappa_{\rm SPT-3G} = -0.93 \pm 0.59,
\end{equation}
While the sign matches the result of \citet{balkenhol21}, our central value is compatible with zero at $1.6\,\sigma$.
We conclude that this test provides no significant evidence that the SPT-3G survey field aligns with a local density anomaly.

\subsection{Effective Number of Neutrino Species, $N_{\rm eff}$}
\label{sec:neff}

Additional relativistic particles in the early universe, e.g., axion-like particles, hidden photons, gravitinos, massless Goldstone bosons, additional neutrino species, as well as other forms of energy injection imprint on the CMB power spectra.
At the parameter level, this modifies the effective number of neutrino species, $\Neff$, which is $3.044$ in the standard model \citep{froustey20, bennett20, brust13, pdg2020, abazajian16}.

We report constraints on the \lcdm{}$+\Neff$ model in Table \ref{tab:ext_param_table_w_planck}, finding
\begin{equation}
\Neff = 3.55 \pm 0.58.
\end{equation}
This result is compatible with the standard model prediction at $0.9\,\sigma$.
The best-fit \lcdm{}$+\Neff$ model does not improve on the good fit to the SPT-3G data achieved by \lcdm{} significantly ($\Delta \chi^2 = -0.2$).

The addition of sample-variance-limited measurements of the damping tail of the \TT{} power spectrum improves on the cosmological constraints achieved by SPT-3G 2018 \EETE{} in this model.
As shown Figure \ref{fig:TT_improv}, the posterior of $\Neff$ tightens by $14\%$ when adding the SPT-3G 2018 \TT{} band powers.
The allowed volume across the full six dimensional parameter space shrinks by a factor of $2.8$.

We find good agreement on \Neff{} between the SPT-3G and \planck{} data with the central values separated by $1.0\,\sigma$.
Comparing all six parameters simultaneously, we find $\chi^2=3.3$, which translates to a PTE value of $77\%$.
The parameter constraints are compatible with statistical fluctuations.

We list joint constraints from SPT-3G 2018 and \planck{} in Table \ref{tab:ext_param_table_w_planck} and report $\Neff = 3.00 \pm 0.18$.
This constraint on the effective number of neutrino species is in excellent agreement with the standard model prediction of $3.044$ ($0.2\,\sigma$).
While the addition of the SPT-3G to the \planck{} data set only leads to a marginal improvement of the $\Neff{}$ constraint ($4\%$), the allowed seven-dimensional parameter volume is reduced by a factor of $1.5$.

\subsection{Effective Number of Neutrino Species and Primordial Helium Abundance, $N_{\rm eff} + Y_P$}
\label{sec:neff_yp}

Varying $\Neff$ alone assumes that any additional relativistic species present at recombination were also present at big-bang nucleosynthesis.
By simultaneously marginalizing over the primordial helium abundance, $\Yp$, we remove this assumption and flexibly probe the relativistic energy density in the early universe \citep{cyburt16, pdg2020}.

We present constraints from SPT-3G 2018 \TTTEEE{} in Table \ref{tab:ext_param_table_w_planck}.
We report
\begin{equation}
\begin{aligned}
\Neff &= 4.7 \pm 1.3,\\
\Yp &= 0.165 \pm 0.058.
\end{aligned}
\end{equation}
The central values of the \Neff{} and \Yp{} constraints are compatible with the standard model predictions at $1.3\,\sigma$ and $1.4\,\sigma$, respectively.
We report no significant improvement to the goodness-of-fit for this model over \lcdm{} ($\Delta \chi^2 = -2.1$ for two additional parameters).

Comparing the determinants of the parameter covariances when using \TTTEEE{} vs. \EETE{} data, we find that the allowed parameter volume is reduced by a factor of $2.4$ through the inclusion of temperature band powers.
The $\Neff{}$ and $\Yp{}$ uncertainties shrink by $5\%$ and $15\%$, respectively, which we show in Figure \ref{fig:TT_improv}.

Again, we find good agreement between SPT-3G and \planck{} data in this model: across the full seven-dimensional parameter space we report $\chi^2=4.5$, which translates to a PTE value of $72\%$.
The \Neff{} and \Yp{} constraints of the two data sets are compatible at $1.4\,\sigma$ and $1.3\,\sigma$, respectively.
We conclude that the differences in parameter constraints are compatible with statistical fluctuations.

Joint constraints from SPT-3G 2018 and \planck{} are given in Table \ref{tab:ext_param_table_w_planck}.
We report $\Neff = 3.09 \pm 0.28$ and $\Yp = 0.238 \pm 0.016$.
The central values of the joint SPT-3G and \planck{} $\Neff{}$ and $\Yp{}$ constraints lie within $0.2\,\sigma$ and $0.5\,\sigma$ of their standard model predictions, respectively, and improve on the \planck{} only results by $9\%$ and $8\%$, respectively.
Across the full eight-dimensional parameter space, the addition of SPT-3G to \planck{} data leads to a reduction of the allowed parameter volume by a factor of $1.8$.

\subsection{Primordial Magnetic Fields}
\label{sec:pmf}

\begin{figure}[ht!]
  \centering
  \includegraphics[width=\linewidth]{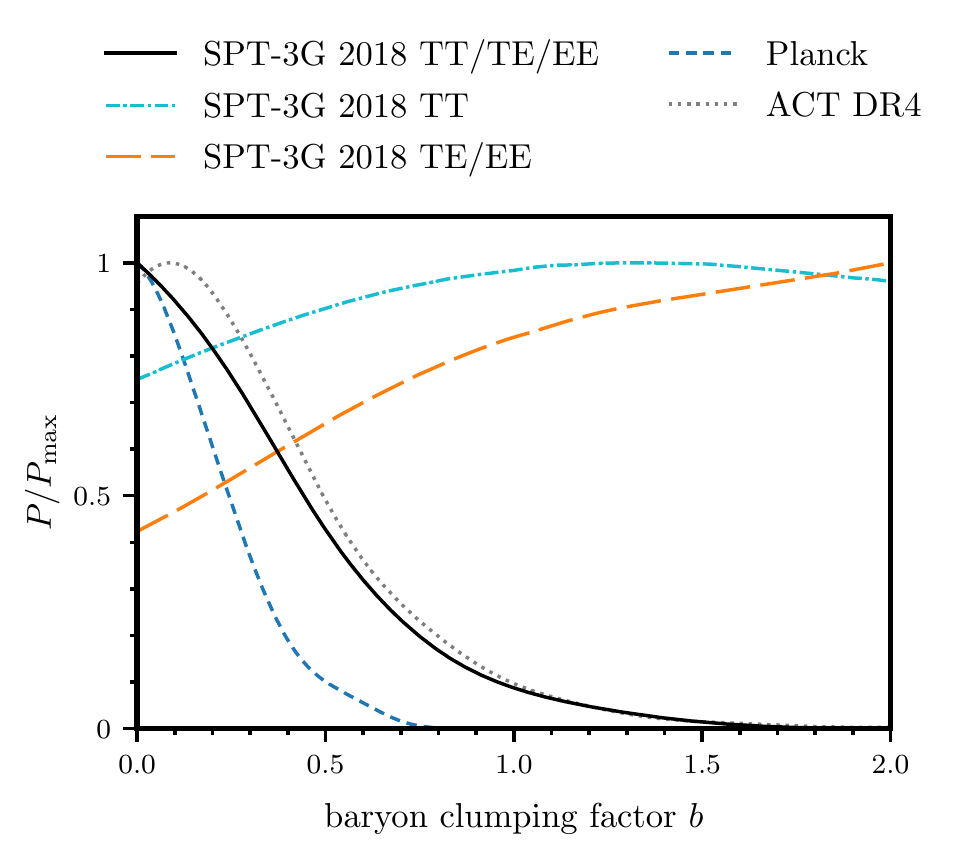}
  \caption{\label{fig:PMF}
  Marginalized one-dimensional posterior distributions for the SPT-3G 2018 \TTTEEE{} (black solid line), \TT{} (light blue dash-dotted line), and \EETE{} (orange long dashed line) on the clumping factor $b$ induced by primordial magnetic fields.
  We also show the constraints from \planck{} primary CMB and lensing data (dark blue short dashed line) and ACT DR4 (gray dotted line).
  The combination of \TT{} and \EETE{} spectra allows us to break degeneracies and set a tight constraint on $b$.
  The SPT-3G and ACT data have similar constraining power.}
\end{figure}

The presence of primordial magnetic fields (PMFs), i.e. magnetic fields prior to recombination, increases the inhomogeneity of the baryon density, $\rho_b$.
This so-called baryon clumping effect is parametrized by $b \equiv (\langle\rho_b^2\rangle - \langle\rho_b\rangle^2) / \langle\rho_b\rangle^2 $, such that $b=0$ corresponds to no PMFs.
With other cosmological parameters fixed, increasing $b>0$ changes the width of the visibility function and shifts it to higher redshifts, i.e. recombination occurs sooner, which leads one to infer higher values of $\Hubble$ from CMB data \citep{jedamzik13, jedamzik19, jedamzik20, galli22}.
Because the distribution of baryons in the early universe is not known precisely, we use the three-zone toy model put forward by \citet{jedamzik13} and \citet{jedamzik20}.

We list constraints on \lcdm{}+$b$ from the SPT-3G 2018 \TTTEEE{} data in Table \ref{tab:PM_table} and show the marginalized one-dimensional posterior for $b$ in Figure \ref{fig:PMF}.
We find a 95\% confidence upper limit of
\begin{equation}
b < 1.0.
\end{equation}
The tight limit on the PMF-induced baryon clumping limits the possibility of resolving the Hubble tension through this model; we find $H_0 = 70.0 \pm 1.9\,\kmsmpc$, which remains $1.3\,\sigma$ below the distance-ladder analysis of \citet{riess22}.
We find no improvement to the goodness-of-fit for this model compared to \lcdm{} ($\Delta\chi^2 = 0$).

\begin{table}[ht!]
\def\arraystretch{1.2}
\footnotesize
\setlength{\tabcolsep}{6pt}
\centering
\begin{tabular}{l D{+}{\,\pm\,}{7.7} D{+}{\,\pm\,}{7.7}}
\hline\hline
\vphantom{\makecell{A\\A\\A}}
& \multicolumn{1}{c}{\makecell{SPT-3G 2018}}
& \multicolumn{1}{c}{\makecell{SPT-3G 2018\\ + \planck{}}}\\
\hline
$\Omega_{\mathrm{b}}h^2$& 0.02216+0.00032& 0.02234+0.00013 \\

$\Omega_{\mathrm{c}}h^2$& 0.1185+0.0039& 0.1210+0.0013 \\

$100\theta_{\mathrm{MC}}$& 1.0475+0.0049& 1.0442+0.0024 \\

$10^9A_{\mathrm{s}}e^{-2\tau}$& 1.87+0.03& 1.8830+0.0097 \\

$n_\mathrm{s}$& 0.964+0.017& 0.9610+0.0043 \\

$b$& \multicolumn{1}{D{+}{<\,}{7.7}}{+1.0}& \multicolumn{1}{D{+}{<\,}{7.7}}{+0.37} \\
\hline
$H_{\mathrm{0}}\,[\kmsmpc{}]$& 70.0+1.9& 68.10+0.74 \\

$\sigma_8$& 0.809+0.017& 0.8137+0.0065 \\

$S_8$& 0.794+0.041& 0.828+0.012 \\

$\Omega_\mathrm{\Lambda}$& 0.710+0.021& 0.6894+0.0076 \\

$\mathrm{Age}/\mathrm{Gyr}$& 13.62+0.14& 13.706+0.071 \\

$100\theta_\ast$& 1.04040+0.00075& 1.04086+0.00029 \\

\hline
\end{tabular}
\caption[PMF constraints]{
Constraints on primordial magnetic fields from SPT-3G 2018 \TTTEEE{} alone and in combination with \planck{} data.
For consistency, we report results for $100\theta_{\mathrm{MC}}$.
However, the assumptions around recombination used in this approximation to the sound horizon fail in this model \citep{hu96b}.
Hence, we also report results for the accurate angular scale of the sound horizon at recombination, $100\theta_\ast$.}
\label{tab:PM_table}
\end{table}

Measurements of the full \TTTEEE{} power spectra are crucial in this model.
\citet{galli22} point out a degeneracy between $b$ and $n_{\mathrm{s}}, 10^9 A_{\mathrm{s}} e^{-2\tau}$ that prohibits meaningful constraints on $b$ if only \TT{} or only \EETE{} power spectrum measurements are available (see Figure 6 therein).
Therefore, while \citet{galli22} report an effective non-constraint on $b$ using the SPT-3G 2018 \EETE{} data set of \citetalias{dutcher21}, the addition of \TT{} data in this work allows for a meaningful constraint, which we visualize in Figure \ref{fig:TT_improv}.

Due to the sensitivity of the $b$ constraint to the $n_{\mathrm{s}}$ values inferred from temperature and polarization data we confirm that our result is consistent with expectations based on simulations.
The upper limit we report for the data is within $20\%$ of what we infer from simulated band powers centered on $b=0$.

We find good agreement between SPT-3G and \planck{} constraints in this model.
Across the full seven-dimensional parameter space we report $\chi^2=2.3$, which translates to a PTE value of $88\%$.
We report joint constraints from SPT-3G 2018 and \planck{} data on \lcdm{}+$b$ in Table \ref{tab:PM_table}.
We find a 95\% confidence upper limit of $b < 0.37$.
The addition of the SPT-3G data to \planck{} tightens the $b$ upper limit by $40\%$ and reduces the volume of the allowed parameter space by a factor of $2.5$.

\section{Conclusion}
\label{sec:conclusion}

In this work, we present a measurement of the CMB temperature power spectrum using SPT-3G data recorded in 2018.
The \TT{} band powers are sample-variance-limited across the reported angular multipole range of $750 < \ell < 3000$.
Together with the already published polarization data \citepalias{dutcher21} from the same observing season, this completes the SPT-3G 2018 \TTTEEE{} data set.
We analyze the internal consistency of the data using a variety of tools: null tests, difference spectra, complement spectra (across frequencies and spectrum types), MV comparisons, and parameter-level subset tests.
We find good agreement across frequencies, spectrum types, and angular multipoles.

We present cosmological parameter constraints from the SPT-3G 2018 \TTTEEE{} band powers.
This is the first analysis using SPT-only measurements of all three primary CMB power spectra and the complete data set provides the strongest constraining power to date from SPT.
The data are well fit by \lcdm{} with a PTE value of $15\%$.
We constrain the expansion rate today to $H_0 = 68.3 \pm 1.5\,\kmsmpc{}$, the combined structure growth parameter to $S_8 = 0.797 \pm 0.042$, and find a preference for $n_{\mathrm{s}} < 1$ at $1.8\,\sigma$.
The addition of the SPT-3G temperature power spectrum measurement to the \EETE{} data improves cosmological parameter constraints by $8-27\%$ and reduces the allowed five-dimensional parameter volume by a factor of $2.7$.
We report excellent agreement between the SPT-3G and \planck{} data with deviations of $<\,1\sigma$ for all cosmological parameters.
Adding the SPT-3G band powers to the \planck{} primary power spectrum measurement leads to a reduction of the allowed six-dimensional parameter volume by a factor of $1.7$.

We consider a series of extensions to the standard model, drawing on the following parameters: the strength of gravitational lensing affecting the primary CMB power spectra, $A_{\mathrm{L}}$, the effective number of neutrino species, $\Neff{}$, the primordial helium abundance, $\Yp{}$, and the baryon-clumping induced by primordial magnetic fields, $b$.
We do not find a preference for any of these extensions over the standard model.
The addition of temperature data to \EETE{} power spectrum measurements leads to significant improvements on cosmological constraints.
For \lcdm{}$+A_{\mathrm{L}}$, \lcdm{}$+\Neff{}$, and \lcdm{}$+\Neff{}+Y_P$, the posterior widths of extension parameters shrink by $5-24\%$ and the multidimensional allowed parameter volume decreases by factors of $2.4-3.1$.
In the case of primordial magnetic fields, the combination of temperature and polarization data is essential to break degeneracies between $b$ and $n_{\mathrm{s}}, 10^9 A_{\mathrm{s}} e^{-2\tau}$ \citep{galli22}.
We find a $95\%$ confidence upper limit on the PMF-induced baryon clumping of $b<1.0$.
Our findings reflect that joint analyses of \TTTEEE{} power spectrum measurements yield a substantial increase in constraining power over \EETE{} alone; this approach is key to distinguishing between significant deviations from the standard model and statistical fluctuations and provides further ways to test the data for systematic effects.

The framework presented here will be used for on-going analyses of SPT-3G data recorded in the 2019 and 2020 observing seasons.
These observations include measurements of the same $\sim\! 1500\,\sqdeg$ survey field used here, but achieve a map noise $\sim\!3.5\times$ smaller.
Moreover, extended survey data from these seasons cover an additional $\sim\!2800\,\sqdeg$, reducing sample variance and improving measurements of the power spectrum on large angular scales.
The combined SPT-3G measurements presented in this work represent a significant improvement for cosmological constraints from ground-based CMB data, and are an important demonstration for future experiments, such as CMB-S4 \citep{abazajian19}.

\acknowledgments
We thank Karsten Jedamzik and Levon Pogosian for their help with models featuring baryon clumping due to primordial magnetic fields.
The South Pole Telescope program is supported by the National Science Foundation (NSF) through the awards OPP-1852617 and OPP-2147371.
Partial support is also provided by the Kavli Institute of Cosmological Physics at the University of Chicago.
Argonne National Laboratory's work was supported by the U.S. Department of Energy, Office of High Energy Physics, under Contract No. DE-AC02-06CH11357.
Work at Fermi National Accelerator Laboratory, a DOE-OS, HEP User Facility managed by the Fermi Research Alliance, LLC, was supported under Contract No. DE-AC02-07CH11359.
The Cardiff authors acknowledge support from the UK Science and Technologies Facilities Council (STFC).
The IAP authors acknowledge support from the Centre National d'\'{E}tudes Spatiales (CNES).
This project has received funding from the European Research Council (ERC) under the European Union’s Horizon 2020 research and innovation programme (grant agreement No 101001897).
This research used resources of the IN2P3 Computer Center (http://cc.in2p3.fr).
M.A. and J.V. acknowledge support from the Center for AstroPhysical Surveys at the
National Center for Supercomputing Applications in Urbana, IL.
J.V. acknowledges support from the Sloan Foundation.
K.F. acknowledges support from the Department of Energy Office of Science Graduate Student Research (SCGSR) Program.
The Melbourne authors acknowledge support from the Australian Research Council's Discovery Project scheme (No. DP210102386).
L.B. acknowledges support from the Albert Shimmins Fund.
The McGill authors acknowledge funding from the Natural Sciences and Engineering Research Council of Canada, Canadian Institute for Advanced Research, and the Fonds de recherche du Qu\'ebec Nature et technologies.
The UCLA and MSU authors acknowledge support from NSF AST-1716965 and CSSI-1835865.
A.S.M. is supported by the MSSL STFC Consolidated Grant.
This research was done using resources provided by the Open Science Grid \citep{pordes07, sfiligoi09}, which is supported by the NSF Award No. 1148698, and the U.S. Department of Energy's Office of Science.
Some of the results in this paper have been derived using the healpy and HEALPix\footnote{\url{ http://healpix.sf.net/}} packages \citep{gorski05, zonca19}.
The data analysis pipeline also uses the scientific python stack \citep{hunter07, jones01, vanDerWalt11}.
\vspace*{-3mm}

\bibliography{spt}

\clearpage

\begin{appendices}
\section*{Appendix}
\subsection{Updates to the Polarization Analysis Pipeline}
\label{app:changes}

We make two key updates to the analysis of the \EETE{} spectra from \citetalias{dutcher21}, which primarily update the covariance matrix.
First, we account for correlated noise across frequencies.
Extending the work in \citetalias{dutcher21}, we take the difference between two half-depth coadded maps at different frequencies.
We divide the power spectrum of this difference map by the square root of the product of the power spectra of the corresponding auto-frequency noise spectra.
This yields an estimate of the correlation coefficient of the noise between two frequency channels.
We find that for intensity the $95\,\mathrm{GHz}$ and $150\,\mathrm{GHz}$ channels, as well as the $95\,\mathrm{GHz}$ and $220\,\mathrm{GHz}$ channels, are moderately correlated with $\rho \approx 0.6$ at $\ell=750$ and $\rho \leq 0.2$ at $\ell \geq 2000$.
The $150\,\mathrm{GHz}$ and $220\,\mathrm{GHz}$ channels are highly correlated with $\rho \approx 0.9$ at $\ell=750$ and $\rho \leq 0.4$ at $\ell \geq 2000$.
This correlation is high compared to past and contemporary ground-based CMB experiments, due to the novel trichoic architecture of SPT-3G pixels \citep{sobrin22}.
The behaviour with $\ell$ matches the expectation that only atmospheric noise is correlated across frequencies, not instrumental noise.
The different degrees of correlation are a consequence of a water emission line at $183\,\mathrm{GHz}$ and an oxygen line at $119\,\mathrm{GHz}$ \citep{lane98, pardo01}.
We use the correlation coefficients derived in this way to update the noise model in the covariance calculation (see \S\RNum{4}H in \citetalias{dutcher21} for details).
We detect no correlated noise across frequencies in polarization or correlated noise between temperature and polarization.

Second, we use a series of $1,000$ simulations to update the mode-coupling model in the covariance calculation.
The Lambert azimuthal equal-area projection does not preserve angles and leads to increasing bin-to-bin correlations at high $\ell$.
For each simulation, we generate a CMB-only HEALPix sky, mask the map using the data apodization mask, and project the curved-sky map into a flat-sky map.
We estimate the correlation matrix using the scatter of the power spectra of the $1,000$ flat-sky maps.
The recovered correlation structure matches the data well and is less noisy than the data estimate due to the increased number of independent realizations.
Following \citetalias{dutcher21}, we fit second-order polynomials to band-diagonal elements of the correlation estimate from simulations and use these fits in the data correlation matrix.
While in principle filtering effects not captured by these simulations lead to off-diagonal elements in the covariance matrix, the correlation structure of the data is completely dominated by the flat-sky projection.

\begin{figure*}[ht!]
  \centering
  \includegraphics[width=\linewidth]{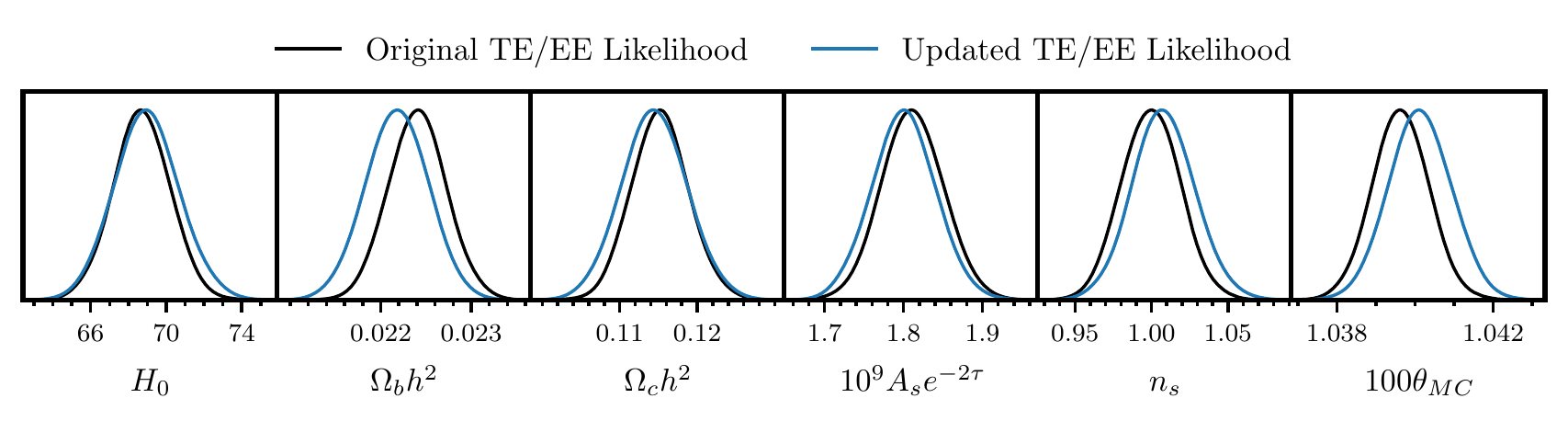}
  \caption{\label{fig:eete_par_shift}
  Marginalized posterior distributions for core \lcdm{} and $\Hubble{}$ from the original (black) and updated (blue) SPT-3G 2018 \EETE{} likelihood.
  The posteriors widen slightly; the largest change is a $15\%$ correction to the $\Omega_{\mathrm{c}} h^2$ uncertainty.
  The shift to the central values of parameter constraints are less than the size of the new error bars.
  }
\end{figure*}

We compare parameter constraints from the original \EETE{} likelihood to the updated version in Table \ref{tab:eete_param_shift_table} and Figure \ref{fig:eete_par_shift}.
The central values of cosmological parameter constraints shift by less than the size of the new error bars.
The parameter uncertainties generally widen with the updated covariance, by at most $15\%$ for $\Omega_{\mathrm{c}} h^2$.
The addition of \TT{} data to the updated covariance allows for as good as or better parameter constraints than reported in \citetalias{dutcher21}. 

\begin{table*}[ht!]
\def\arraystretch{1.2}
\footnotesize
\setlength{\tabcolsep}{12pt}
\centering
\begin{tabular}{l @{\hskip 20pt} D{+}{\,\pm\,}{7.7} D{+}{\,\pm\,}{7.7}}
\hline\hline
& \multicolumn{1}{l}{\text{SPT-3G 2018 \EETE{} (Original)}}
& \multicolumn{1}{c}{\text{\text{SPT-3G 2018 \EETE{} (Updated)}}} \\
\hline
$\Omega_{\mathrm{b}}h^2$ &
  0.02241+0.00032&
  0.02218+0.00035\\
$\Omega_{\mathrm{c}}h^2$ &
  0.1152+0.0037&
  0.1145+0.0043\\
$100\theta_{\mathrm{MC}}$ &
  1.03963+0.00073&
  1.04013+0.00081\\
$10^9 A_{\mathrm{s}} e^{-2\tau}$ &
  1.811+0.040&
  1.800+0.041\\
$n_{\mathrm{s}}$ &
  1.000+0.019&
  1.008+0.021\\
\hline
$H_{\mathrm{0}}\,[\kmsmpc{}]$ &
  68.7+1.5&
  69.0+1.7\\
$\sigma_8$ &
  0.788+0.016&
  0.786+0.018\\
$S_8 \equiv \sigma_8 \sqrt{\Omega_{\mathrm{m}}/0.3}$ &
  0.779+0.042&
  0.772+0.047\\
$\Omega_{\mathrm{\Lambda}}$ &
  0.706+0.021&
  0.710+0.023\\
${\mathrm{Age}}/{\mathrm{Gyr}}$ &
  13.809+0.049&
  13.813+0.052\\
\hline
\end{tabular}
\caption[
\EETE{} shift.
]{
Comparison of marginalized constraints and 68\% errors of $\Lambda$CDM free and derived parameters from SPT-3G 2018 \EETE{} data using the original and updated likelihood.
}
\label{tab:eete_param_shift_table}
\end{table*}

\subsection{Baseline Priors}
\label{app:priors}

We present the baseline priors used in the likelihood analysis and the best-fit values of nuisance parameters in \lcdm{} in Table \ref{tab:priors}.

We briefly present updates made to the galactic dust prior calculation of \citetalias{dutcher21} here.
We model the spectral dependence of galactic dust using a modified black-body spectrum and retain the angular dependence of \citetalias{dutcher21}, i.e. using a power law.
The spectra are normalized at $150\,\mathrm{GHz}$ and $\ell=80$.
We fit combinations of the cross-spectra of the $143\,\mathrm{GHz}$, $217\,\mathrm{GHz}$, $353\,\mathrm{GHz}$, and $545\,\mathrm{GHz}$ \planck{} PR3 half-mission maps \citep{planck18-3} calculated on the SPT-3G survey field to the best-fit \planck{} CMB spectrum plus galactic dust and extragalactic foregrounds.
We ensure the resulting constraints on the galactic dust parameters are robust with respect to the modelling of extragalactic foregrounds and the bin width of the cross-spectrum band powers.
We conservatively widen the constraints the data provide on the galactic dust amplitudes by a factor of three before adopting them as priors in our cosmological analysis.
The baseline priors on galactic dust are listed in Table \ref{tab:priors}.

\begin{table*}[ht!]
\def\arraystretch{1.4}
\footnotesize
\setlength{\tabcolsep}{12pt}
\centering
\begin{tabular}{l @{\hskip 30pt} l @{\hskip 20pt} p{7cm}}
\hline\hline
Parameter & Prior & Description\\
\hline
\multicolumn{3}{l}{\text{\!\!\!\!General}}\\
$\tau$ & $\mathcal{N}(0.0540, 0.0074)$ & Optical depth to reionization\\
$100\kappa$ & $\mathcal{N}(0, 0.045)\,[0.0]$ & Super-sample lensing convergence\\
\hline
\multicolumn{3}{l}{\text{\!\!\!\!Temperature}}\\
 $A^{\mathrm{cirrus}}_{80}$ & $\mathcal{N}(1.88, 0.48)\,[1.93]$ & Galactic cirrus amplitude\\
 $\alpha^{\mathrm{cirrus}}$ & $\mathcal{N}(-2.53, 0.05)\,[-2.53]$ & Galactic cirrus power law index\\
 $\beta^{\mathrm{cirrus}}$ & $\mathcal{N}(1.48, 0.02)\,[1.48]$ & Galactic cirrus spectral index\\
 $D^\mathrm{Poisson, \TT}_{3000, 95\times95}$ &  $\mathcal{N}(51.3, 9.4)\,[62.61]$ & \TT{} Poisson power for $95\times95\,\mathrm{GHz}$\\
 $D^\mathrm{Poisson, \TT}_{3000, 95\times150}$ & $\mathcal{N}(22.4, 7.1)\,[27.9]$ & \TT{} Poisson power for $95\times150\,\mathrm{GHz}$\\
 $D^\mathrm{Poisson, \TT}_{3000, 95\times220}$ & $\mathcal{N}(20.7, 5.9)\,[24.3]$ & \TT{} Poisson power for $95\times220\,\mathrm{GHz}$\\
 $D^\mathrm{Poisson, \TT}_{3000, 150\times150}$ & $\mathcal{N}(15.3, 4.1)\,[16.7]$ & \TT{} Poisson power for $150\times150\,\mathrm{GHz}$\\
 $D^\mathrm{Poisson, \TT}_{3000, 150\times220}$ & $\mathcal{N}(28.4, 4.2)\,[28.6]$ & \TT{} Poisson power for $150\times220\,\mathrm{GHz}$\\
 $D^\mathrm{Poisson, \TT}_{3000, 220\times220}$ & $\mathcal{N}(76.0, 14.9)\,[78.5]$ & \TT{} Poisson power for $220\times220\,\mathrm{GHz}$\\
 $A^{\mathrm{CIB-cl.}}_{80}$ & $\mathcal{N}(3.2, 1.8)\,[5.2]$ & CIB clustering amplitude\\
 $\beta^{\mathrm{CIB-cl.}}$ & $\mathcal{N}(2.26, 0.38)\,[1.85]$ & CIB clustering spectral index\\
 $A^{\mathrm{tSZ}}$ & $\mathcal{N}(3.2, 2.4)\,[4.7]$ & tSZ amplitude\\
 $\xi$ & $\mathcal{N}(0.18, 0.33)\,[0.09]$ & tSZ-CIB correlation\\
 $A^{\mathrm{kSZ}}$ & $\mathcal{N}(3.7, 4.6)\,[3.7]$ & kSZ amplitude\\
\hline
\multicolumn{3}{l}{\text{\!\!\!\!Polarization}}\\
 $D^\mathrm{Poisson, \EE}_{3000, 95\times95}$ &  $\mathcal{N}(0.041, 0.012)\,[0.041]$ & \EE{} Poisson power for $95\times95\,\mathrm{GHz}$\\
  $D^\mathrm{Poisson, \EE}_{3000, 95\times150}$ & $\mathcal{N}(0.0180, 0.0054)\,[0.0177]$ & \EE{} Poisson power for $95\times150\,\mathrm{GHz}$\\
 $D^\mathrm{Poisson, \EE}_{3000, 95\times220}$ & $\mathcal{N}(0.0157, 0.0047)\,[0.0157]$ & \EE{} Poisson power for $95\times220\,\mathrm{GHz}$\\
 $D^\mathrm{Poisson, \EE}_{3000, 150\times150}$ & $\mathcal{N}(0.0115, 0.0034)\,[0.0115]$ & \EE{} Poisson power for $150\times150\,\mathrm{GHz}$\\
 $D^\mathrm{Poisson, \EE}_{3000, 150\times220}$ & $\mathcal{N}(0.0190, 0.0057)\,[0.0188]$ & \EE{} Poisson power for $150\times220\,\mathrm{GHz}$\\
 $D^\mathrm{Poisson, \EE}_{3000, 220\times220}$ & $\mathcal{N}(0.048, 0.014)\,[0.048]$ & \EE{} Poisson power for $220\times220\,\mathrm{GHz}$\\
 $A^{\TE}_{80}$ & $\mathcal{N}(0.120, 0.051)\,[0.138]$ & \TE{} amplitude of polarized galactic dust\\
 $\alpha_{\TE}$ & $\mathcal{N}(-2.42, 0.04)\,[-2.42]$ & \TE{} power law index of polarized galactic dust\\
 $\beta_{\TE}$ & $\mathcal{N}(1.51, 0.04)\,[1.51]$ & \TE{} spectral index of polarized galactic dust\\
 $A^{\EE}_{80}$ & $\mathcal{N}(0.05, 0.022)\,[0.052]$ & \EE{} amplitude of polarized galactic dust\\
 $\alpha_{\EE}$ & $\mathcal{N}(-2.42, 0.04)\,[-2.42]$ & \EE{} power law index of polarized galactic dust\\
 $\beta_{\EE}$ & $\mathcal{N}(1.51, 0.04)\,[1.51]$ & \EE{} spectral index of polarized galactic dust\\
\hline
\multicolumn{3}{l}{\text{\!\!\!\!Calibration}}\\
 $T_{\rm cal}^{\rm \,95\,GHz}$ & $\mathcal{N}(1.0, 0.0056)\,[1.0]$ & Temperature calibration at $95\,\mathrm{GHz}$\\
 $T_{\rm cal}^{\rm \,150\,GHz}$ & $\mathcal{N}(1.0, 0.0056)\,[0.9975]$ & Temperature calibration at $150\,\mathrm{GHz}$\\
 $T_{\rm cal}^{\rm \,220\,GHz}$ & $\mathcal{N}(1.0, 0.0075)\,[0.9930]$ & Temperature calibration at $220\,\mathrm{GHz}$\\
 $E_{\rm cal}^{\rm \,95\,GHz}$ & $\mathcal{N}(1.0, 0.0087)\,[1.0009]$ & Polarization calibration at $95\,\mathrm{GHz}$\\
 $E_{\rm cal}^{\rm \,150\,GHz}$ & $\mathcal{N}(1.0, 0.0082)\,[1.0020]$ & Polarization calibration at $150\,\mathrm{GHz}$\\
 $E_{\rm cal}^{\rm \,220\,GHz}$ & $\mathcal{N}(1.0, 0.016)\,[1.019]$ & Polarization calibration at $220\,\mathrm{GHz}$\\
\hline
\end{tabular}
\caption[
Baseline priors
]{
Overview of nuisance parameters in the SPT-3G 2018 likelihood and baseline priors. Gaussian priors are listed as $\mathcal{N}(\mu, \sigma)$, where $\mu$ is the mean and $\sigma$ the standard deviation. Best-fit values for nuisance parameters are given in brackets. All amplitude parameters are in units of $\mathrm{\mu K^2}$. The best-fit values of all nuisance parameters lie within $1.2\,\sigma$ of the central values of their priors. The prior on the optical depth to reionization is not used when including \planck{} data in the analysis.
}
\label{tab:priors}
\end{table*}


\subsection{Multifrequency Band Powers}
\label{app:bandpowers}

We present the full multifrequency power spectrum measurements in tables \ref{tab:tt_bandpowers_table}, \ref{tab:te_bandpowers_table}, and \ref{tab:ee_bandpowers_table} below.

\begin{table*}[h!]
\footnotesize
\setlength{\tabcolsep}{2.5pt}
\def\arraystretch{1.2}
\centering
\begin{tabular}{|c c | D{.}{.}{1} D{.}{.}{-1} | D{.}{.}{1} D{.}{.}{-1} | D{.}{.}{1} D{.}{.}{-1} | D{.}{.}{1} D{.}{.}{-1} | D{.}{.}{1} D{.}{.}{-1} | D{.}{.}{1} D{.}{.}{-1}|}
\hline
\rule{0pt}{3ex} \multirow{2}{*}{$\ell$ Range} & \multirow{2}{*}{$\ell_\mathrm{eff}$} & \multicolumn{2}{c|}{$\mathrm{95\times95\,GHz}$} & \multicolumn{2}{c|}{$\mathrm{95\times150\,GHz}$} & \multicolumn{2}{c|}{$\mathrm{95\times220\,GHz}$} & \multicolumn{2}{c|}{$\mathrm{150\times150\,GHz}$} & \multicolumn{2}{c|}{$\mathrm{150\times220\,GHz}$} & \multicolumn{2}{c|}{$\mathrm{220\times220\,GHz}$} \\\cline{3-14}
 & & D_b & \sigma_b & D_b & \sigma_b & D_b & \sigma_b & D_b & \sigma_b & D_b & \sigma_b & D_b & \sigma_b \\[2pt]
\hline
750 -- 800 & 775 & 2549.3 & 83.9 & 2556.1 & 84.1 & 2583.5 & 92.3 & 2567.7 & 85.3 & - & - & - & - \\
800 -- 850 & 825 & 2673.5 & 79.2 & 2682.0 & 79.3 & 2682.8 & 87.4 & 2694.6 & 80.5 & - & - & - & - \\
850 -- 900 & 874 & 2191.5 & 73.7 & 2185.9 & 73.9 & 2185.8 & 80.9 & 2187.5 & 75.1 & - & - & - & - \\
900 -- 950 & 925 & 1594.4 & 53.1 & 1602.2 & 53.2 & 1641.3 & 59.4 & 1618.7 & 54.2 & - & - & - & - \\
950 -- 1000 & 974 & 1215.8 & 39.5 & 1211.3 & 39.6 & 1213.4 & 44.6 & 1211.0 & 40.4 & - & - & - & - \\
1000 -- 1050 & 1024 & 1024.5 & 34.4 & 1014.3 & 34.2 & 1009.6 & 38.7 & 1009.8 & 34.8 & 1014.1 & 40.4 & 1050.0 & 52.3 \\
1050 -- 1100 & 1074 & 1244.8 & 35.8 & 1237.7 & 35.5 & 1247.4 & 39.6 & 1236.6 & 36.0 & 1254.6 & 41.1 & 1291.8 & 51.9 \\
1100 -- 1150 & 1124 & 1243.4 & 37.1 & 1238.0 & 36.7 & 1223.3 & 40.7 & 1240.4 & 37.2 & 1236.7 & 42.0 & 1266.1 & 52.1 \\
1150 -- 1200 & 1174 & 1223.5 & 37.7 & 1214.2 & 37.3 & 1212.3 & 40.8 & 1211.2 & 37.7 & 1216.1 & 41.9 & 1239.0 & 50.9 \\
1200 -- 1250 & 1224 & 940.3 & 29.0 & 926.8 & 28.6 & 943.7 & 32.2 & 921.5 & 29.1 & 950.9 & 33.3 & 1011.0 & 42.4 \\
1250 -- 1300 & 1274 & 792.3 & 23.5 & 780.9 & 22.9 & 766.8 & 26.1 & 778.6 & 23.2 & 776.3 & 26.9 & 798.8 & 35.9 \\
1300 -- 1350 & 1324 & 753.0 & 21.8 & 744.8 & 21.3 & 746.0 & 24.6 & 744.1 & 21.7 & 757.8 & 25.4 & 809.8 & 33.9 \\
1350 -- 1400 & 1374 & 797.3 & 24.8 & 786.4 & 24.2 & 775.9 & 26.9 & 782.8 & 24.4 & 783.7 & 27.6 & 811.3 & 35.1 \\
1400 -- 1450 & 1424 & 828.7 & 24.7 & 818.0 & 24.1 & 819.6 & 26.8 & 818.0 & 24.4 & 833.5 & 27.6 & 881.0 & 34.9 \\
1450 -- 1500 & 1474 & 774.5 & 22.4 & 766.8 & 21.7 & 772.6 & 24.5 & 766.1 & 22.1 & 783.2 & 25.2 & 825.6 & 32.6 \\
1500 -- 1550 & 1524 & 653.0 & 19.6 & 643.3 & 19.0 & 656.6 & 21.5 & 642.3 & 19.2 & 666.9 & 22.0 & 724.2 & 29.2 \\
1550 -- 1600 & 1574 & 517.8 & 14.8 & 501.6 & 14.1 & 497.6 & 16.7 & 495.4 & 14.2 & 503.3 & 17.2 & 550.5 & 24.9 \\
1600 -- 1650 & 1624 & 436.4 & 13.7 & 421.1 & 13.1 & 412.1 & 15.5 & 416.6 & 13.3 & 421.8 & 16.0 & 467.8 & 23.6 \\
1650 -- 1700 & 1674 & 426.5 & 11.9 & 412.9 & 11.3 & 411.4 & 14.2 & 407.7 & 11.6 & 420.2 & 14.7 & 473.6 & 22.8 \\
1700 -- 1750 & 1724 & 424.4 & 13.2 & 413.2 & 12.6 & 412.0 & 15.3 & 411.3 & 12.9 & 422.7 & 15.8 & 484.8 & 23.6 \\
1750 -- 1800 & 1775 & 408.9 & 12.0 & 395.3 & 11.4 & 404.6 & 14.2 & 394.3 & 11.7 & 417.0 & 14.7 & 477.5 & 22.5 \\
1800 -- 1850 & 1824 & 390.5 & 11.0 & 372.1 & 10.3 & 362.0 & 12.9 & 365.4 & 10.5 & 370.0 & 13.3 & 415.1 & 21.0 \\
1850 -- 1900 & 1874 & 309.6 & 9.9 & 288.4 & 9.1 & 283.1 & 11.7 & 280.1 & 9.3 & 291.1 & 11.9 & 356.4 & 19.6 \\
1900 -- 1950 & 1925 & 264.0 & 8.7 & 250.3 & 7.9 & 253.7 & 10.4 & 247.9 & 8.0 & 265.5 & 10.5 & 317.3 & 18.4 \\
1950 -- 2000 & 1974 & 279.4 & 8.5 & 256.6 & 7.6 & 241.7 & 10.1 & 248.9 & 7.7 & 253.0 & 10.2 & 319.1 & 18.2 \\
2000 -- 2100 & 2051 & 264.0 & 4.5 & 245.8 & 4.0 & 242.7 & 5.4 & 241.3 & 4.1 & 253.9 & 5.5 & 317.3 & 10.0 \\
2100 -- 2200 & 2152 & 215.1 & 4.1 & 192.8 & 3.6 & 189.5 & 5.0 & 186.1 & 3.7 & 198.8 & 5.0 & 251.4 & 9.8 \\
2200 -- 2300 & 2250 & 170.0 & 3.4 & 146.2 & 2.7 & 145.3 & 4.3 & 141.5 & 2.8 & 157.5 & 4.2 & 240.8 & 9.3 \\
2300 -- 2400 & 2350 & 158.6 & 3.2 & 138.8 & 2.5 & 132.4 & 4.2 & 135.5 & 2.5 & 155.4 & 3.9 & 230.5 & 9.3 \\
2400 -- 2500 & 2451 & 142.6 & 3.0 & 117.9 & 2.3 & 120.5 & 3.9 & 111.4 & 2.2 & 132.1 & 3.6 & 216.0 & 9.2 \\
2500 -- 2600 & 2550 & 128.0 & 2.9 & 98.7 & 2.1 & 93.8 & 3.8 & 91.8 & 2.0 & 107.7 & 3.4 & 178.6 & 9.3 \\
2600 -- 2700 & 2649 & 122.7 & 2.9 & 91.6 & 1.9 & 84.8 & 3.8 & 85.7 & 1.9 & 106.1 & 3.2 & 191.6 & 9.4 \\
2700 -- 2800 & 2750 & 118.5 & 2.9 & 85.1 & 1.9 & 75.1 & 3.9 & 74.6 & 1.7 & 87.9 & 3.2 & 183.0 & 9.7 \\
2800 -- 2900 & 2850 & 107.3 & 2.9 & 74.8 & 1.8 & 71.8 & 3.9 & 64.9 & 1.6 & 89.7 & 3.1 & 207.5 & 10.1 \\
2900 -- 3000 & 2947 & 109.5 & 3.1 & 70.5 & 1.8 & 59.8 & 4.1 & 58.4 & 1.6 & 75.6 & 3.1 & 154.3 & 10.5 \\
\hline
\end{tabular}
\caption[\TT{} multifrequency band powers]{
\TT{} multifrequency band power measurements, $D_b$, and associated uncertainties, $\sigma_b$, (both in units of $\mu$K$^2$) for a given angular multipole range and the window function-weighted multipole $\ell_\mathrm{eff}$.}
\label{tab:tt_bandpowers_table}
\end{table*}

\begin{table*}[h!]
\footnotesize
\setlength{\tabcolsep}{2.5pt}
\def\arraystretch{1.2}
\centering
\begin{tabular}{|c c | D{.}{.}{1} D{.}{.}{-1} | D{.}{.}{1} D{.}{.}{-1} | D{.}{.}{1} D{.}{.}{-1} | D{.}{.}{1} D{.}{.}{-1} | D{.}{.}{1} D{.}{.}{-1} | D{.}{.}{1} D{.}{.}{-1}|}
\hline
\rule{0pt}{3ex} \multirow{2}{*}{$\ell$ Range} & \multirow{2}{*}{$\ell_\mathrm{eff}$} & \multicolumn{2}{c|}{$\mathrm{95\times95\,GHz}$} & \multicolumn{2}{c|}{$\mathrm{95\times150\,GHz}$} & \multicolumn{2}{c|}{$\mathrm{95\times220\,GHz}$} & \multicolumn{2}{c|}{$\mathrm{150\times150\,GHz}$} & \multicolumn{2}{c|}{$\mathrm{150\times220\,GHz}$} & \multicolumn{2}{c|}{$\mathrm{220\times220\,GHz}$} \\\cline{3-14}
 & & D_b & \sigma_b & D_b & \sigma_b & D_b & \sigma_b & D_b & \sigma_b & D_b & \sigma_b & D_b & \sigma_b \\[2pt]
\hline
300 -- 350 & 326 & 88.7 & 12.0 & 93.3 & 12.2 & 99.7 & 13.9 & 101.1 & 12.7 & 110.2 & 14.4 & 113.2 & 20.3 \\
350 -- 400 & 375 & 43.8 & 8.8 & 42.5 & 8.7 & 36.6 & 10.7 & 42.7 & 9.2 & 40.7 & 11.3 & 39.9 & 17.2 \\
400 -- 450 & 425 & -44.9 & 7.6 & -45.6 & 7.3 & -43.0 & 9.2 & -47.8 & 7.5 & -47.0 & 9.5 & -43.2 & 15.0 \\
450 -- 500 & 475 & -69.1 & 6.7 & -69.0 & 6.3 & -64.9 & 7.9 & -70.0 & 6.4 & -64.4 & 8.0 & -53.0 & 13.2 \\
500 -- 550 & 525 & -34.1 & 5.5 & -34.7 & 5.0 & -48.2 & 6.7 & -34.8 & 5.2 & -46.6 & 6.7 & -57.9 & 12.1 \\
550 -- 600 & 575 & 11.9 & 6.2 & 11.3 & 5.9 & 15.2 & 7.4 & 10.5 & 6.1 & 15.5 & 7.5 & 20.7 & 12.3 \\
600 -- 650 & 625 & 24.2 & 7.0 & 23.9 & 6.7 & 21.5 & 8.2 & 24.5 & 7.0 & 23.0 & 8.3 & 21.3 & 12.7 \\
650 -- 700 & 674 & -63.6 & 7.7 & -63.4 & 7.4 & -58.0 & 8.7 & -63.1 & 7.5 & -59.1 & 8.8 & -59.7 & 12.9 \\
700 -- 750 & 725 & -119.9 & 7.3 & -121.2 & 6.9 & -114.0 & 8.3 & -122.8 & 7.0 & -115.7 & 8.3 & -104.7 & 12.6 \\
750 -- 800 & 774 & -121.7 & 7.3 & -120.7 & 6.7 & -124.1 & 8.3 & -121.4 & 6.8 & -126.0 & 8.2 & -124.1 & 12.8 \\
800 -- 850 & 824 & -52.8 & 5.6 & -50.6 & 4.8 & -43.2 & 6.8 & -48.6 & 5.0 & -39.9 & 6.7 & -25.5 & 12.0 \\
850 -- 900 & 874 & 41.2 & 5.8 & 38.5 & 5.1 & 38.5 & 6.9 & 36.7 & 5.3 & 37.2 & 6.9 & 36.6 & 11.8 \\
900 -- 950 & 924 & 54.7 & 5.5 & 56.1 & 4.9 & 58.9 & 6.6 & 56.9 & 5.1 & 61.3 & 6.6 & 70.1 & 11.2 \\
950 -- 1000 & 974 & 12.5 & 5.3 & 13.1 & 4.9 & 14.4 & 6.3 & 13.9 & 5.0 & 13.7 & 6.3 & 17.9 & 10.6 \\
1000 -- 1050 & 1024 & -52.2 & 5.6 & -51.9 & 5.2 & -55.4 & 6.5 & -51.8 & 5.4 & -55.7 & 6.5 & -56.4 & 10.5 \\
1050 -- 1100 & 1074 & -75.8 & 5.3 & -74.7 & 4.7 & -71.9 & 6.2 & -73.7 & 4.9 & -72.0 & 6.1 & -69.8 & 10.4 \\
1100 -- 1150 & 1124 & -48.4 & 4.6 & -52.8 & 3.9 & -58.4 & 5.6 & -55.9 & 4.1 & -60.2 & 5.5 & -65.8 & 10.1 \\
1150 -- 1200 & 1174 & -9.7 & 4.2 & -10.1 & 3.4 & -6.9 & 5.2 & -10.8 & 3.6 & -7.1 & 5.1 & -1.9 & 9.9 \\
1200 -- 1250 & 1224 & 4.9 & 4.1 & 4.3 & 3.4 & 4.2 & 5.1 & 4.3 & 3.6 & 4.3 & 5.0 & 8.3 & 9.7 \\
1250 -- 1300 & 1274 & -15.4 & 4.1 & -15.8 & 3.4 & -17.2 & 5.1 & -16.1 & 3.6 & -16.7 & 4.9 & -16.3 & 9.5 \\
1300 -- 1350 & 1324 & -47.3 & 4.2 & -48.2 & 3.5 & -43.6 & 5.1 & -49.1 & 3.6 & -42.8 & 5.0 & -39.5 & 9.5 \\
1350 -- 1400 & 1374 & -62.0 & 4.3 & -62.0 & 3.5 & -55.3 & 5.3 & -63.0 & 3.7 & -56.7 & 5.1 & -47.3 & 9.9 \\
1400 -- 1450 & 1424 & -41.2 & 4.1 & -41.9 & 3.1 & -41.2 & 5.2 & -42.9 & 3.3 & -41.0 & 5.0 & -30.7 & 10.1 \\
1450 -- 1500 & 1474 & -10.9 & 3.9 & -11.8 & 2.8 & -8.6 & 5.0 & -13.0 & 3.0 & -9.9 & 4.7 & -4.2 & 10.0 \\
1500 -- 1550 & 1524 & 8.5 & 3.6 & 9.1 & 2.6 & 4.8 & 4.7 & 10.2 & 2.8 & 5.9 & 4.5 & -7.3 & 9.7 \\
1550 -- 1600 & 1574 & -3.8 & 3.5 & -0.8 & 2.6 & -4.2 & 4.5 & 1.1 & 2.8 & 0.3 & 4.3 & -5.1 & 9.4 \\
1600 -- 1650 & 1624 & -13.9 & 3.4 & -15.4 & 2.6 & -15.7 & 4.4 & -14.5 & 2.7 & -13.3 & 4.1 & -8.0 & 9.3 \\
1650 -- 1700 & 1674 & -31.1 & 3.3 & -32.0 & 2.4 & -32.4 & 4.3 & -33.1 & 2.5 & -31.7 & 4.0 & -32.9 & 9.4 \\
1700 -- 1750 & 1724 & -22.0 & 3.4 & -24.0 & 2.3 & -25.9 & 4.4 & -26.0 & 2.5 & -26.7 & 4.1 & -25.0 & 9.7 \\
1750 -- 1800 & 1775 & -15.8 & 3.3 & -15.2 & 2.2 & -17.6 & 4.4 & -14.7 & 2.4 & -17.4 & 4.0 & -21.4 & 9.9 \\
1800 -- 1850 & 1824 & -14.2 & 3.2 & -10.0 & 2.1 & -7.1 & 4.3 & -8.4 & 2.2 & -7.3 & 3.9 & 3.4 & 9.8 \\
1850 -- 1900 & 1874 & -3.9 & 3.1 & -3.3 & 2.0 & -5.1 & 4.1 & -3.4 & 2.2 & -3.3 & 3.8 & -12.6 & 9.7 \\
1900 -- 1950 & 1924 & -11.9 & 3.0 & -11.2 & 2.0 & -10.8 & 4.1 & -11.3 & 2.1 & -10.9 & 3.7 & -13.9 & 9.7 \\
1950 -- 2000 & 1975 & -15.1 & 3.1 & -16.4 & 2.0 & -17.8 & 4.1 & -16.4 & 2.1 & -17.2 & 3.7 & -18.6 & 10.0 \\
2000 -- 2100 & 2050 & -16.1 & 1.7 & -14.2 & 1.0 & -14.6 & 2.3 & -13.7 & 1.1 & -13.9 & 2.0 & -17.7 & 5.6 \\
2100 -- 2200 & 2150 & -5.4 & 1.6 & -4.8 & 1.0 & -9.1 & 2.3 & -4.3 & 1.1 & -5.8 & 2.0 & 3.5 & 5.9 \\
2200 -- 2300 & 2250 & -7.7 & 1.6 & -6.4 & 0.9 & -3.9 & 2.3 & -5.0 & 1.0 & -3.6 & 1.9 & -8.8 & 6.1 \\
2300 -- 2400 & 2350 & -8.9 & 1.7 & -8.8 & 0.9 & -10.5 & 2.4 & -9.3 & 1.0 & -10.5 & 1.9 & -19.6 & 6.4 \\
2400 -- 2500 & 2450 & -7.5 & 1.7 & -4.7 & 0.9 & -5.7 & 2.4 & -2.3 & 0.9 & -0.4 & 1.9 & 0.3 & 6.7 \\
2500 -- 2600 & 2550 & -0.9 & 1.7 & -4.2 & 0.9 & -4.0 & 2.5 & -3.6 & 0.9 & -5.1 & 1.9 & -14.1 & 7.0 \\
2600 -- 2700 & 2649 & -4.9 & 1.8 & -3.3 & 0.9 & -6.6 & 2.6 & -3.2 & 0.9 & -3.7 & 1.9 & -2.4 & 7.4 \\
2700 -- 2800 & 2749 & 1.5 & 1.9 & -2.1 & 1.0 & 5.2 & 2.8 & -3.8 & 0.9 & 1.9 & 2.0 & 16.4 & 7.9 \\
2800 -- 2900 & 2849 & 2.5 & 2.1 & 0.2 & 1.0 & -0.2 & 3.0 & -0.7 & 1.0 & -5.4 & 2.1 & -3.9 & 8.4 \\
2900 -- 3000 & 2946 & -7.8 & 2.3 & -2.3 & 1.1 & -5.5 & 3.2 & -2.2 & 1.0 & 0.8 & 2.2 & 16.9 & 9.0 \\
\hline
\end{tabular}
\caption[\TE{} multifrequency band powers]{
\TE{} multifrequency band power measurements, $D_b$, and associated uncertainties, $\sigma_b$, (both in units of $\mu$K$^2$) for a given angular multipole range and the window function-weighted multipole $\ell_\mathrm{eff}$.
The data have been minorly updated from \citetalias{dutcher21}.}
\label{tab:te_bandpowers_table}
\end{table*}

\begin{table*}[h!]
\footnotesize
\setlength{\tabcolsep}{2.5pt}
\def\arraystretch{1.2}
\centering
\begin{tabular}{|c c | D{.}{.}{1} D{.}{.}{-1} | D{.}{.}{1} D{.}{.}{-1} | D{.}{.}{1} D{.}{.}{-1} | D{.}{.}{1} D{.}{.}{-1} | D{.}{.}{1} D{.}{.}{-1} | D{.}{.}{1} D{.}{.}{-1}|}
\hline
\rule{0pt}{3ex} \multirow{2}{*}{$\ell$ Range} & \multirow{2}{*}{$\ell_\mathrm{eff}$} & \multicolumn{2}{c|}{$\mathrm{95\times95\,GHz}$} & \multicolumn{2}{c|}{$\mathrm{95\times150\,GHz}$} & \multicolumn{2}{c|}{$\mathrm{95\times220\,GHz}$} & \multicolumn{2}{c|}{$\mathrm{150\times150\,GHz}$} & \multicolumn{2}{c|}{$\mathrm{150\times220\,GHz}$} & \multicolumn{2}{c|}{$\mathrm{220\times220\,GHz}$} \\\cline{3-14}
 & & D_b & \sigma_b & D_b & \sigma_b & D_b & \sigma_b & D_b & \sigma_b & D_b & \sigma_b & D_b & \sigma_b \\[2pt]
\hline
300 -- 350 & 325 & 13.1 & 1.1 & 12.7 & 1.1 & 11.9 & 1.3 & 13.0 & 1.1 & 12.5 & 1.3 & 11.7 & 2.0 \\
350 -- 400 & 375 & 19.7 & 1.3 & 20.4 & 1.3 & 18.7 & 1.5 & 20.9 & 1.3 & 19.5 & 1.5 & 17.5 & 2.3 \\
400 -- 450 & 425 & 19.0 & 1.2 & 18.7 & 1.1 & 17.7 & 1.3 & 18.9 & 1.1 & 18.1 & 1.3 & 17.2 & 2.1 \\
450 -- 500 & 475 & 11.2 & 0.7 & 11.9 & 0.7 & 11.0 & 0.9 & 12.4 & 0.7 & 10.9 & 0.9 & 9.2 & 1.7 \\
500 -- 550 & 524 & 7.1 & 0.5 & 7.2 & 0.4 & 7.5 & 0.6 & 6.9 & 0.4 & 8.1 & 0.6 & 9.1 & 1.5 \\
550 -- 600 & 575 & 11.1 & 0.7 & 11.2 & 0.6 & 12.1 & 0.9 & 11.7 & 0.7 & 11.6 & 0.9 & 11.2 & 1.9 \\
600 -- 650 & 624 & 29.1 & 1.3 & 29.3 & 1.2 & 28.7 & 1.5 & 29.8 & 1.2 & 29.2 & 1.4 & 33.3 & 2.5 \\
650 -- 700 & 674 & 39.0 & 1.5 & 38.9 & 1.3 & 38.9 & 1.7 & 38.5 & 1.4 & 39.0 & 1.7 & 39.7 & 2.9 \\
700 -- 750 & 725 & 33.7 & 1.4 & 34.2 & 1.3 & 32.6 & 1.7 & 34.7 & 1.3 & 33.5 & 1.6 & 31.5 & 2.9 \\
750 -- 800 & 774 & 21.2 & 1.1 & 20.7 & 0.9 & 21.7 & 1.3 & 20.2 & 0.9 & 20.9 & 1.2 & 22.2 & 2.7 \\
800 -- 850 & 824 & 13.2 & 0.8 & 13.3 & 0.6 & 13.0 & 1.0 & 13.6 & 0.6 & 13.1 & 0.9 & 13.2 & 2.5 \\
850 -- 900 & 874 & 16.9 & 0.9 & 17.1 & 0.7 & 17.6 & 1.2 & 16.9 & 0.8 & 17.4 & 1.1 & 18.6 & 2.9 \\
900 -- 950 & 924 & 31.8 & 1.3 & 31.3 & 1.1 & 30.3 & 1.6 & 31.3 & 1.1 & 31.7 & 1.5 & 28.8 & 3.4 \\
950 -- 1000 & 974 & 41.3 & 1.6 & 40.2 & 1.4 & 40.1 & 2.0 & 40.3 & 1.4 & 39.1 & 1.9 & 35.8 & 3.9 \\
1000 -- 1050 & 1024 & 39.4 & 1.6 & 38.2 & 1.3 & 38.7 & 2.0 & 38.1 & 1.4 & 36.6 & 1.9 & 39.6 & 4.1 \\
1050 -- 1100 & 1075 & 26.1 & 1.3 & 26.1 & 1.0 & 24.6 & 1.7 & 26.1 & 1.1 & 24.8 & 1.5 & 19.8 & 3.9 \\
1100 -- 1150 & 1124 & 15.5 & 1.0 & 15.1 & 0.7 & 14.4 & 1.4 & 14.8 & 0.7 & 13.6 & 1.2 & 10.4 & 3.8 \\
1150 -- 1200 & 1174 & 13.1 & 1.0 & 12.2 & 0.7 & 10.7 & 1.4 & 12.5 & 0.7 & 11.8 & 1.2 & 12.2 & 4.0 \\
1200 -- 1250 & 1224 & 20.6 & 1.3 & 21.7 & 0.9 & 23.6 & 1.7 & 21.9 & 1.0 & 21.9 & 1.5 & 17.5 & 4.5 \\
1250 -- 1300 & 1275 & 29.9 & 1.5 & 29.0 & 1.1 & 28.1 & 2.0 & 29.3 & 1.2 & 26.4 & 1.8 & 26.1 & 5.0 \\
1300 -- 1350 & 1325 & 31.2 & 1.6 & 30.7 & 1.1 & 28.1 & 2.1 & 31.8 & 1.2 & 27.9 & 1.9 & 23.7 & 5.4 \\
1350 -- 1400 & 1374 & 24.1 & 1.4 & 22.3 & 1.0 & 21.8 & 2.0 & 22.0 & 1.0 & 24.5 & 1.7 & 38.9 & 5.6 \\
1400 -- 1450 & 1424 & 14.2 & 1.3 & 12.9 & 0.8 & 11.7 & 1.9 & 12.4 & 0.8 & 11.1 & 1.5 & 5.3 & 5.7 \\
1450 -- 1500 & 1474 & 10.9 & 1.3 & 10.1 & 0.7 & 11.3 & 2.0 & 10.3 & 0.8 & 13.2 & 1.5 & 18.7 & 6.1 \\
1500 -- 1550 & 1524 & 15.0 & 1.4 & 15.3 & 0.8 & 12.4 & 2.2 & 14.0 & 0.9 & 10.9 & 1.7 & 7.8 & 6.5 \\
1550 -- 1600 & 1574 & 22.1 & 1.6 & 20.8 & 1.0 & 21.8 & 2.4 & 20.9 & 1.0 & 23.7 & 2.0 & 23.1 & 7.0 \\
1600 -- 1650 & 1624 & 17.6 & 1.7 & 19.9 & 1.0 & 20.2 & 2.6 & 20.5 & 1.1 & 21.3 & 2.1 & 23.3 & 7.4 \\
1650 -- 1700 & 1674 & 19.2 & 1.7 & 18.3 & 1.0 & 14.4 & 2.6 & 17.9 & 1.0 & 18.5 & 2.0 & 12.6 & 7.7 \\
1700 -- 1750 & 1724 & 7.4 & 1.7 & 10.1 & 0.9 & 10.7 & 2.6 & 10.4 & 0.9 & 13.9 & 1.9 & 0.3 & 8.1 \\
1750 -- 1800 & 1775 & 10.1 & 1.7 & 8.7 & 0.9 & 11.1 & 2.7 & 8.4 & 0.9 & 7.9 & 1.9 & 14.5 & 8.6 \\
1800 -- 1850 & 1825 & 8.3 & 1.8 & 9.0 & 0.9 & 5.7 & 2.9 & 9.5 & 0.9 & 5.3 & 2.1 & -0.4 & 9.1 \\
1850 -- 1900 & 1874 & 9.7 & 2.0 & 9.7 & 1.0 & 9.5 & 3.1 & 9.7 & 1.0 & 12.8 & 2.3 & 13.8 & 9.7 \\
1900 -- 1950 & 1924 & 12.7 & 2.1 & 12.8 & 1.1 & 17.9 & 3.3 & 11.8 & 1.1 & 7.6 & 2.4 & 0.6 & 10.3 \\
1950 -- 2000 & 1975 & 12.4 & 2.2 & 10.1 & 1.1 & 8.8 & 3.4 & 11.3 & 1.1 & 13.7 & 2.5 & 6.0 & 10.9 \\
2000 -- 2100 & 2049 & 6.7 & 1.2 & 6.2 & 0.6 & 7.7 & 2.0 & 6.3 & 0.6 & 6.1 & 1.4 & 4.7 & 6.4 \\
2100 -- 2200 & 2148 & 5.3 & 1.3 & 5.5 & 0.7 & 1.0 & 2.2 & 5.3 & 0.6 & 5.2 & 1.5 & 9.1 & 7.2 \\
2200 -- 2300 & 2249 & 7.4 & 1.5 & 7.6 & 0.7 & 6.6 & 2.5 & 5.9 & 0.7 & 7.0 & 1.7 & 8.6 & 8.1 \\
2300 -- 2400 & 2349 & 1.2 & 1.7 & 2.6 & 0.8 & 4.1 & 2.8 & 4.8 & 0.7 & 1.0 & 1.8 & 13.0 & 8.8 \\
2400 -- 2500 & 2449 & 6.6 & 1.9 & 4.0 & 0.9 & 5.2 & 3.0 & 2.6 & 0.8 & 5.1 & 1.9 & -0.9 & 9.7 \\
2500 -- 2600 & 2549 & 2.7 & 2.1 & 2.5 & 0.9 & 0.4 & 3.3 & 2.6 & 0.9 & 3.0 & 2.1 & -2.5 & 10.6 \\
2600 -- 2700 & 2649 & 5.8 & 2.3 & 0.5 & 1.0 & 0.0 & 3.7 & 2.3 & 0.9 & 2.2 & 2.3 & 10.3 & 11.6 \\
2700 -- 2800 & 2749 & -0.8 & 2.6 & 0.8 & 1.2 & 9.1 & 4.1 & 2.0 & 1.0 & 3.5 & 2.6 & -5.3 & 12.8 \\
2800 -- 2900 & 2849 & 0.9 & 3.0 & 3.1 & 1.3 & 4.6 & 4.6 & 0.5 & 1.2 & -3.2 & 2.9 & -6.2 & 14.0 \\
2900 -- 3000 & 2946 & -2.0 & 3.4 & -2.6 & 1.5 & -7.2 & 5.1 & 1.0 & 1.3 & 7.3 & 3.2 & -4.2 & 15.5 \\
\hline
\end{tabular}
\caption[\EE{} multifrequency band powers]{
\EE{} multifrequency band power measurements, $D_b$, and associated uncertainties, $\sigma_b$, (both in units of $\mu$K$^2$) for a given angular multipole range and the window function-weighted multipole $\ell_\mathrm{eff}$.
The data have been minorly updated from \citetalias{dutcher21}.}
\label{tab:ee_bandpowers_table}
\end{table*}

\subsection{Difference Spectra}
\label{app:diff_specs}

We follow \citet{planck15-11} and form difference spectra, $\Delta \hat{D}^{\nu\mu;\kappa\tau} = \hat{D}^{\nu\mu} - \hat{D}^{\kappa\tau}$, where $\hat{D}^{\nu\mu}$ are foreground-subtracted multifrequency band powers.
The covariance of a difference spectrum is $\mathcal{C}^{\Delta \nu\mu ; \kappa\tau} = A\mathcal{C}^{\nu\mu ; \kappa\tau}A^T$, where $\mathcal{C}^{\nu\mu;\kappa\tau}$ is the $2\times 2$ matrix of the relevant covariance blocks and $A=\left(\mathbb{I}, -\mathbb{I}\right)$.

\begin{figure*}[ht!]
  \centering
  \includegraphics[width=\linewidth]{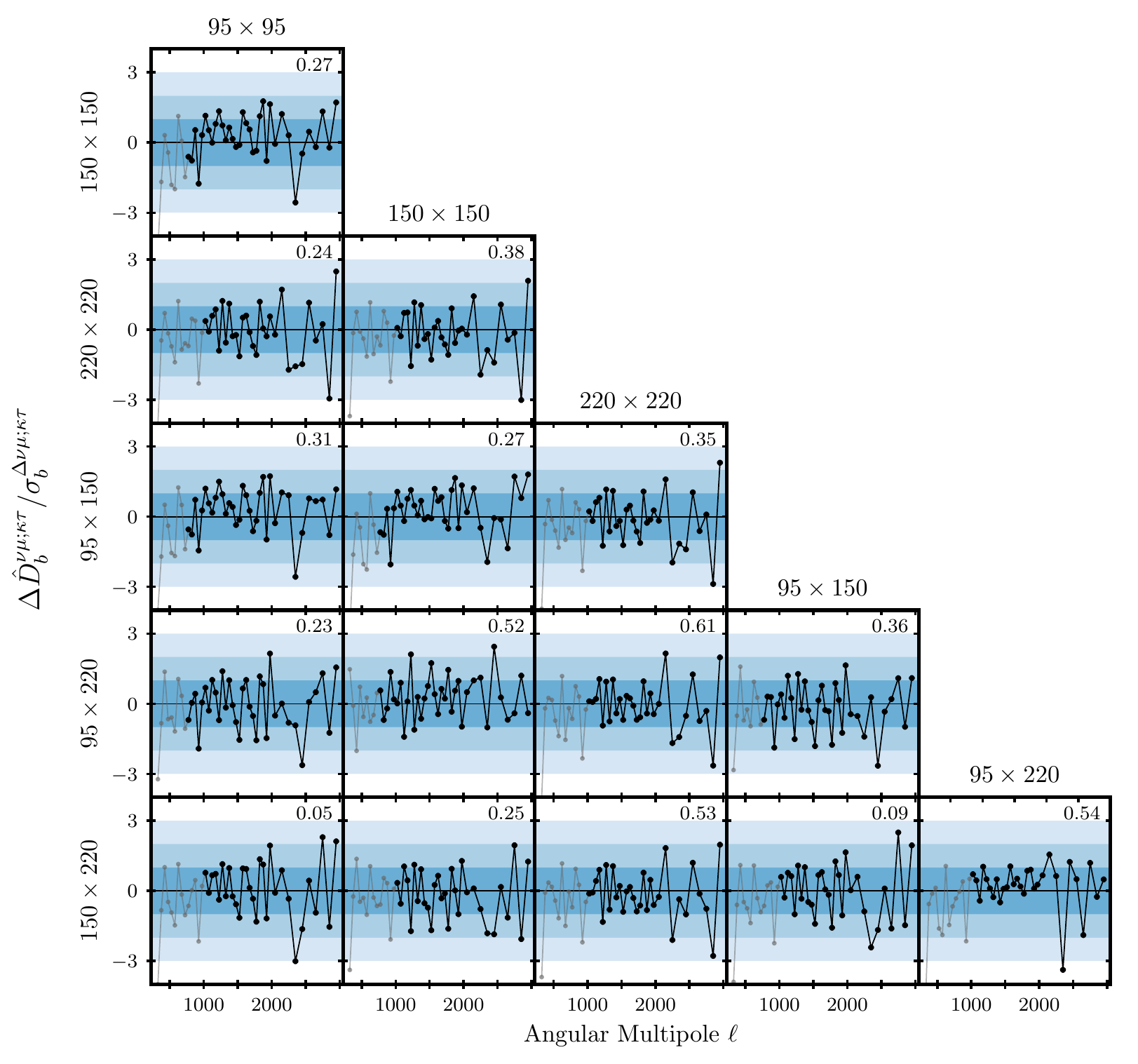}
  \caption{\label{fig:TT_diff_specs}
  Relative \TT{} difference spectra as indicated by the row and column labels, i.e. difference spectra $\Delta\hat{D}_b^{\nu\mu;\kappa\tau}$ divided by the square root of the associated covariance, $\sigma_b^{\Delta\nu\mu;\kappa\tau}$.
  The blue shading indicates the range of $1-3\,\sigma$ fluctuations, while gray indicates data excluded in the analysis.
  We conservatively exclude all \TT{} data at $\ell < 750$.
  This is motivated by the shape of the transfer function, which slowly rises and plateaus at $\ell \approx 750$; the common-mode filter removes \TT{} power on large and intermediate angular scales.
  We further exclude $150\times 220\,\mathrm{GHz}$ and $220\times 220\,\mathrm{GHz}$ \TT{} spectra at $\ell < 1000$, based on our model for correlated atmospheric noise.
  The PTE values are indicated in the top right corner of each panel.
   All PTE values are in the 95th percentile and the multifrequency spectra are in good agreement with one another.
  }
\end{figure*}

\begin{figure*}[ht!]
  \centering
  \includegraphics[width=\linewidth]{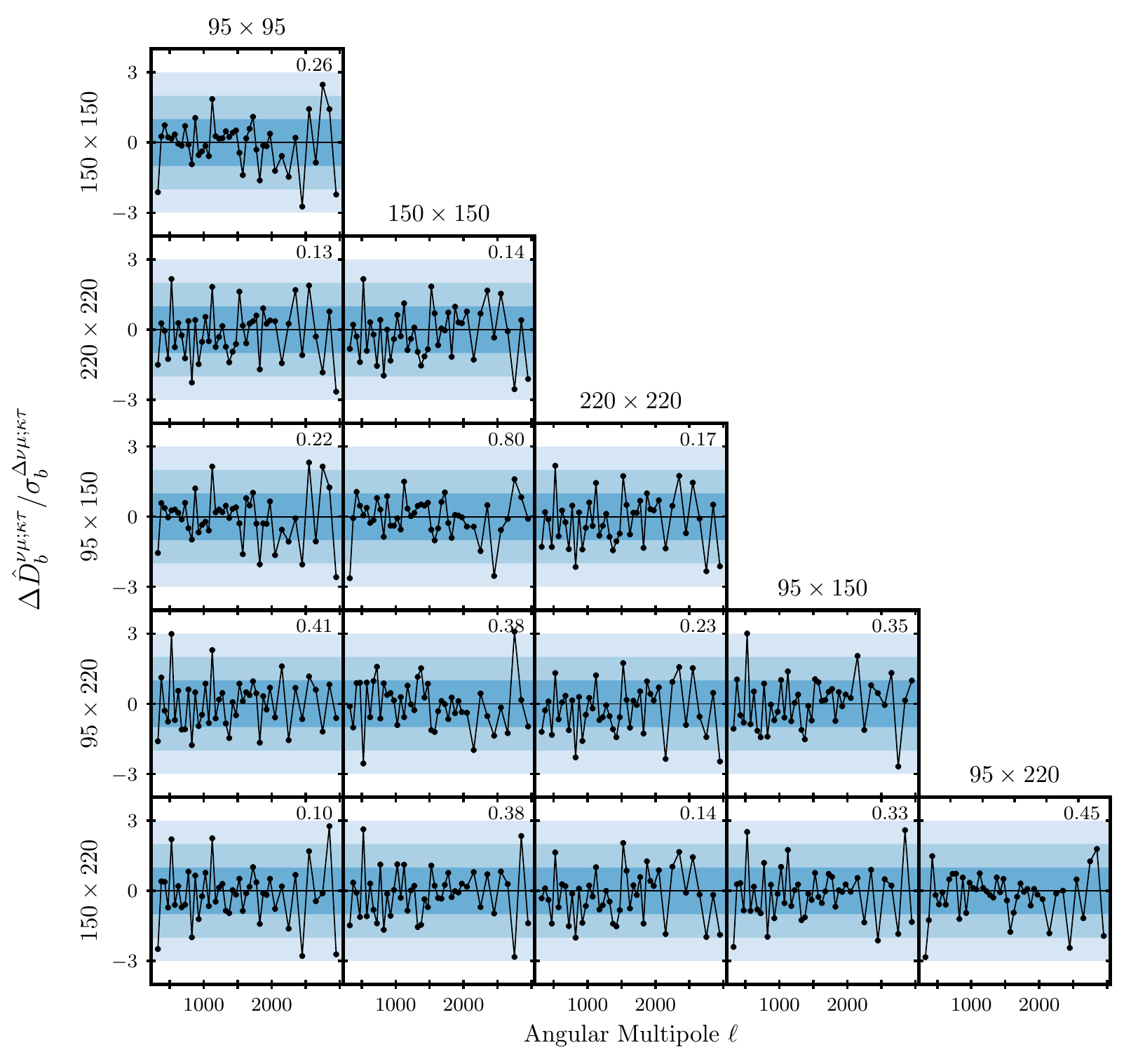}
  \caption{\label{fig:TE_diff_specs}
  Relative \TE{} difference spectra as indicated by the row and column labels, i.e. difference spectra $\Delta\hat{D}_b^{\nu\mu;\kappa\tau}$ divided by the square root of the associated covariance, $\sigma_b^{\Delta\nu\mu;\kappa\tau}$.
  The blue shading indicates the range of $1-3\,\sigma$ fluctuations and PTE values are given in the top right corner of each panel.
   All PTE values are in the 95th percentile and the multifrequency spectra are in good agreement with one another.
  }
\end{figure*}

\begin{figure*}[ht!]
  \centering
  \includegraphics[width=\linewidth]{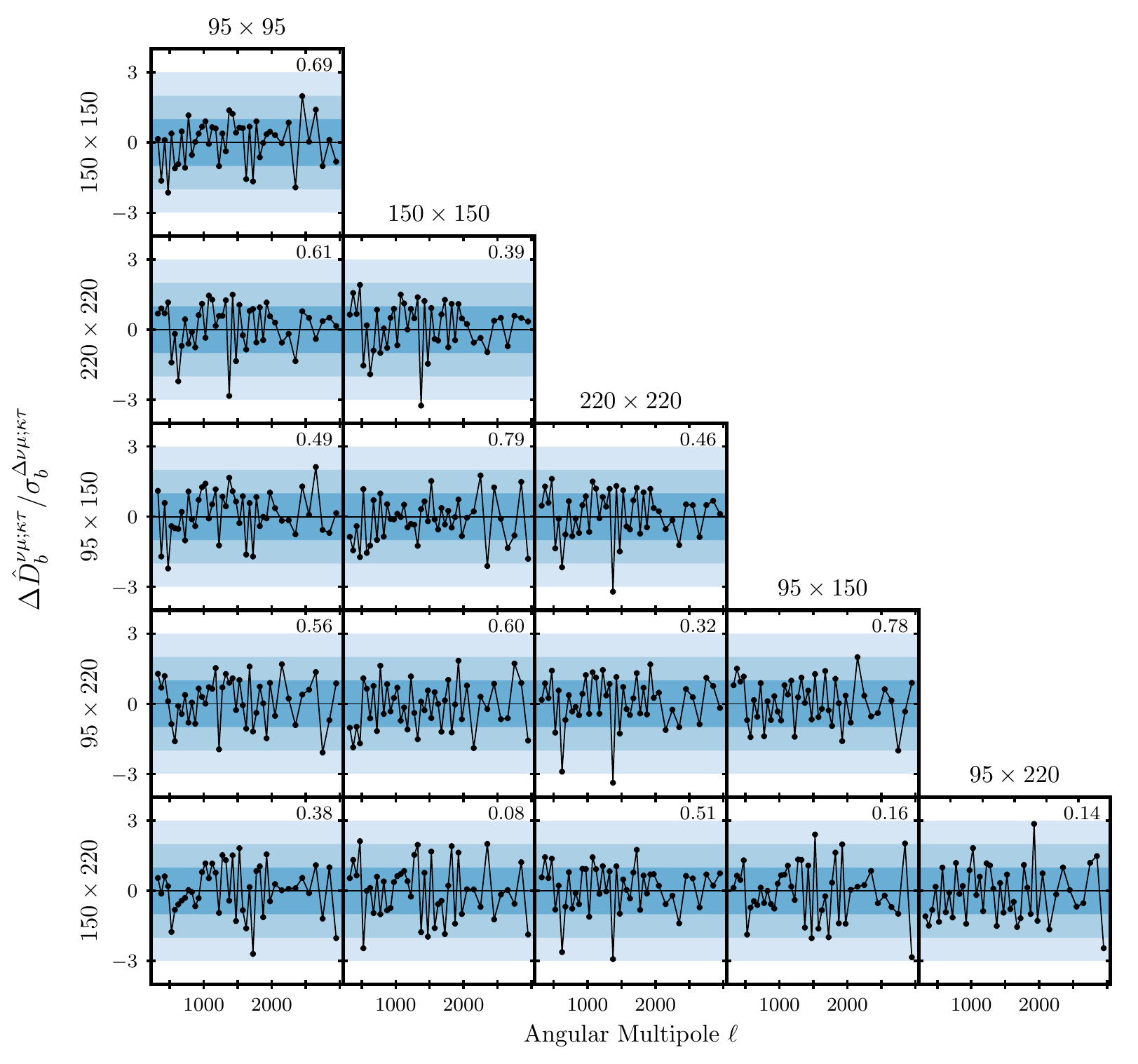}
  \caption{\label{fig:EE_diff_specs}
  Relative \EE{} difference spectra as indicated by the row and column labels, i.e. difference spectra $\Delta\hat{D}_b^{\nu\mu;\kappa\tau}$ divided by the square root of the associated covariance, $\sigma_b^{\Delta\nu\mu;\kappa\tau}$.
  The blue shading indicates the range of $1-3\,\sigma$ fluctuations.
   The multifrequency spectra are in good agreement with one another, as evidenced by the PTE values (given in the top right corner of each panel) which all lie in the 95th percentile.
  }
\end{figure*}

We show the \TT{}, \TE{}, and \EE{} difference spectra in figures \ref{fig:TT_diff_specs}, \ref{fig:TE_diff_specs}, and \ref{fig:EE_diff_specs}, respectively.
While we observe no significant features, such as slopes, constant offsets, or signal leakage, the \TT{} difference spectra show a dip at $\ell\approx 2350$.
This is caused by a bifurcation of the multifrequency spectra over a region of $\Delta\ell\approx 300$ width, with higher frequencies seeing a stronger signal.
This feature is not present in the polarization spectra.
It is not clear what is causing this bifurcation; for unmodelled foreground contamination, we expect to see a slope in the difference spectra, rather than a well-localized feature.
Ultimately, this feature is not statistically significant: comparing the $45$ difference spectra to zero using a $\chi^2$ statistic, the lowest PTE value is $5\%$ ($150\times 220\,\mathrm{GHz} - 95\times 95\,\mathrm{GHz}$ \TT{}).
We conclude that the difference spectra are consistent with zero and take this as further evidence that the multifrequency band powers are consistent with measuring the same underlying signal.

\subsection{Multifrequency Residuals}
\label{app:full_resids}

\begin{figure*}[ht!]
  \centering
  \includegraphics[width=\linewidth]{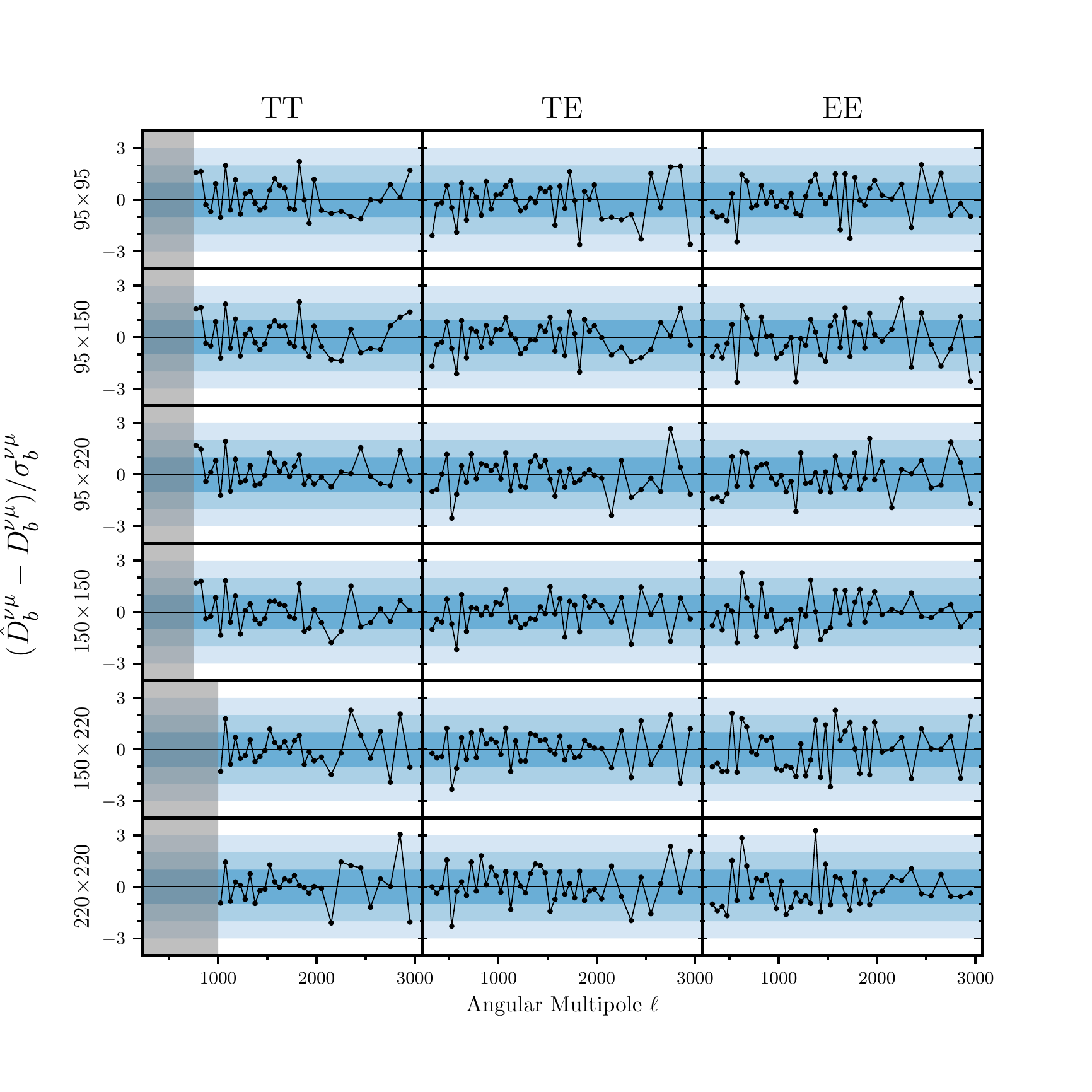}
  \caption{\label{fig:full_resids}
  Relative residuals of the SPT-3G 2018 \TTTEEE{} multifrequency band powers to the best-fit \lcdm{} model, i.e. difference between the SPT-3G data and the model prediction scaled by the error bar of the band powers measurement.
  The blue shading indicates the range of $1-3\,\sigma$ fluctuations.
  Note that the SPT-3G band powers are correlated by up to $40\%$ for neighboring bins.
  The residuals are consistent with zero and the standard model provides a good fit to the data.
  }
\end{figure*}

We show the residuals of the SPT-3G 2018 \TTTEEE{} multifrequency band powers to the best-fit \lcdm{} model in Figure \ref{fig:full_resids}.

\end{appendices}

\end{document}